\documentclass[twoside,twocolumn,9pt]{article}
\usepackage{extsizes}
\usepackage[super,sort&compress,comma]{natbib} 
\usepackage[version=3]{mhchem}
\usepackage[left=1.5cm, right=1.5cm, top=1.785cm, bottom=2.0cm]{geometry}
\usepackage{balance}
\usepackage{mathptmx}
\usepackage{sectsty}
\usepackage{graphicx} 
\usepackage{lastpage}
\usepackage[format=plain,justification=justified,singlelinecheck=false,font={stretch=1.125,small,sf},labelfont=bf,labelsep=space]{caption}
\usepackage{float}
\usepackage{fancyhdr}
\usepackage{fnpos}
\usepackage[english]{babel}
\addto{\captionsenglish}{%
  
}
\usepackage{array}
\usepackage{droidsans}
\usepackage{charter}
\usepackage[T1]{fontenc}
\usepackage[usenames,dvipsnames]{xcolor}
\usepackage{setspace}
\usepackage[compact]{titlesec}
\usepackage{hyperref}
\usepackage{epstopdf}%This line makes .eps figures into .pdf - please comment out if not required.

\usepackage{xcolor}

\definecolor{cream}{RGB}{222,217,201}
\graphicspath{{figures/}}
\usepackage{subcaption}
\usepackage{amsmath,bm}
\usepackage{amsbsy}
\usepackage[utf8]{inputenc}
\begin{document}

\pagestyle{fancy}
\thispagestyle{plain}
\fancypagestyle{plain}{
%%%HEADER%%%
\renewcommand{\headrulewidth}{0pt}
}
%%%END OF HEADER%%%

%%%PAGE SETUP - Please do not change any commands within this section%%%
\makeFNbottom
\makeatletter
\renewcommand\LARGE{\@setfontsize\LARGE{15pt}{17}}
\renewcommand\Large{\@setfontsize\Large{12pt}{14}}
\renewcommand\large{\@setfontsize\large{10pt}{12}}
\renewcommand\footnotesize{\@setfontsize\footnotesize{7pt}{10}}
\makeatother

\renewcommand{\thefootnote}{\fnsymbol{footnote}}
\renewcommand\footnoterule{\vspace*{1pt}% 
\color{cream}\hrule width 3.5in height 0.4pt \color{black}\vspace*{5pt}} 
\setcounter{secnumdepth}{5}

\makeatletter 
\renewcommand\@biblabel[1]{#1}            
\renewcommand\@makefntext[1]% 
{\noindent\makebox[0pt][r]{\@thefnmark\,}#1}
\makeatother 
\renewcommand{\figurename}{\small{Fig.}~}
\sectionfont{\sffamily\Large}
\subsectionfont{\normalsize}
\subsubsectionfont{\bf}
\setstretch{1.125} %In particular, please do not alter this line.
\setlength{\skip\footins}{0.8cm}
\setlength{\footnotesep}{0.25cm}
\setlength{\jot}{10pt}
\titlespacing*{\section}{0pt}{4pt}{4pt}
\titlespacing*{\subsection}{0pt}{15pt}{1pt}
%%%END OF PAGE SETUP%%%

%%%FOOTER%%%
\fancyfoot{}
\fancyfoot[LO,RE]{\vspace{-7.1pt}\includegraphics[height=9pt]{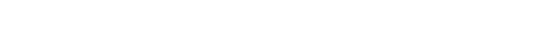}}
\fancyfoot[CO]{\vspace{-7.1pt}\hspace{13.2cm}\includegraphics{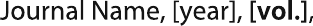}}
\fancyfoot[CE]{\vspace{-7.2pt}\hspace{-14.2cm}\includegraphics{head_foot/RF}}
\fancyfoot[RO]{\footnotesize{\sffamily{1--\pageref{LastPage} ~\textbar  \hspace{2pt}\thepage}}}
\fancyfoot[LE]{\footnotesize{\sffamily{\thepage~\textbar\hspace{3.45cm} 1--\pageref{LastPage}}}}
\fancyhead{}
\renewcommand{\headrulewidth}{0pt} 
\renewcommand{\footrulewidth}{0pt}
\setlength{\arrayrulewidth}{1pt}
\setlength{\columnsep}{6.5mm}
\setlength\bibsep{1pt}
%%%END OF FOOTER%%%

%%%FIGURE SETUP - please do not change any commands within this section%%%
\makeatletter 
\newlength{\figrulesep} 
\setlength{\figrulesep}{0.5\textfloatsep} 

\newcommand{\topfigrule}{\vspace*{-1pt}% 
\noindent{\color{cream}\rule[-\figrulesep]{\columnwidth}{1.5pt}} }

\newcommand{\botfigrule}{\vspace*{-2pt}% 
\noindent{\color{cream}\rule[\figrulesep]{\columnwidth}{1.5pt}} }

\newcommand{\dblfigrule}{\vspace*{-1pt}% 
\noindent{\color{cream}\rule[-\figrulesep]{\textwidth}{1.5pt}} }

\makeatother
%%%END OF FIGURE SETUP%%%

%%%TITLE, AUTHORS AND ABSTRACT%%%
\twocolumn[
  \begin{@twocolumnfalse}
% {\includegraphics[height=30pt]{head_foot/SM}\hfill\raisebox{0pt}[0pt][0pt]{\includegraphics[height=55pt]{head_foot/RSC_LOGO_CMYK}}\\[1ex]
% \includegraphics[width=18.5cm]{head_foot/header_bar}}\par
% \vspace{1em}
% \sffamily
% \begin{tabular}{m{4.5cm} p{13.5cm} }

% \includegraphics{head_foot/DOI} & 
\noindent\LARGE{Continuum elastic models for force transmissions in biopolymer gels} \\%Article title goes here instead of the text "This is the title"
 \vspace{0.3cm} \\
%  & \noindent\large{Haiqin Wang\textit{$^{a,b}$}, and Xinpeng Xu\textit{$^{b,a}$}$^{\ast}$} \\%Author names go here instead of "Full name", etc.
\noindent\large{Haiqin Wang\textit{$^{a,b}$}, and Xinpeng Xu\textit{$^{b,a}$}$^{\ast}$} 
\\%Author names go here instead of "Full name", \emph{etc}.

%% \ast \ddag

% \includegraphics{head_foot/dates} & 
\noindent\normalsize{We review continuum elastic models for the transmission of both \textcolor{black}{external forces and internal active cellular forces} in biopolymer gels, and relate them to recent experiments. Rather than being exhaustive, we focus on continuum elastic models for small affine deformations and intend to provide a systematic continuum method and some analytical  perspectives to the study of force transmissions in biopolymer gels. We start from a very brief review of the nonlinear mechanics of individual biopolymers and a summary of constitutive models for the nonlinear elasticity of biopolymer gels. We next show that the simple 3-chain model can give predictions that well fit the shear experiments of some biopolymer gels, including the effects of strain-stiffening and negative normal stress. We then review continuum models for the transmission of \textcolor{black}{internal active forces} that are induced by a spherically contracting cell embedded in a three-dimensional biopolymer gel. Various scaling regimes for the decay of cell-induced displacements are identified for linear isotropic and anisotropic materials, and for biopolymer gels with nonlinear compressive-softening and strain-stiffening elasticity, respectively. After that, we present (using an energetic approach) the generic and unified continuum theory proposed in [Ben-Yaakov \emph{et al., Soft Matter}, 2015, \textbf{11}, 1412] about how the transmission of forces in the biogel matrix can mediate long-range interactions between cells with mechanical homeostasis. We show the predictions of the theory in a special hexagonal multicellular array, and relate them to recent experiments. Finally, we conclude this paper with comments on the limitations and outlook of continuum modeling, and highlight the needs of complementary theoretical approaches such as discrete network simulations to force transmissions in biopolymer gels and phenomenological active gel theories for multicellular systems.
} \\
%The abstract goes here instead of the text "The abstract should be...Any references in the abstract should be written out in full \textit{\emph{e.g.}}\ [Surname \textit{et al., Journal Title}, 2000, \textbf{35}, 3523]."

%\end{tabular}

 \end{@twocolumnfalse} \vspace{0.6cm}

  ]
%%%END OF TITLE, AUTHORS AND ABSTRACT%%%

%%%FONT SETUP - please do not change any commands within this section
\renewcommand*\rmdefault{bch}\normalfont\upshape
\rmfamily
\section*{}
\vspace{-1cm}

%%%FOOTNOTES%%%

\footnotetext{\textit{$^{a}$~Technion -- Israel Institute of Technology, Haifa, 32000, Israel.}}
\footnotetext{\textit{$^{b}$~Physics Program, Guangdong Technion -- Israel Institute of Technology, Shantou, Guangdong 515063, China.}}
\footnotetext{\textit{$^{\ast}$~Correspondence author, E-mail: xu.xinpeng@gtiit.edu.cn}}

%Please use \dag to cite the ESI in the main text of the article.
%If you article does not have ESI please remove the the \dag symbol from the title and the footnotetext below.
%\footnotetext{\dag~Electronic Supplementary Information (ESI) available: [details of any supplementary %information available should be included here]. See DOI: 10.1039/cXsm00000x/}
%additional addresses can be cited as above using the lower-case letters, c, d, e... If all authors are %from the same address, no letter is required

%\footnotetext{\ddag~Additional footnotes to the title and authors can be included \textit{\emph{e.g.}}\ `Present %address:' or `These authors contributed equally to this work' as above using the symbols: \ddag, %\textsection, and \P. Please place the appropriate symbol next to the author's name and include a %\texttt{\textbackslash footnotetext} entry in the the correct place in the list.}

%%%END OF FOOTNOTES%%%

\newcommand{\gsim}{\raisebox{-0.13cm}{~\shortstack{$>$ \\[-0.07cm] $\sim$}}~} 
\newcommand{\lsim}{\raisebox{-0.13cm}{~\shortstack{$<$ \\[-0.07cm] $\sim$}}~} 

%\newcommand{\gsim}{{\;\raise0.3ex\hbox{$>$\kern-0.75em\raise-1.1ex\hbox{$\sim$}}\;}}
%\newcommand{\lsim}{{\;\raise0.3ex\hbox{$<$\kern-0.75em\raise-1.1ex\hbox{$\sim$}}\;}}

%%%MAIN TEXT%%%%
% The main text of the article\cite{Mena2000} should appear here.
%============Main text==================================%
\section{Introduction}\label{sec:Introduction}
%=======================================================%
% Crosslinked semi-flexible polymer networks are ubiquitous in both the cytoskeleton and
% the extracellular matrix. 

\begin{figure}
  \centering
  \includegraphics[clip=true, viewport=1 1 550 550, keepaspectratio, width=0.3\textwidth]{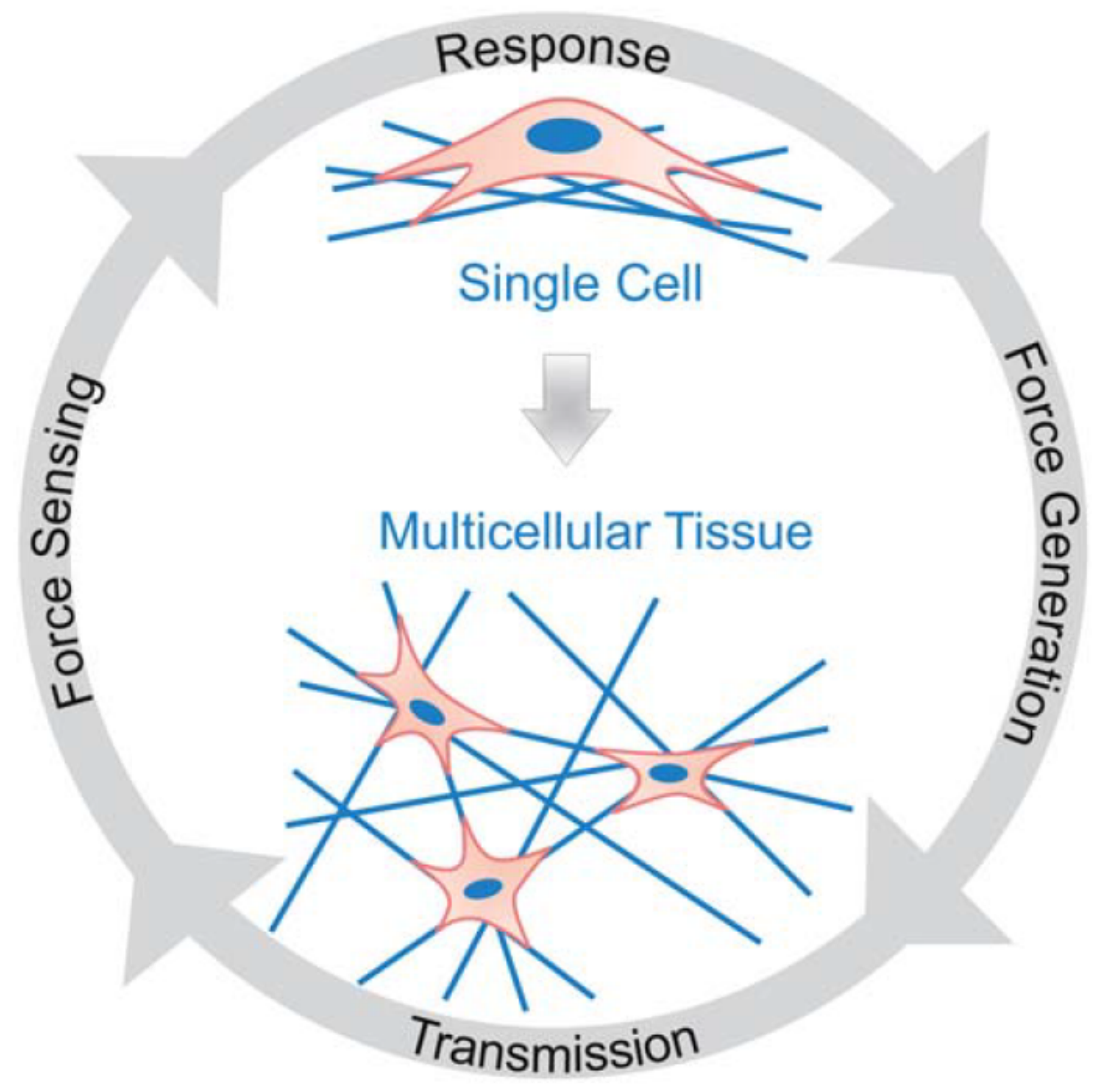}
    \caption{(Color online) Schematic illustration of mechanical interactions between cell and matrix, and matrix-mediated cell-cell interactions. In a multicellular tissue, cells are linked by cell-cell contacts or via the ECM. The self-organization of cells strongly depends on the properties of the surrounding matrix as well as cell contractility and cell density. The forces generated by individual cells are transmitted across the matrix and modify the structure and mechanical properties of the whole tissue, increasing the interaction range beyond that of a single cell. Other cells sense these changes and respond accordingly, leading to a mechanical feedback loop, self-organization and collective behaviors. Reproduced from Kollmannsberger \emph{et al.} \cite{Fratzl2011} with permission from the Royal Society of Chemistry. }
    \label{fig:introduction-cellmatrixcell}
\end{figure}

Cells in animal tissues are surrounded by a complex multicomponent biopolymer gel, composed of water, proteins and polysaccharides and known as extracellular matrix (ECM)\cite{Alberts2007,Frantz2010}. Some animal cells can actively adhere to the ECM and are mechanically connected to it by some discrete protein complexes, the so-called focal adhesions\cite{Alberts2007,Fratzl2011,Mohammadi2014,Sam2013a}. Cells can generate physical forces by the actomyosin cytoskeleton, for example, via myosin motor activity (the contractile forces), and/or actin polymerization (the protrusion forces)\cite{Harris1981,Fratzl2011,Mohammadi2014,Sam2013a}. When these forces are transmitted into ECM through focal adhesions, they can induce deformations and/or flows, alter the structure and mechanical properties of the ECM in the proximity of the cell \cite{Janmey2013,Fratzl2011,Mohammadi2014,Jones2015}, such as fiber alignment and matrix compaction, etc. The deformation fields and structural or other mechanical changes that are induced either by cells themselves or by external mechanical perturbations can, in turn, trigger cells' special sensory systems that enable them to sense and respond to the mechanical signals in their surrounding matrix\cite{Janmey2005,Janmey2013,Fratzl2011,Mohammadi2014,Jones2015}. As such, cell-ECM mechanical interactions play important roles at both single cell level and multicellular tissue level (as schematically shown in Fig.~\ref{fig:introduction-cellmatrixcell}), such as in cell migration\cite{Lo2000,Wang2012}, proliferation\cite{Wang2012,Lesman2014,Xu2017}, differentiation\cite{Lo2000,Xu2017,Discher2006,Janmey2005}, cancer invasion\cite{Friedl1997,Liphardt2014,Shenoy2017}, mineral deposition\cite{Duncan1995}, and self-organization of cells in tissues \cite{Fratzl2011,Janmey2019Tissue,Stefan2019}.  

The physical, topological, and biochemical composition of ECM are not only complex, tissue-specific, but are also markedly heterogeneous\cite{Alberts2007,Frantz2010}. In most \emph{in vitro} experiments for the study of force transmission in the matrix, only one or two types of constituting proteins (such as collagen, fibrin, and elastin) are extracted from ECM to form cross-linked biopolymer gels\cite{Alberts2007,Frantz2010,Janmey2013,Koenderink2010,Koenderink2019}. Recent experiments have shown that the displacements and structure changes induced by cells adhered to biopolymer gels can reach a distance of tens of cell diameters away\cite{Harris1981,Janmey2009,Shenoy2016,NotbohmLesman2015,Korff1999}, in contrast to the distance of several cell diameters reached by displacements induced by cells on linear synthetic gels. Such phenomena support long-range cell-cell mechanical communication, a process that can mechanically couple distant cells and coordinate processes such as capillary sprouting\cite{Korff1999} and synchronous beating\cite{Tzlil2016}. The long-range transmission of cellular forces is usually attributed to the unique nonlinear mechanics and fibrous nature of the biopolymer gels\cite{Janmey2013,Janmey2009,Qi2013,MacKintosh2014}. 

\begin{figure}[htbp]
  \centering
  \includegraphics[width=1\linewidth]{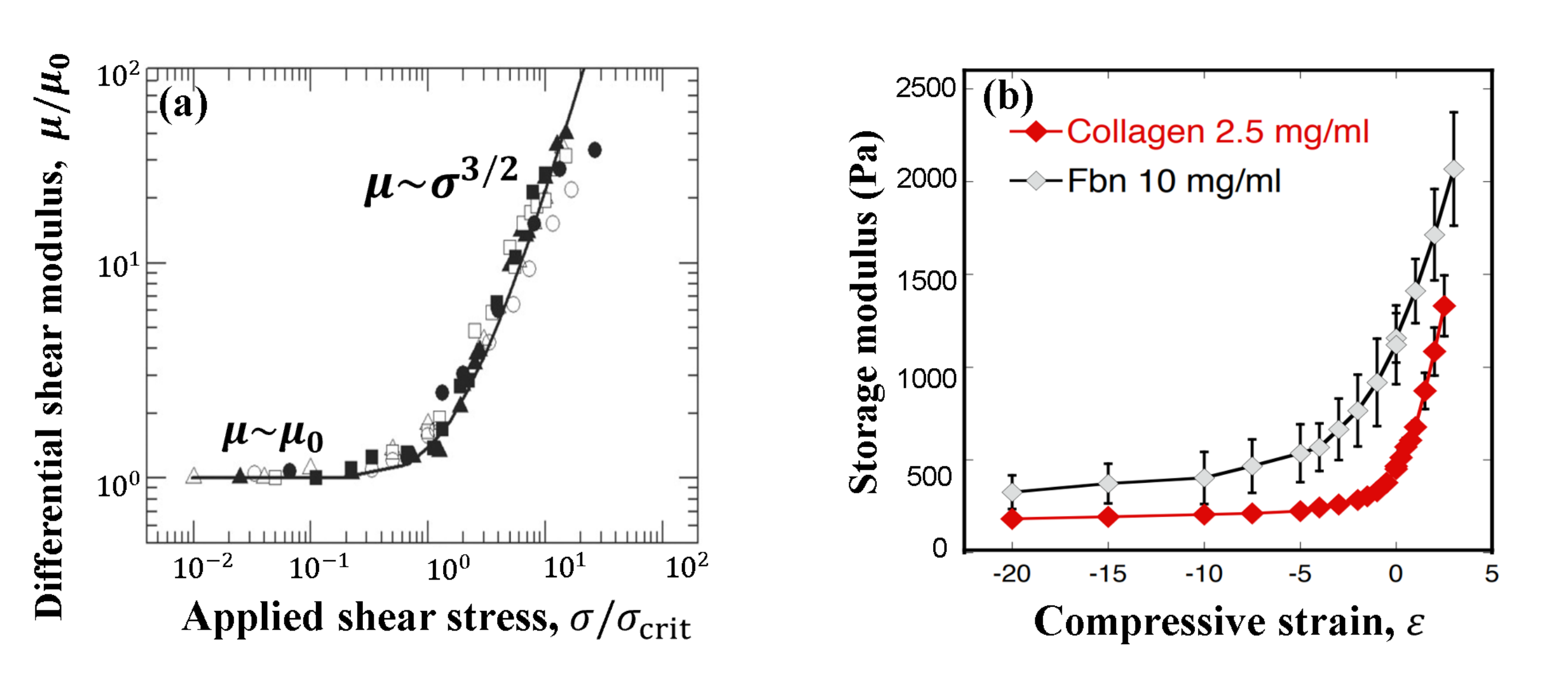} 
  \caption{(Color online) Nonlinear elasticity of biopolymer gels. (a) Shear-stiffening of biopolymer gels. The differential shear modulus $\mu$ (normalized by zero-strain modulus $\mu_0$) as a function of applied steady shear stress $\sigma$ (normalized by the critical stress $\sigma_{\rm crit}$). Below $\sigma_{\rm crit}$, the gel is linear and $\mu\sim \mu_0$ is almost constant; above $\sigma_{\rm crit}$, $\mu$ increases dramatically and stiffens by following a $3/2$-power law (\emph{i.e.}, $\mu \sim \sigma^{3/2}$) until the gel breaks. Reproduced from Gardel \emph{et al}.\cite{GardelMacKintosh2004} with permission from AAAS. (b) Compressive-softening of biopolymer gels. The storage modulus of collagen and fibrin gels as a function of axial strain. The gels show significant softening response at very small compression. Reproduced from Oosten \emph{et al}.\cite{Janmey2016} with permission from Nature Publishing Group. }
\label{fig:introduction-GelNonlinearity}   
\end{figure}

A striking feature of biopolymer gels is their asymmetric elastic response to extension (or shear) and compression\cite{MacKintosh2014}: they stiffen (increase shear modulus) when they are increasingly extended or sheared\cite{MacKintosh2014,GardelMacKintosh2004,Lubensky2005}, but soften (decrease shear modulus) when they are compressed\cite{Janmey2016,MacKintoshJanmey2016,Kim2014,Kim2016,Xu2017PRE}. For shear stresses that exceed a small critical value (corresponding to small strains about 5\%-10\%\cite{MacKintosh2014,GardelMacKintosh2004,Lubensky2005}), biopolymer gels show a power-law stiffening of the elastic modulus with increasing stress, where the elastic modulus in some of these gels increases as the 3/2 power of the applied stress\cite{MacKintosh2014,GardelMacKintosh2004,Lubensky2005} as shown in Fig.~\ref{fig:introduction-GelNonlinearity}(a). Recently, collagen-I gels are found to stiffen linearly (instead of 3/2 power) with the applied stress\cite{MacKintosh2015}. In contrast to their behavior under shear, biopolymer gels show nonlinear strain softening upon compression \cite{Janmey2016,MacKintoshJanmey2016,Kim2014,Kim2016}. For compressive stresses exceeding a very small critical value, biopolymer gels can completely lose their resistance to shear stress\cite{Janmey2016,MacKintoshJanmey2016,Kim2014,Kim2016,Xu2017PRE} as shown in Fig.~\ref{fig:introduction-GelNonlinearity}(b). Such nonlinear shear-stiffening and compressive-softening can be attributed to the microstructural nonlinearities of the constituent, semiflexible filaments which stiffen under extension and soften (due to buckling) under compression\cite{MacKintosh2014,Xu2017PRE}. 

When cells are embedded in such fibrous biopolymer gels, they pull on the gel and produce non-equilibrium forces by myosin motors that consume adenosine triphosphate (ATP). These contractile cells can be represented as force dipoles~\cite{Sam2013a} that \textcolor{black}{deform} the gel to produce strains and stresses, which result in effective elastic interactions with other cells. The \textcolor{black}{\emph{internal active forces}} applied by embedded living cells on elastic biopolymer gels are in contrast to the usual engineering view of elastic materials with forces exerted externally on their macroscopic boundaries. Note that the contractile cellular forces applied to biopolymer gels can induce anisotropy in their elasticity, for example, due to fiber stiffening\cite{MacKintosh2014,GardelMacKintosh2004,Lubensky2005}, buckling\cite{MacKintosh2014,NotbohmLesman2015,Notbohm2015,Lenz2015,Xu2015PRE,Xu2017PRE} and collective network responses (\emph{e.g.}, fiber alignment\cite{NotbohmLesman2015,Mahadevan2009,Mao2015,Sander2013}). This strain-induced elastic anisotropy (\emph{i.e.}, unequal principal stiffnesses) has recently been identified as an important mechanism for long-range force transmission and cell-cell mechanical signaling\cite{Xu2015PRE,Xu2020BJ}.

In this paper, we review continuum models for the transmission of both \textcolor{black}{external forces} and \textcolor{black}{internal active cellular forces} in biopolymer gels. We only consider the elastic responses of biopolymer gels~\cite{NotbohmLesman2015,Sam2013b} and focus on small affine deformations. Note that biopolymer gels are in general viscoelastic or viscoplastic, however at short time scales and small deformations, biopolymer gels behave as elastic materials, and particularly for gels with high connectivity, affine deformation can be further assumed~\cite{Lubensky2011,Xu2015PRE,Xu2020BJ}. We expect that the continuum elastic models reviewed in this work on force transmission in biopolymer gels to be most relevant for environments sparse in cells, such as connective tissues and engineered tissue scaffolds \cite{Janmey2019Tissue,Stefan2019}. \textcolor{black}{In these cases, there are usually rare cell-cell junctions and the effects of excluded volume and molecular signaling can often be neglected.} The interactions between cells that are mostly mediated by \textcolor{black}{the forces transmitting in the matrices} (the biopolymer gels) and regulated by the mechanical homeostasis of individual cells are the dominant mechanisms for the self-organization of the tissue.

This paper is organized as follows. Sec.~\ref{sec:Introduction} introduces mechanical interactions between cells and matrix, and the nonlinear elasticity of biopolymer gels. Sec.~\ref{sec:Single} is devoted to a brief summary of the nonlinear force-strain relations of individual semiflexible biopolymers. In Sec.~\ref{sec:ExtForce}, we then review the classical theory of linear isotropic and anisotropic elasticity, and show their extension based on affine deformations of nonlinear biopolymers to modeling the nonlinear elasticity of biopolymer gels. The nonlinear elastic responses of biopolymer gels to \emph{external forces} such as shear and tensile stresses are calculated and compared to experiments. In Sec.~\ref{sec:IntForce}, we show that the above continuum elastic models can be used to calculate the transmission range of \emph{internal cellular forces} in linear isotropic and anisotropic materials, and in highly nonlinear biopolymer gels. The theoretical predictions are then connected with recent quantitative experiments and finite-element simulations on the decay of displacements induced by spherical cells or cell aggregates in biopolymer gels. In Sec.~\ref{sec:CellCellInt}, a generic and unified continuum theory is reviewed to show how the force transmission in biopolymer gels can mediate long-range cell-cell interactions. The calculations from the ideal spherically symmetric geometry give predictions that are consistent at least quantitatively with experiments about matrix-mediated interactions between cells at various mechanical homeostatic states. Finally, this review is concluded in Sec.~\ref{sec:Conclusion} with a brief summary and a few general remarks and outlook.

% \begin{figure}[htbp]
%   \centering
%   \includegraphics[width=0.83\linewidth]{GelNonlinearity1.pdf}  
%   \caption{Nonlinear shear-stiffening elasticity of biopolymer gels. The differential shear modulus $\mu$ (normalized by zero-strain modulus $\mu_0$) as a function of applied steady shear stress $\sigma$ (normalized by the critical stress $\sigma_{\rm crit}$). Below $\sigma_{\rm crit}$, the gel is linear and $\mu\sim \mu_0$ is almost constant; above $\sigma_{\rm crit}$, $\mu$ increases dramatically and stiffens by following a $3/2$-power law (\emph{i.e.}, $\mu \sim \sigma^{3/2}$) until the gel breaks. Reproduced from Gardel \emph{et al}.\cite{GardelMacKintosh2004} with permission from AAAS.} 
% \label{fig:introduction-GelNonlinearity}(a)   
% \end{figure}

% \begin{figure}[htbp]
%   \centering
%   \includegraphics[width=0.9\linewidth]{GelNonlinearity2.pdf}  
%   \caption{(Color online) Nonlinear compressive-softening elasticity of biopolymer gels. The storage modulus of collagen and fibrin gels as a function of axial strain. The gels show significant softening response even for very small compression. Reproduced from Oosten \emph{et al}.\cite{Janmey2016} with permission from Nature Publishing Group.  } 
% \label{fig:introduction-GelNonlinearity}(b)   
% \end{figure}

\section{Nonlinear elasticity of stiff semiflexible biopolymers} \label{sec:Single}

Biopolymers such as those making up extracellular matrix (a complex biopolymer gel) usually have complex hierarchical structures in contrast to most synthetic polymers\cite{Koenderink2010,MacKintosh2014,Koenderink2019}. They typically consist of globular proteins that are often arranged into bundles of filaments, \emph{e.g.}, collagen and fibrin\cite{Koenderink2010,Koenderink2019}. In comparison to most synthetic polymers, biopolymers are usually far more stiff to bend and their stiffness can be characterized by a dimensionless stiffness parameter $c\equiv \ell_p/\ell_c$. Here $\ell_c$ is the polymer contour length and $\ell_p=\kappa/k_BT$ is the persistence length with $\kappa$ being the bending modulus, $k_B$ the Boltzmann constant, and $T$ the temperature. The persistence length $\ell_p$ can be regarded as the contour length at which significant thermal bending fluctuations occur\cite{MacKintosh2014}. 

In this review, we focus only on biopolymers that are stiff to thermal bending with large stiffness $c>1$ or $\ell_p>\ell_c$. In order to model the transmission of forces in biopolymer gels later, we firstly summarize, in this section, the mechanical responses of individual biopolymers to both stretch and compression. Two theoretical models of biopolymers are often employed, according to the degree of coupling between constituting filaments \cite{MacKintosh2014,Xu2017PRE}:  

(i) \textit{Athermal elastic rod model: $c\gg 1$ or $\ell_p\gg \ell_c$}. The internal monomer-monomer interactions dominate over the configuration entropic effects, and the biopolymer can be treated as an athermal elastic rod. This applies to tightly bundled biopolymers, \emph{e.g.}, collagen. In this case, thermal undulations of the constituent filaments are suppressed due to their close packing arrangement. The bending modulus, $\kappa$, of the biopolymer scales as $\kappa \sim Y a^4$ with $Y$ being the Young's modulus of the elastic rod and $a$ being the bundle radius \cite{Landau1986}. The response to tensile (stretching) and compressive forces are purely elastic with force-strain relation given by
\begin{align}\label{eq:Single-AthermalRod-tauepsilon}
f = 
\begin{cases} 
k\ell_c \epsilon,  &\mbox{if} \quad f >-f_b \\
-f_b + \rho_0 k\ell_c (\epsilon+\epsilon_b), & \mbox{if} \quad f <-f_b
\end{cases},
\end{align}
in which $k \sim Y a^2/\ell_c$ is the biopolymer stiffness, and 
\begin{equation}\label{eq:Single-AthermalRod-EulerInstab}
f_b\equiv \kappa \pi^2/\ell_c^2, \quad \epsilon_b\equiv f_b/k\ell_c \sim \pi^2 a^2/\ell_c^2
\end{equation}
are the magnitude of the Euler critical force and strain for buckling instability \cite{Landau1986}, respectively. Note that the post-buckling behaviors is described, to a good approximation, by a linear force-compression relation\cite{Bazant2010,Xu2015PRE,Xu2017PRE} with $0<\rho_0 < 1$ indicating much smaller stiffness of post-buckling rods than that before buckling. 

\begin{figure}
  \centering
  \includegraphics[clip=true, viewport=1 1 550 360, keepaspectratio, width=0.35\textwidth]{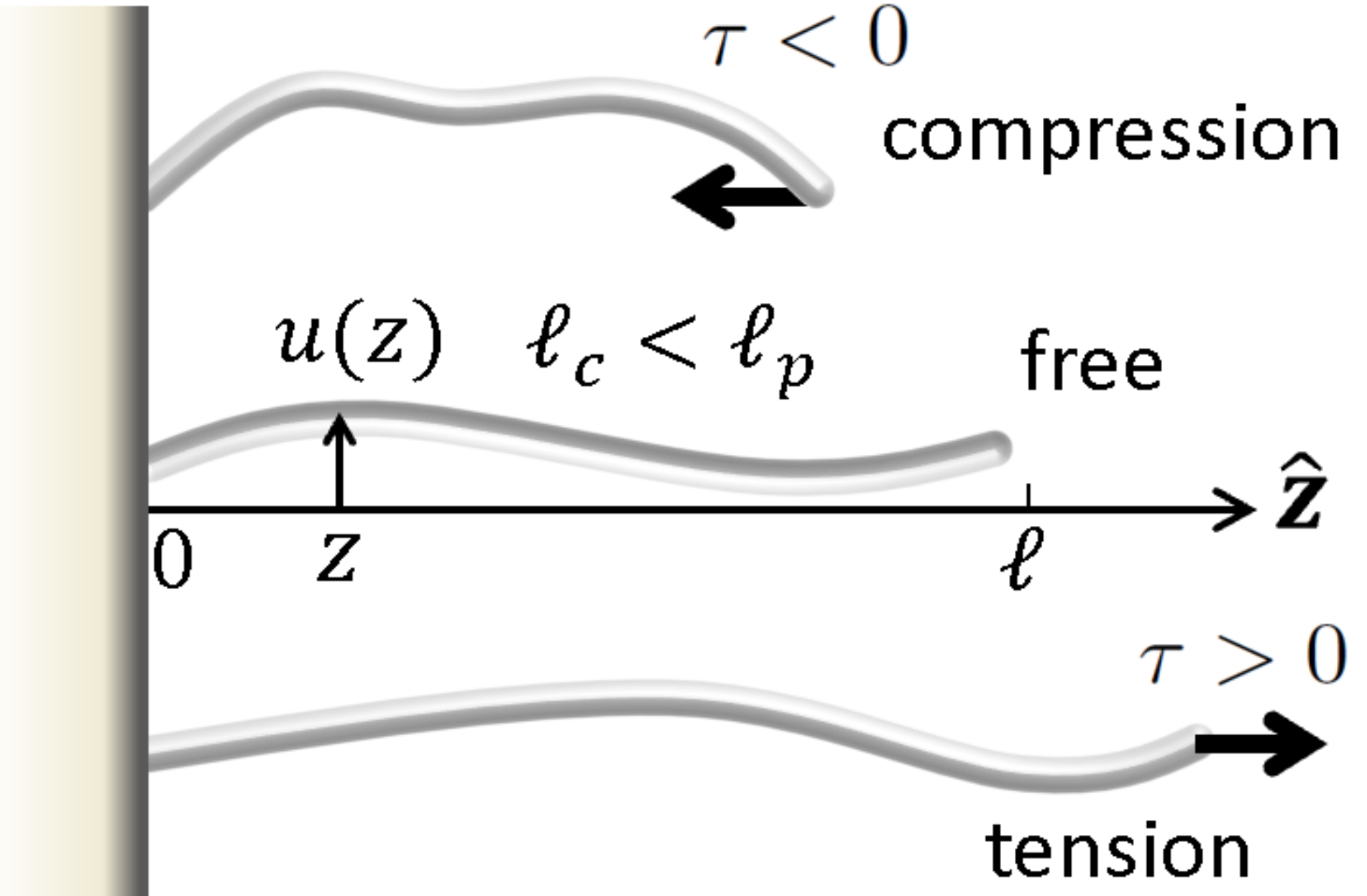}
    \caption{Schematic illustration of a stiff semiflexible biopolymer along the $\hat{\mathbf z}$ direction with dimensionless stiffness parameter $c=\ell_p/\ell_c\gg 1$. Here $\ell_{c}$ is the contour length, $\ell_{p}$ is the persistence length, and $\ell$ is the end-to-end distance of the biopolymer. The transverse biopolymer fluctuation is described by one transverse coordinate $u(z)$ in two dimensions. Compressive forces over a critical value around $f_b=\kappa \pi^2/\ell_c^2$ as shown in Eq.~(\ref{eq:Single-AthermalRod-EulerInstab}) will induce buckling instability and softening of the biopolymer. Tensile forces will stretch out transverse fluctuations and reduce conformational entropy. Restoring forces are then generated, tending to draw the stretched biopolymer back to equilibrium states with larger entropy, and the inextensibility of the biopolymer results in its significant stiffening when the biopolymer length approaches its contour length. }
    \label{fig:Single-semiflexibleChain}
\end{figure}

(ii) \textit{Inextensible, stiff semiflexible or wormlike chain model: $c\gsim 1$ or $\ell_p\gsim \ell_c$}. This applies to biopolymers in bundles with significant solvent, \emph{e.g.}, fibrin. In this case, the polymer elasticity results from the enthalpy arising from bending as well as entropy of conformation thermal fluctuations \cite{Koenderink2010,MacKintosh2014}. The response (stretch and compression) of such biopolymers to applied longitudinal forces is briefly reviewed as follows. For more detailed calculations, we suggest the nice reviews on semiflexible polymers and their networks by Broedersz and MacKintosh\cite{MacKintosh2014} and by Meng and Terentjev\cite{Meng2017}.  

\subsection{Stretching stiffening of stiff biopolymers}\label{sec:Single-stretch}

An inextensible, stiff, semiflexible biopolymer is nearly straight with only small transverse thermal fluctuations\cite{Rubinstein2003, MacKintosh2014}. We define the $z$-axis as the average orientation of the biopolymer and consider the simple case with only one transverse coordinate $u(z)$ describing the transverse chain fluctuations \cite{MacKintosh2014} as schematically shown in Fig.~\ref{fig:Single-semiflexibleChain}. For such a semiflexible chain under a tensile force $f>0$, the Hamiltonian is given by
\begin{equation}\label{eq:Single-Stretch-H}
{\cal H} = \frac{\kappa}{2} \int_0^{\ell} dz \left(\frac{\partial^2 u}{\partial z^2} \right)^2 + \frac{f}{2} \int_0^{\ell} dz \left(\frac{\partial u}{\partial z}\right)^2,
\end{equation}
with $\ell$ being the end-to-end distance of the biopolymer. Note that a finite resistance to bending is the essence of the general worm-like chain model (WLC) for semiflexible polymers\cite{Rubinstein2003, MacKintosh2014}. Tensile forces $f>0$ applied on biopolymers will stretch out transverse fluctuations and reduce conformational entropy and hence increase the free energy. This will result in restoring forces that balance $f$ and try to restore the biopolymers back to their equilibrium states with largest entropy and lowest free energy. 

Applying equipartition to each bending mode in Fourier space, using the mean-field constraint on chain inextensibility, and performing the summation of Fourier modes, the normalized end-to-end distance of the biopolymer $x\equiv \ell/\ell_c$ is expressed as a function of the applied tensile force $\tau\equiv f/f_b>0$ (normalized by Euler buckling force $f_b$): 
\begin{equation}\label{eq:Single-Stretch-xphi}
x = 1-\frac{1}{2c}\frac{\pi\sqrt{\tau}\coth(\pi\sqrt{\tau})-1}{\pi^2\tau}.
\end{equation}
with $f_b$ being defined in Eq.~(\ref{eq:Single-AthermalRod-EulerInstab}), 
%\begin{equation} 
%x = \begin{cases} 1-\frac{1}{2c}\frac{\pi\sqrt{\tau}\coth(\pi\sqrt{\tau})%-1}{\pi^2\tau}, &\mbox{if}\quad \tau>0 \\
%1-\frac{1}{2c}\frac{\pi\sqrt{-\tau}\cot(\pi\sqrt{-\tau})-1}{-\pi^2\tau}, %& \mbox{if} \quad \tau<0 \end{cases},
%\end{equation}
and the (normalized) force-free average polymer end-to-end distance $x_{0}=\ell_{0}/\ell_{c}$ is non-zero given by $x_0=1-1/6c$ when $\tau=0$.

Two typically limiting regimes of the extension-force relation (\ref{eq:Single-Stretch-xphi}) can be identified according to the magnitude of $\tau$ as follows. 

(i) \textit{Small tensile forces: $\tau \ll 1$}. A linear force-extension relation is obtained from the expansion of Eq.~(\ref{eq:Single-Stretch-xphi}) in $\tau$ to the leading order as
\begin{equation}\label{eq:Single-Stretch-xphiLin}
x - x_0 \approx \frac{\pi^2}{90 c} \tau = \frac{\ell_c^3}{90\ell_p \kappa}f.
\end{equation}
That is, in the linear regime (at small deformations around $x_0$), the biopolymer responds with the effective spring constant, $k= df/d(x \ell_c)= 90\kappa \ell_p/\ell_c^4$.

(ii) \textit{Large tensile forces: $\tau \gg 1$}. An expansion of Eq.~(\ref{eq:Single-Stretch-xphi}) in $1/\tau$ to the leading order gives
\begin{equation}\label{eq:Single-Stretch-xphiStiffen}
x \approx 1 - \frac{1}{2c}\frac{1}{\pi \sqrt{\tau}}.
\end{equation} 
As always for theories based on worm-like chain model, the finite extension limit gives the divergent force scaling: $f \sim (1-x)^{-2}$. 

As noted above, the force-extension relation, $\tau(x)$, can be obtained by numerically inverting Eq.~(\ref{eq:Single-Stretch-xphi}). In practice, however, it is often preferable to use more tractable interpolations to the exact force-extension relation\footnote[2]{The idea and its importance of such interpolations of force-extension relation are analogous to the famous van der Waals' equation of state (pressure-volume relation), which interpolates between the two limits of ideal gases and incompressible fluids\cite{Couture2000}}, as is done by Marko and Siggia \cite{MarkoSiggia1995} for relatively soft semiflexible polymers such as DNA with $c\ll 1$ or $\ell_p\ll \ell_c$. Some popular interpolations for inextensible and stiff biopolymers with $c>1$ or $\ell_p> \ell_c$ are summarized as follows. 
%Marko-Siggia's interpolation
%\[
%\tau=\frac{1}{c^{2}}\left[\frac{1}{4\left(1-x\right)^{2}}-\frac{1}{4}+x\r%ight]
%\]
\begin{figure}[htbp]
  \centering
  \includegraphics[width=0.7\linewidth]{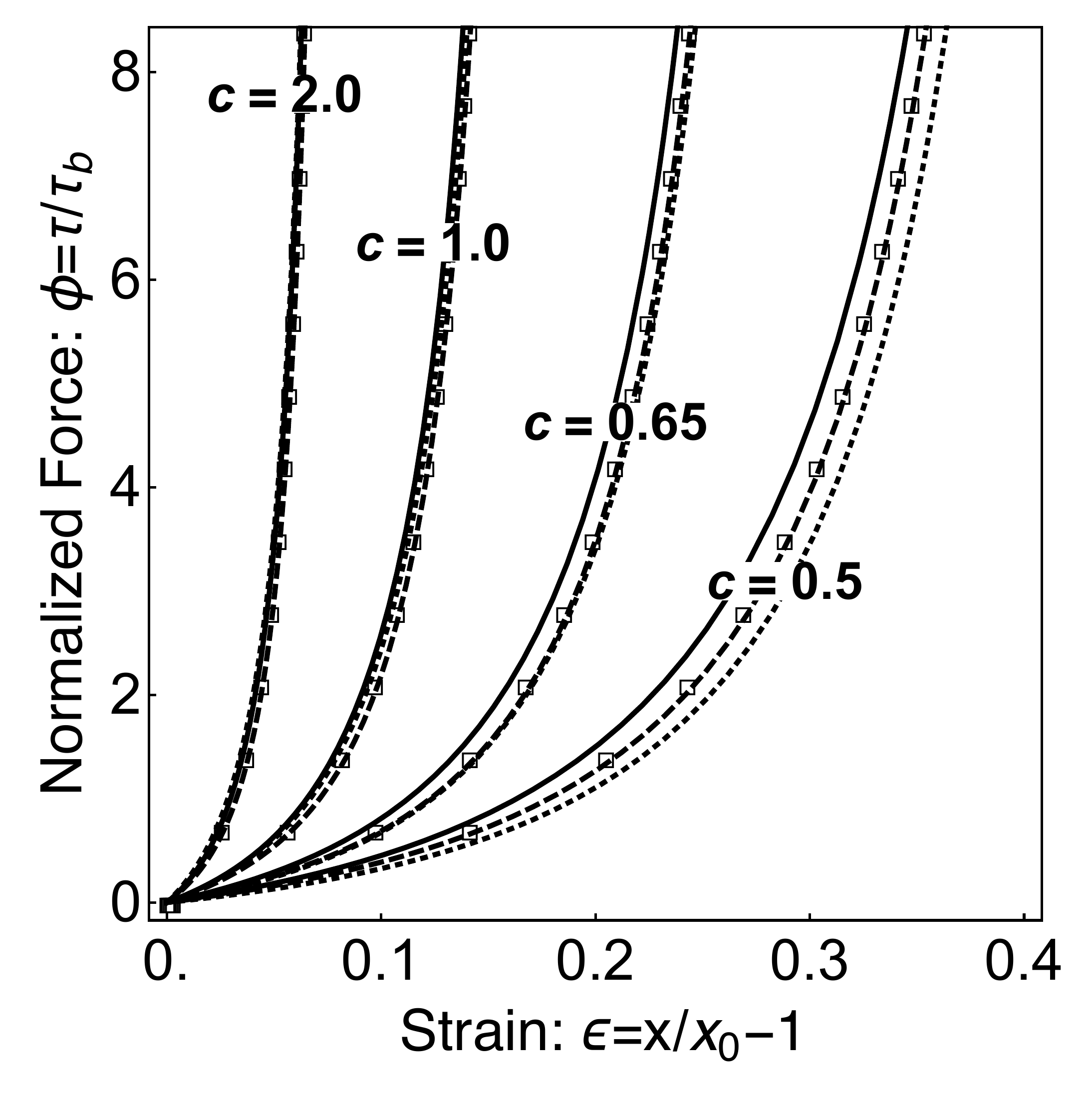}
  \caption{Force-stretch relations of a single semiflexible biopolymer with various stiffness parameters $c=\ell_p/\ell_c$. The discrete data points (open squares) are taken from the analytical relation (\ref{eq:Single-Stretch-xphi}). The three interpolations in Eqs.~(\ref{eq:Single-Stretch-MacKintosh})--(\ref{eq:Single-Stretch-Terentjev}) are represented by the solid lines, the dashed lines and the dotted lines, respectively. The deviation of each line from discrete analytical points shows the accuracy of the corresponding interpolation. For MacKintosh's and Palmer-Boyce's interpolations, the chosen stiffness parameters $c=2.0$, $1.0$, $0.65$, and $0.5$, correspond to stiffening strain $\epsilon_s=0.09$, $0.20$, $0.34$, and $0.50$, respectively. For Terentjev's interpolation, the same $\epsilon_s$ are chosen corresponding to $c=2.25$, $1.18$, $0.80$, and $0.65$, respectively. The nonlinear stiffening becomes significant around $\tau \sim 1$.} 
\label{fig:Single-force-stretch}
\end{figure}

(i) \textit{MacKintosh's interpolation \cite{MacKintosh2014}}: An approximate force-extension relation can be obtained from the above asymptotic limits in Eqs.~(\ref{eq:Single-Stretch-xphiLin}) and (\ref{eq:Single-Stretch-xphiStiffen}) as: 
\begin{equation}\label{eq:Single-Stretch-MacKintosh}
\tau=\frac{9}{\pi^2}\left[\frac{\left(1-x_{0}\right)^{2}}{\left(1-x\right)^{2}}-1-\frac{1}{3}\frac{x-x_{0}}{1-x_{0}}\right]
\end{equation}
with $x_0=1-1/6c$.
%The strain energy is given by the work $w(\epsilon) = %\int_{\ell_0}^{\ell} f(\ell') d\ell' = %\ell_0\int_{0}^{\epsilon} f(\epsilon') d\epsilon'$:
%\begin{equation}\label{eq:single-energy1}
%w(\epsilon)= \frac{1}{2}\alpha\epsilon^2 %\left[\frac{6}{5}(1-b\epsilon)^{-1}- \frac{1}{5}\right],
%\end{equation}
%with $\alpha = 90 k_BT{\ell_p^2\ell_0^2}/{\ell_c^4} \approx %90 k_BT{\ell_p^2}/{\ell_c^2}$.

(ii) \textit{Palmer-Boyce's interpolation}: Palmer and Boyce \cite{PalmerBoyce2008} offer a rather accurate analytical interpolation of the exact relation (\ref{eq:Single-Stretch-xphi}) using $Pad\Grave{e}$ approximation: 
\begin{equation}\label{eq:Single-Stretch-PalmerBoyce}
\tau=\frac{1}{4\pi^2 c^{2}}\frac{1}{\left(1-x\right)^{2}}\frac{1-6c\left(1-x\right)}{1-2c\left(1-x\right)}    
\end{equation}   
with $x_0=1-1/6c$.

(iii) \textit{Terentjev's interpolation}: Within the mean field approximation of global inextensibility, Blundell and Terentjev \cite{Terentjev2009} have found a simple algebraic expression for the force-extension relation as 
\begin{equation}\label{eq:Single-Stretch-Terentjev} 
\tau=x \left(\frac{1-x_0^{2}}{1-x^{2}}\right)^{2}-x.
\end{equation}
Here the (normalized) force-free average polymer end-to-end distance $x_0$ is given by $x_{0}^{2}=1-2/\pi^{3/2}c$, valid in a large range of $c$, which is different from $x_0=1-1/6c$ that is used in Eqs.~(\ref{eq:Single-Stretch-xphi}), (\ref{eq:Single-Stretch-MacKintosh}), and (\ref{eq:Single-Stretch-PalmerBoyce}) and valid only for stiff semiflexible biopolymers with $c\gsim 1$. 
Terentjev's interpolation has an advantage of being fully analytical and captures the right physics across the full range between the flexible limit (Gaussian chains) and stiff limit (rigid elastic rods), as well as in stretch (tension) and compression/buckling regimes up to the highly bent elastica limit. 

Note that for stiff semiflexible biopolymers, the average end-to-end distance, $\ell_0$ (or $x_0$) is non-zero and the strain of the deformed biopolymer can be defined by $\epsilon\equiv (\ell-\ell_{0})/{\ell_{0}}$, which is related to the extension $x=\ell/\ell_c$ by
\begin{equation}\label{eq:Single-Stretch-xepsilon}
\epsilon=\frac{x}{x_{0}}-1, \quad \rm{or}, \quad
x=x_0(1+\epsilon).
\end{equation}  
The maximal \textit{stiffening} strain $\epsilon_s$ at $x=1$ or $\ell=\ell_c$ that accounts for the biopolymer inextensibility \cite{MacKintosh2014} is then given by $\epsilon_s= (\ell_c-\ell_0)/{\ell_0}={1}/{x_0}-1$.  
Substituting the relation (\ref{eq:Single-Stretch-xepsilon}) into Eqs.~(\ref{eq:Single-Stretch-xphi}), (\ref{eq:Single-Stretch-MacKintosh})--(\ref{eq:Single-Stretch-Terentjev}) respectively, we then obtain the force-strain relations $f(\epsilon)$ of stiff semiflexible biopolymers. These relations are plotted in Fig.~\ref{fig:Single-force-stretch} for various stiffening strain $\epsilon_s$. It is shown in Fig.~\ref{fig:Single-force-stretch} that under external stretch with $\epsilon>0$, the interpolated force-strain relations obtained from  (\ref{eq:Single-Stretch-MacKintosh})--(\ref{eq:Single-Stretch-Terentjev}) all yield the small linear strain limit (\ref{eq:Single-Stretch-xphiLin}) and large nonlinear stretch-stiffening limit (\ref{eq:Single-Stretch-xphiStiffen}). For small $\epsilon_s$ (corresponding to large $c>1$), the interpolations (\ref{eq:Single-Stretch-MacKintosh}) and (\ref{eq:Single-Stretch-PalmerBoyce}) fit the analytical force-strain relation from Eq.~(\ref{eq:Single-Stretch-xphi}) quite well by deviation less than $15\, \%$. They also overlap the more general Terentjev's interpolation formula (\ref{eq:Single-Stretch-Terentjev}) at $c>1$, but the deviation between them increases (from $\sim 5\,\% $ to $\sim  30\,\%$) as $c$ decreases (or $\epsilon_s$ increases, from $0.06$ to $0.5$) to be smaller than $1$, \emph{i.e.}, when $\ell_p<\ell_c$.  

\begin{figure}[htbp]
  \centering
  \includegraphics[width=0.75\linewidth]{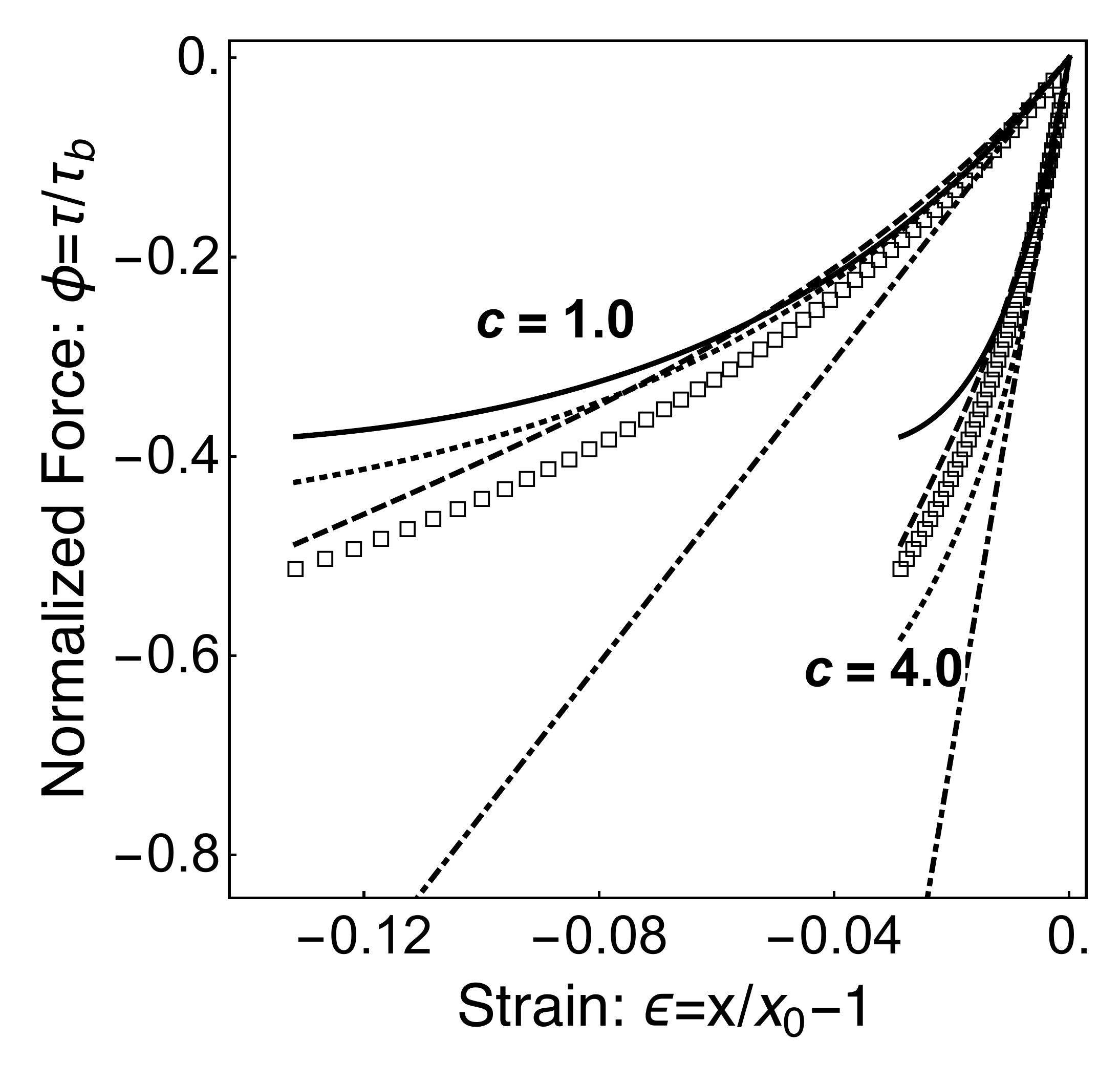}
  \caption{Force-compression relations for a semiflexible biopolymer with $c=1.0$ and $c=4.0$ at small compression before buckling. The discrete  points (open squares) are taken from the analytical relation (\ref{eq:Single-Compress-xphi}). The interpolations  (\ref{eq:Single-Stretch-MacKintosh})--(\ref{eq:Single-Stretch-Terentjev}) are represented by the solid lines, the dashed lines and the dotted lines, respectively. The dash-dotted lines are the linear interpolation (\ref{eq:Single-Compress-xphiLin}).} 
\label{fig:Single-force-compression}   
\end{figure}
 
\subsection{Compression softening due to buckling}\label{sec:Single-Compress}
 
We now consider the response of stiff biopolymers to longitudinal compression\cite{MacKintosh2014,Odijk1998,Emanuel2007,Lipowsky2010,Terentjev2009} as schematically shown in Fig.~\ref{fig:Single-semiflexibleChain}. According to the magnitude of the applied compressive force, $|f|$, two regimes can be identified as follows. 

(i) \textit{Small compressive forces: $|f|<f_b$, or, $\tau>-1$}. For small compression, stiff biopolymer responds by compressing longitudinally along its average orientation, \emph{i.e.}, the $z$-axis. One can follow the same method introduced previously for stretched biopolymers with $\tau>0$ to obtain $x(\tau)$ for compressed biopolymers with $\tau<0$ as
\begin{equation}\label{eq:Single-Compress-xphi} 
x = 
1+\frac{1}{2c}\frac{\pi\sqrt{-\tau}\cot(\pi\sqrt{-\tau})-1}{-\pi^2\tau},
\end{equation} 
which is in contrast to Eq.~(\ref{eq:Single-Stretch-xphi}) for stretched biopolymers. 
The exact force-extension relation, $\tau(x)$, can then be obtained by numerically inverting Eq.~(\ref{eq:Single-Compress-xphi}). As in the case of stretched biopolymers, some interpolated force-extension relations have been proposed, for example, a simple linear interpolation:
\begin{equation}\label{eq:Single-Compress-xphiLin} 
\tau = \frac{90c}{\pi^2}(x-x_0).
\end{equation} 
In addition, the interpolations  (\ref{eq:Single-Stretch-MacKintosh})--(\ref{eq:Single-Stretch-Terentjev}) are also applicable to compressed biopolymers before buckling.

(ii) \textit{Large compressive forces: $|f|\gsim f_b$, or, $\tau\gsim -1$}. As the applied compressive force increases to $|f|\sim f_b$, a stiff semiflexible biopolymer undergoes a buckling instability  \cite{Odijk1998,Emanuel2007,Lipowsky2010, Terentjev2009} as schematically shown in Fig.~\ref{fig:Single-semiflexibleChain}. This is analogous to the classical Euler buckling instability occurring in compressed athermal elastic rods as explained near Eq.~(\ref{eq:Single-AthermalRod-EulerInstab}). It has been shown \cite{Odijk1998,Emanuel2007,Lipowsky2010,Terentjev2009} that thermal fluctuations in semiflexible polymers modify the sharp, purely mechanical Euler buckling force $f_b$. In the presence of thermal fluctuations, the critical buckling force is \textit{decreased} in three dimensions \cite{Odijk1998, Emanuel2007} where small thermal forces help in triggering buckling, but is \textit{increased} in dimensions smaller than three\cite{Lipowsky2010} because in this case, the energy gain by deforming a biopolymer with a force decreases if the biopolymer has already shortened by thermal fluctuations. Once the biopolymer is buckled, it loses stiffness to additional compression. To a good approximation, the post-buckling behaviors can be described by a linear force-compression relation\cite{Bazant2010,Xu2015PRE,Xu2017PRE}:
\begin{equation} \label{eq:Single-Compress-xphiSoften} 
\tau =\frac{90c}{\pi^2}\left[(x-x_0)-(1-\rho_0)(x-x_b)\right]= -1+\frac{90c}{\pi^2}\rho_0(x-x_b),
\end{equation}
in which the buckling extension $x_b=x_0-\pi^2/90c$ and buckling strain $-\epsilon_b=x_b/x_0-1=-\pi^2/90cx_0$, obtained from Eq.~(\ref{eq:Single-Compress-xphiLin}) when $\tau=-1$, and $0<\rho_0 \ll 1$ indicating much smaller stiffness of post-buckling biopolymers than that before buckling. 

In Fig.~\ref{fig:Single-force-compression}, we plot the force-strain relations $f(\epsilon)$ for compressed semiflexible polymers, which are obtained by substituting the relation (\ref{eq:Single-Stretch-xepsilon}) into the corresponding force-extension relations in Eqs.~(\ref{eq:Single-Stretch-MacKintosh})--(\ref{eq:Single-Stretch-Terentjev}), and  (\ref{eq:Single-Compress-xphi}), respectively. It is shown that for large $c$, the interpolations (\ref{eq:Single-Stretch-MacKintosh})--(\ref{eq:Single-Stretch-Terentjev}) fit the analytical force-strain relation from Eq.~(\ref{eq:Single-Compress-xphi}) well up to buckling $|\tau| \sim 1$, around which nonlinear compressive softening becomes significant (with much smaller slope magnitude than the linear relation) even before the occurrence of buckling instability. 

Before ending this subsection, we summarize that the nonlinear elastic response of inextensible, stiff, semiflexible biopolymers to both extension and compression can be well described by the following piecewise interpolation of force-strain relation\cite{Xu2015PRE,Xu2017PRE}:
\begin{equation}\label{eq:Single-xphiSummary} 
\tau(x) =
  \begin{cases}
  \frac{9}{\pi^2}\left[\frac{\left(1-x_{0}\right)^{2}}{\left(1-x\right)^{2}}-1-\frac{1}{3}\frac{x-x_{0}}{1-x_{0}}\right], & \text{for} \quad x>x_{0} \\
    \frac{90c}{\pi^2}\left[(x-x_0)-(1-\rho)(x-x_b)\right],      & \text{for} \quad x<x_{0}
  \end{cases},
\end{equation}  
which is analogous to the piecewise force-strain relation for athermal rods in Eq.~(\ref{eq:Single-AthermalRod-tauepsilon}). 
Here $\rho=\rho_0+(1-\rho_0)\Theta(\epsilon +\epsilon_b)$ with $0 \leq \rho_0 \ll 1$ characterizing the nonlinear softening due to the microbuckling of stiff biopolymer \cite{Terentjev2009, Lipowsky2010, MacKintosh2014} for compression over $-\epsilon_b$ and $\Theta(x)$ being the Heaviside step function of $x$.
From the force-extension relations (\ref{eq:Single-xphiSummary}) using Eq.~(\ref{eq:Single-Stretch-xepsilon}), we calculate the interpolated strain energy $w(\epsilon)$ of a single semiflexible biopolymer by $w_{\rm{chain}} = \ell_0\int_{0}^{\epsilon} f(\epsilon') d\epsilon'$ as the piecewise continuous form:
\begin{equation}\label{eq:Single-wChain} 
w_{\rm{chain}}(\epsilon) =
  \begin{cases}
  \frac{1}{2}\alpha\epsilon^2 \left[\frac{6}{5}(1-\epsilon/\epsilon_s)^{-1}- \frac{1}{5}\right] , & \text{for} \quad \epsilon>0 \\
    \frac{1}{2}\alpha\tilde{\rho}\epsilon^2,      & \text{for} \quad \epsilon<0
  \end{cases}
\end{equation}  
with $\alpha=k\ell_0^2=90k_BT\ell_0^2 \ell_p^2/\ell_c^4$ and $\tilde{\rho}=1-(1-\rho) (1+\epsilon_b/\epsilon)^2$ varying between $1$ and $\rho_0$.

%================================================================%
\section{Continuum models for the elastic responses of biopolymer gels to externally applied forces}\label{sec:ExtForce}
%================================================================%
As mentioned in the introduction section \ref{sec:Introduction}, biopolymer gels composed of crosslinked semiflexible biopolymers have highly nonlinear elasticity (see Fig.~\ref{fig:introduction-GelNonlinearity}): stiffen upon shear or tension and soften upon compression. In this section, we review continuum models for the elastic responses of biopolymer gels to externally applied forces, such as simple shear stress and uniaxial tensile stress. We first review the classical theories for linear isotropic and anisotropic materials very briefly, and then discuss the extension of these classical theories to modeling nonlinear biopolymer gels at both small and large deformations. We assume that the deformations of the biopolymer gels upon external forces are homogeneous (uniform in the whole gel) and affine. In this case, the nonlinear elasticity of biopolymer gels is fully attributed to the nonlinear mechanics (or force-strain relation) of constituting biopolymers as reviewed in the previous section. 

%============================%
\subsection{Theory of linear isotropic elasticity}\label{sec:ExtForce-liniso}
\label{subsec: theory of isotropic}
%============================%
For linear isotropic elastic materials\cite{Landau1986}, the deformation free energy density is given by
\begin{equation}\label{eq:ExtForce-liniso-FmuK}
F=  \mu_0\tilde{\epsilon}_{ik}^2+\frac{1}{2}K\epsilon_{ll}^2,
\end{equation}
or equivalently,
\begin{equation}\label{eq:ExtForce-liniso-FnuE}
    F=\frac{E_0}{2(1+\nu_0)}\left(\epsilon_{ik}^2+ \frac{\nu_0}{1-2\nu_0}\epsilon_{ll}^2\right),
\end{equation} 
with $\tilde{\epsilon}_{ik}\equiv \epsilon_{ik}-\epsilon_{ll}\delta_{ik}/3$ being the deviatoric (traceless) strain tensor. Here, the elastic constants, $\mu_0$, $K$, $E_0$, and $\nu_0$ are the shear modulus, bulk modulus, Young's modulus and Poisson's ratio, respectively. 
From the free energy in Eq.~(\ref{eq:ExtForce-liniso-FnuE}) and using $\sigma_{ik}=\partial F/\partial \epsilon_{ik}$ (applicable also for nonlinear materials), we obtain the stress-strain relation or constitutive relation (Hooke's law) as
\begin{equation}\label{eq:ExtForce-liniso-sigmaepsilon} 
    \sigma_{ik}=\frac{E_0}{1+\nu_0}\left(\epsilon_{ik} + \frac{\nu_0}{1-2\nu_0}\epsilon_{ll}\delta_{ik}\right),
\end{equation}
% Alternatively, we can define a complementary energy in terms of stress components as
% \begin{equation}\label{eq:ExtForce-liniso-Ftilde}
%     \tilde{F}=\frac{1+\nu_0}{2E_0}\left(\sigma_{ik}^2+ \frac{\nu_0}{1+\nu_0}\sigma_{ll}^2\right),
% \end{equation} 
% from which we obtain the strain field using $\epsilon_{ik}=\partial \tilde{F}/\partial \sigma_{ik}$ (applicable only for linear materials):
or inversely, 
\begin{equation}\label{eq:ExtForce-liniso-epsilonsigma}  
    \epsilon_{ik}=\frac{1}{E_0}\left[ (1+\nu_0)\sigma_{ik}-\nu_0\sigma_{ll}\delta_{ik}\right].
\end{equation}
Note that for linear isotropic materials, there are only \textit{two} independent elastic constants, either $K$ and $\mu_0$, or $E_0$ and $\nu_0$, and they are related by $\mu_0={E_0}/{2(1+\nu_0)}$, and $K={E_0}/{3(1-2\nu_0)}$.

Particularly, for a deformation where the principal deformation directions are along the coordinate axes $\hat{\mathbf{x}}_i$ with $i=1,2,3$, the strain tensor is diagonalized and the deformation energy density (\ref{eq:ExtForce-liniso-FmuK}) reduces to 
\begin{align}\label{eq:ExtForce-liniso-Fepsilon}
F= 
\mu_0(\epsilon_1^2+\epsilon_{2}^2+\epsilon_{3}^2) +\frac{\tilde{K}}{2}(\epsilon_{1}+\epsilon_{2}+\epsilon_{3})^2,
\end{align} 
where $\tilde{K}\equiv K-2\mu_0/3$ is the modified bulk modulus, and hence
\begin{equation}\label{eq:ExtForce-liniso-E0nu0}
E_0=2\mu_0(1+\nu_0), \quad 
\nu_0=\frac{\tilde{K}}{2(\tilde{K}+\mu_0)}
\end{equation}  
with $0\le\nu_0\le 1/2$ for positive $\tilde{K}$. From the energy (\ref{eq:ExtForce-liniso-Fepsilon}), we obtain the three principal stress components
\begin{align}\label{eq:ExtForce-liniso-sigmaepsilon}
\sigma_i&=2\mu_0\epsilon_i + \tilde{K}(\epsilon_1+\epsilon_2+\epsilon_3), \quad {\rm{with}}\quad i=1,\,2,\,3.
\end{align}
%============================%
\subsection{Theory of linear anisotropic elasticity}\label{sec:ExtForce-linaniso}
%============================%
We consider a typical linear anisotropic material -- transversely isotropic (or briefly \textit{transtropic}) material\cite{Lekhnitskii1981}. A transtropic material is one with physical properties that are symmetric about an axis that is normal to a plane of isotropy, for example hexagonal close-packed crystals\cite{Landau1986} and nematic elastomers\cite{Terentjev2007}. In this case, the deformation free energy density is given, based on symmetry considerations, by\cite{Landau1986}
\begin{align}\label{eq:ExtForce-linaniso-Fc}
F= 
&\frac{1}{2}c_{1}\epsilon_{11}^2
+ 2c_{2}(\epsilon_{22}+\epsilon_{33})^2 +c_{3}[(\epsilon_{22}-\epsilon_{33})^2+4\epsilon_{23}^2]\\ \nonumber
&+2c_{4}\epsilon_{11}(\epsilon_{22}+\epsilon_{33})
+2c_{5}(\epsilon_{12}^2+\epsilon_{13}^2).
\end{align} 
Here we have taken $\hat{\mathbf{x}}_1$ as the axis of symmetry, $\hat{\mathbf{x}}_2$ and $\hat{\mathbf{x}}_3$ span the plane of isotropy. Note that for transtropic materials, there are \textit{five} independent elastic constants, $c_i$ ($i=1,...,5$). 
From the free energy (\ref{eq:ExtForce-linaniso-Fc}), we obtain the stress-strain relations (in comparison to Eq.~(\ref{eq:ExtForce-liniso-sigmaepsilon}))
\begin{subequations}\label{eq:ExtForce-linaniso-sigmaepsilon}
\begin{align}
   \sigma_{11}&= c_{1}\epsilon_{11} +2c_{4}\epsilon_{22} +2c_{4}\epsilon_{33}\\
   \sigma_{22}&= 2c_{4}\epsilon_{11} +(4c_{2}+2c_{3})\epsilon_{22} +(4c_{2}-2c_{3})\epsilon_{33}\\
   \sigma_{33}&= 2c_{4}\epsilon_{11} +(4c_{2}-2c_{3})\epsilon_{22} +(4c_{2}+2c_{3})\epsilon_{33}\\
   \sigma_{12}&= 2c_{5}\epsilon_{12},\quad \sigma_{13}=2c_{5}\epsilon_{13},\quad \sigma_{23}=2c_{5}\epsilon_{23},
\end{align} 
\end{subequations}
or inversely in a more illustrative form (in comparison to Eq.~(\ref{eq:ExtForce-liniso-epsilonsigma})), 
\begin{subequations}\label{eq:ExtForce-linaniso-epsilonsigma} 
\begin{align}
    \epsilon_{11}&= \frac{1}{E_{1}}\sigma_{11} -\frac{\nu_{21}}{E_{2}}\sigma_{22} -\frac{\nu_{21}}{E_{2}}\sigma_{33}\\
    \epsilon_{22} &= -\frac{\nu_{12}}{E_{1}}\sigma_{11} +\frac{1}{E_{2}}\sigma_{22} -\frac{\nu_{23}}{E_{2}}\sigma_{33}\\
    \epsilon_{33}&= -\frac{\nu_{12}}{E_{1}}\sigma_{11}-\frac{\nu_{23}}{E_{2}}\sigma_{22} +\frac{1}{E_{2}}\sigma_{33}\\ 
    \epsilon_{12}&= \frac{1}{2\mu_{12}}\sigma_{12},\quad \epsilon_{13}=\frac{1}{2\mu_{12}}\sigma_{13},\quad
    \epsilon_{23}=\frac{1}{2\mu_{23}}\sigma_{23}.
\end{align} 
\end{subequations}
Here $E_1$ and $E_2$ are the Young's moduli along the $\hat{\mathbf{x}}_1$-axis of symmetry and in the isotropic ($\hat{\mathbf{x}}_2$--$\hat{\mathbf{x}}_3$) plane, respectively.
$\nu_{ij} \equiv - {\partial \epsilon_j}/{\partial \epsilon_i}$  (with $i\neq j$ and $i,j=1,2,3$) are the differential Poisson's ratios for tensile stress applied along $i$-direction and contraction in $j$-direction, following the general notation in anisotropic materials\cite{Lekhnitskii1981}. $\mu_{12}$, $\mu_{23}$ are the shear moduli in the $\hat{\mathbf{x}}_1$--$\hat{\mathbf{x}}_2$ plane and the isotropic $\hat{\mathbf{x}}_2$--$\hat{\mathbf{x}}_3$ plane, respectively.
Note that these elastic constants are not all independent, satisfying the relations: ${\nu_{21}}/{E_{2}}={\nu_{12}}/{E_{1}}$ and $\mu_{23}= {E_{2}}/{2(1+\nu_{23})}$. Therefore, there are only five independent elastic constants, for example, $E_1$, $E_2$, $\nu_{21}$, $\nu_{23}$, and $\mu_{12}$, which are related to $c_i$ (see Appendix \ref{sec:app-Transtropic-ElasticCoeff}) by  
\begin{equation}\label{eq:ExtForce-linaniso-cEnu} 
c_{1}=\frac{E_{1}}{m}(1-\nu_{23}),  \,
c_{2}=\frac{E_{2}}{8m}, \,
c_{3}=\frac{1}{2}\mu_{23}, \, c_{4}=\frac{E_{2}\nu_{12}}{2m}, \,
c_{5}=\mu_{12} 
\end{equation}
with $m\equiv 1-\nu_{23}-2\nu_{12}^2E_2/E_1$.

\begin{figure}[h]
\centering
\includegraphics[width=1.0\linewidth]{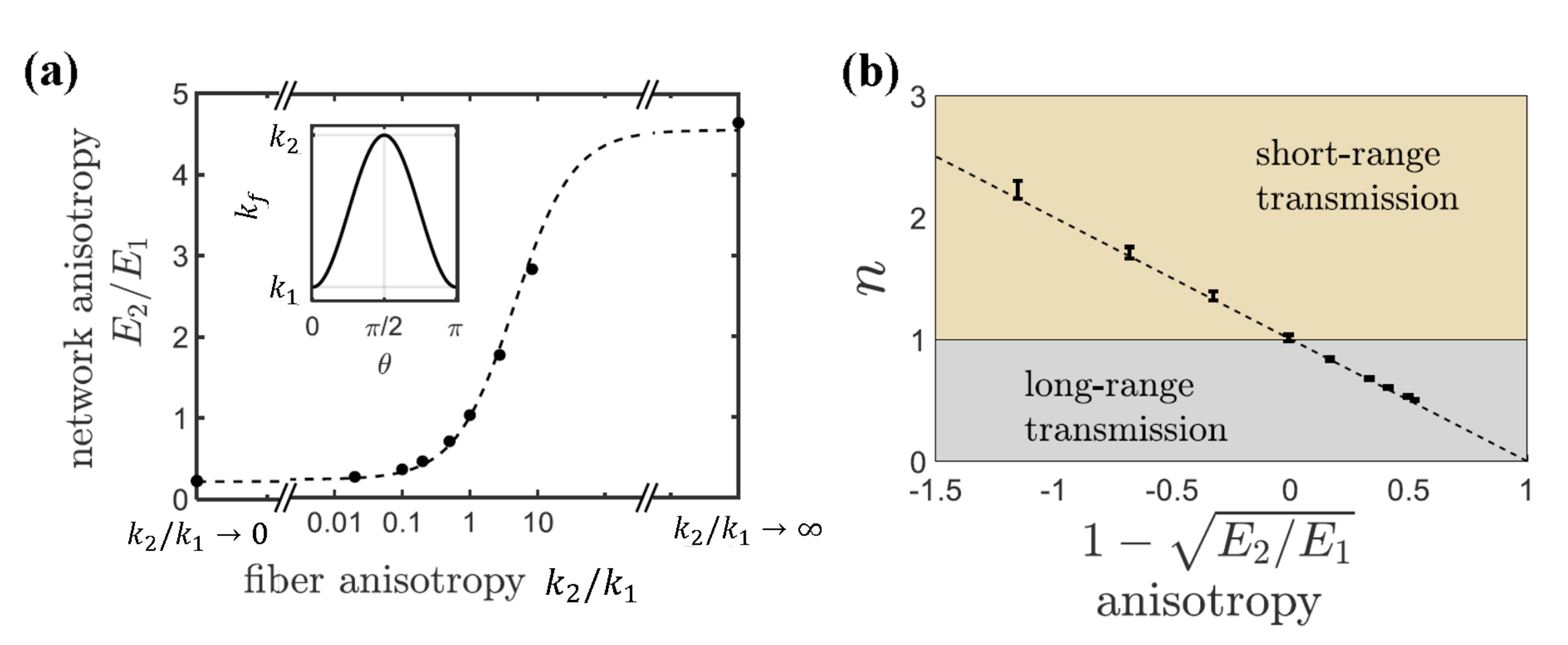}
  \caption{The decay of cell-induced displacements in linear anisotropic networks. (a) An anisotropic network is constructed by introducing an orientation-dependent fiber modulus in Eq.~(\ref{eq:ExtForce-linaniso-kf}). The elastic anisotropy is predicted by affine theory in Eq.~(\ref{eq:app-Transtropic-UniFiberOrient-E2E1ratio}). (b) The near-field decay of displacement follows the power-law $\tilde{u}\sim \tilde{r}^{-n}$ in Eq.~(\ref{eq:IntForce-linaniso-urscaling}). The exponent $n$ is plotted a a function of network anisotropy at infinitesimal cell contractions. A very good linear fitting to $n=\sqrt{E_2/E_1}$ is obtained (the dashed line) in both regions: $E_2>E_1$ (with $n>1$ indicating fast displacement decay) and $E_2<E_1$ (with $n<1$ indicating slow displacement decay). Reproduced from Goren \emph{et al.} \cite{Xu2020BJ} with permission from Elsevier.} 
\label{fig:ExtForce-LinearAnisotropy}
\end{figure}

Now we give some remarks on transtropic elastic materials as follows. 

(i) We consider a particular deformation where one principal deformation direction is along the symmetry axis $\hat{\mathbf{x}}_1$ and the other two along $\hat{\mathbf{x}}_2$ and $\hat{\mathbf{x}}_3$, respectively, in the isotropic plane. In this case, the strain tensor is diagonalized and the deformation energy density in Eq.~(\ref{eq:ExtForce-linaniso-Fc}) reduces to
\begin{align}\label{eq:ExtForce-linaniso-Fc2}
F= 
\frac{1}{2}c_1\epsilon_{1}^2
+ 2c_2(\epsilon_{2}+\epsilon_{3})^2 +c_3(\epsilon_{2}-\epsilon_{3})^2+2c_4\epsilon_{1}(\epsilon_{2}+\epsilon_{3}),
\end{align}
in which $\epsilon_{1,2,3}$ are the three principal strain components, and the four elastic coefficients, $c_i$ ($i=1,...,4$) are, in general, all independent and related to elastic parameters in Eq.~(\ref{eq:ExtForce-linaniso-cEnu}). 
But particularly for linear isotropic materials, only two of $c_{i}$ are independent.
% and given by  
% \begin{equation}\label{eq:ExtForce-liniso-cE0nu0}
% c_{1}=\frac{E_0}{m_0}(1-\nu_0),  \quad
% c_{2}=\frac{E_0}{8m_0}, \quad
% c_{3}=\frac{1}{2}\mu_0, \quad c_{4}=\frac{E_0\nu_0}{2m_0},
% \end{equation}
% with $m_0\equiv 1-\nu_0 -2\nu_0^2$ and $\mu_0$ given in Eq.~(\ref{eq:ExtForce-liniso-E0nu0}). 

(ii) It is interesting to note that a fiber network with anisotropic elastic properties of transtropic materials can be constructed at least in finite element simulations. Goren \emph{et al.}\cite{Xu2020BJ} have constructed a two-dimensional transtropic fibrous network composed of linear fibers that are uniformly distributed in orientation and have orientation-dependent stiffness, $k_f$, as shown in Fig.~\ref{fig:ExtForce-LinearAnisotropy}(a):
\begin{equation}\label{eq:ExtForce-linaniso-kf} 
k_f=k_1\cos^2\theta +k_2\sin^2\theta
\end{equation}
where $\theta\in[0,\pi)$ is the angle of fiber with respect to the axis of symmetry (say, $\hat{\mathbf{x}}_1$-axis). $k_1$ and $k_2$ are the two extrema of $k_f$ along longitudinal directions (\emph{i.e.}, along the symmetry axis with $\theta=0$) and transverse directions (with $\theta=\pi/2$), respectively. Such a network is anisotropic in elasticity but not in geometry (without collective fiber alignment). Physically, such a transtropic network can be a model of fiber-reinforced composites or fibril bundles; it can also be generated either by homogeneous plastic deformations of an isotropic network under uniaxial tension or locally by stretching of cells. The deformation energy can be calculated if affine deformation is assumed (see the Appendix~\ref{sec:app-Transtropic-UniFiberOrient}) and it takes the form of Eq.~(\ref{eq:ExtForce-linaniso-Fc}) from which one obtain Poisson's ratios and Young's moduli. The degree of elastic anisotropy of the transtropic fiber network can be measured by the ratio of the two Young's moduli, which can be varied by changing the fiber anisotropy $k_2/k_1$ as shown in Fig.~\ref{fig:ExtForce-LinearAnisotropy}(a)). 
% When $k_2/k_1$ varies from $0$ to $\infty$, we obtain that $8/15 < E_2/E_1< 32/9$, $1/12< \nu_{12}< 3/8$, $2/45< \nu_{21}< 4/3$, and $\nu_{23}$ varies from $1/3$ to $-3/4$.

\subsection{Continuum models of nonlinear biopolymer gels at small affine deformation \label{sec:ExtForce-nonlinsmall}}  

Based on the nonlinear behaviors of individual biopolymer filaments, several continuum models of biopolymer gels have been proposed in analogy to those for rubber elasticity. Such continuum models are often named unit-cell models\cite{Treloar1975,MacKintosh2014,PalmerBoyce2008,Xu2015PRE,Meng2017}, where a biopolymer gel is treated as an effective continuum composed by periodically repeating blocks or cells, \emph{e.g.},  1-chain sphere models, 3- and 8-chain cubic lattice models, and 4-chain tetrahedra model. In these unit-cell chain models, the macroscopic elastic free energy of the gel is obtained by adding the free energy of individual blocks that are defined affinely. The non-affine deformation, if allowed, is assumed to occur only for chains within each block.

In this subsection, we first review the continuum model proposed in 2015 by Xu and Safran \cite{Xu2015PRE} for fibrous biopolymer gels where the nonlinear elasticity shows up at small deformation. In contrast to perfect crystalline solids or the isotropic homogeneous rubbery networks, in fibrous biogels the non-affine deformations cannot strictly be avoided \cite{Xu2017PRE}. That is, network local structure would relax on the scale of a single mesh unit to lower the local energy and achieve global equilibrium. Although there is no a priori reason to believe the affine approximation is valid, recent theoretical and experimental studies suggest that it is a good approximation for densely cross-linked filaments of high molecular weight \cite{MacKintosh2003,GardelMacKintosh2004,Lubensky2005}.  

\subsubsection{Elastic deformation energy: Xu-Safran 3-chain model \label{sec:ExtForce-nonlinsmall-XuSafran3chain}} 

Among the existing continuum models for biopolymer gels \cite{MacKintosh2014,Meng2017}, the 3-chain models are easier to handle analytically and are found to best fit the experimental data for the elastic deformation of various biopolymer gels\cite{PalmerBoyce2008,Xu2015PRE,Meng2017}. In a typical 3-chain model for cross-linked polymer networks, a primitive cubic block is constructed with lattice points representing the cross-linking sites, and the three chain segments between sites at the block edges are aligned along three principal directions of deformation \cite{Treloar1975,PalmerBoyce2008}. The elasticity of the chain segments represents the emergent response of constituent polymers to applied forces. 
The mesh size of the primitive blocks is usually assumed to be mono-disperse, denoted by $\ell_0$, and then for affine deformation, the lengths of each chain become $\lambda_{1}\ell_0=(1+\epsilon_1)\ell_0$, $\lambda_{2}\ell_0=(1+\epsilon_2)\ell_0$, and $\lambda_{3}\ell_0=(1+\epsilon_3)\ell_0$, respectively. 

In the 3-chain model for biopolymer gels, it is usually simply assumed that the chain segments have the same nonlinear elasticity as single semiflexible biopolymers, see discussions in Sec.~\ref{sec:Single}. In this case, the strain energy density $F$ for a weakly compressible (\emph{i.e.}, almost incompressible) biopolymer gel at small affine deformations is given by
\begin{equation}\label{eq:ExtForce-nonlinsmall-F}
F=\frac{1}{3}n_f\sum_{i=1}^3 w_{\rm{chain}}(\epsilon_i) + \frac{1}{2}\tilde{K}  (\epsilon_1+\epsilon_2+\epsilon_3)^2,
\end{equation}
in terms of the three principal strain components $\epsilon_i$. 
Here $n_f$ is the density of biopolymer segments or crosslinkers. $\tilde{K}$ is the modified bulk modulus (as defined in Eq.~(\ref{eq:ExtForce-liniso-Fepsilon}) for linear isotropic materials); it is related to osmotic bulk modulus that describes the osmotic stress when the entire biopolymer network is isotropically compressed. Since many biopolymer gels are weakly compressible, we assume the compression is small and treat it linearly. $w_{\rm{chain}}(\epsilon_i)$ is the interpolated strain energy of a single semiflexible biopolymer and is given in Eq.~(\ref{eq:Single-wChain}). Note that, in analogy to Eqs.~(\ref{eq:ExtForce-liniso-FmuK}) in the theory of linear elasticity, there are two separate contributions to the energy (\ref{eq:ExtForce-nonlinsmall-F}): the deformation energy (the first term in Eq.~(\ref{eq:ExtForce-nonlinsmall-F})) of the semiflexible biopolymer segments and the energy (the second term in Eq.~(\ref{eq:ExtForce-nonlinsmall-F})) associated with the compressibility of the retained water in the gel. 
% It can cause the elastic modulus to be discontinuous (unless $\rho = 1$) in the neighborhood of $\epsilon=0$ (the undeformed, stress-free state).

For the special case of $\epsilon_{1}>0$ and $\epsilon_{2},\epsilon_{3}\leq 0$, the energy density (\ref{eq:ExtForce-nonlinsmall-F}) can be written as  
\begin{equation}\label{eq:ExtForce-nonlinsmall-F2}
F=\mu_0\left[ \epsilon_1^2 \left(\frac{6}{5}\frac{1}{1- \epsilon_1/\epsilon_s} -\frac{1}{5} \right)+ \tilde{\rho}_2\epsilon_2^2+ \tilde{\rho}_3\epsilon_3^2\right] + \frac{\tilde{K}}{2}(\epsilon_1+\epsilon_2+\epsilon_3)^2,
\end{equation}
where $\mu_0$ is the linear shear modulus, given by $\mu_0 \equiv n_f\alpha/6=15n_fk_BT\ell_0^2 \ell_p^2/\ell_c^4$, $\tilde{\rho}_{2,3}=1-(1-\rho(\epsilon_{2,3})) (1+\epsilon_b/\epsilon_{2,3})^2$ with $\rho(\epsilon)=\rho_0+(1-\rho_0)\Theta(\epsilon +\epsilon_b)$, $\Theta$ being the Heaviside step function, and $0 \leq \rho_0 \ll 1$. Then the three principal stress components are obtained from $\sigma_{i}=\partial F/\partial \epsilon_{i}$ as
\begin{subequations}\label{eq:ExtForce-nonlinsmall-sigmaepsilon} 
\begin{equation}\label{eq:ExtForce-nonlinsmall-sigmaepsilon-1} 
\sigma_1=\frac{6}{5}\mu_0\epsilon_s \left[\frac{1}{(1-\epsilon_1/\epsilon_s)^2}-1 -\frac{\epsilon_1}{3\epsilon_s} \right] + \tilde{K}(\epsilon_1+\epsilon_2+\epsilon_3),
\end{equation}
\begin{equation}\label{eq:ExtForce-nonlinsmall-sigmaepsilon-23} 
\sigma_{2,3}=2\mu_0[\epsilon_{2,3}-(1-\rho)(\epsilon_{2,3}+\epsilon_b)]+ \tilde{K}(\epsilon_1+\epsilon_2+\epsilon_3).
\end{equation}
% \begin{equation}\label{eq:fs-sigma3} 
% \sigma_3= 2\mu_0[\epsilon_3-(1-\rho)(\epsilon_3+\epsilon_b)]+ \tilde{K}(\epsilon_1+\epsilon_2+\epsilon_3).
% \end{equation}
\end{subequations}
Note that there is no residual stress at the undeformed state with $\epsilon_i=0$ and hence $\rho=1$, $\sigma_i=0$ for all $i=1,2,3$ from Eq.~(\ref{eq:ExtForce-nonlinsmall-sigmaepsilon}).

We now consider two interesting limits of the energy (\ref{eq:ExtForce-nonlinsmall-F}) according to the relative magnitude of strain components $\epsilon_i$ to $\epsilon_s$ and $\epsilon_b$. 
\begin{itemize}
    \item \textit{Linear isotropic limit: $\epsilon_1/\epsilon_s\ll 1$ and  $|\epsilon_{2,3}|/\epsilon_b \ll 1$}. The energy density (\ref{eq:ExtForce-nonlinsmall-F2}) and the stress components (\ref{eq:ExtForce-nonlinsmall-sigmaepsilon}) reduce to Eqs.~(\ref{eq:ExtForce-liniso-FmuK}) and (\ref{eq:ExtForce-liniso-sigmaepsilon}), respectively. The linear Young's modulus and Poisson's ratio are then, respectively, given by Eq.~(\ref{eq:ExtForce-liniso-E0nu0}). The relative magnitude of $\tilde{K}$ to $\mu_0$ determines the magnitude of linear Poisson ratio $\nu_0$. The following two limiting cases of $\nu_0$ will be discussed later about the effects of compressibility on the range of force transmission: 
(i) \textit{Limit of small osmotic modulus ($\tilde{K}\ll \mu_0$) and zero Poisson ratio $\nu_0\to 0$}. Water can freely flow out of the gel as the polymer network is deformed by applied stresses. In this limit, the elasticity of the gel is purely due to that of the polymer network. 
(ii) \textit{Limit of weak compressibility with large osmotic modulus ($\tilde{K}\gg \mu_0$) and Poisson ratio $\nu_0\to 1/2$}. For small time scales, the water does not have time to flow out of the gel, but is deformed as the polymer network is distorted. 

\item \textit{Linear anisotropic limit: $\epsilon_1/\epsilon_s \ll 1$ and  $|\epsilon_{2,3}|/\epsilon_b >1$}. In this case, the energy density (\ref{eq:ExtForce-nonlinsmall-F2}) reduces to
\begin{equation}\label{eq:ExtForce-nonlinsmall-Faniso} 
F=\mu_0\left( \epsilon_1^2 + \rho_0\epsilon_2^2+ \rho_0\epsilon_3^2\right) + \frac{\tilde{K}}{2}(\epsilon_1+\epsilon_2+\epsilon_3)^2,
\end{equation} 
which can be casted into the general energy form of Eq.~(\ref{eq:ExtForce-linaniso-Fc2}) with coefficients given by
\begin{equation}\label{eq:ExtForce-nonlinsmall-cmu0K} 
c_1=2\mu_0+\tilde{K}, \quad
c_2=\frac{1}{4}(\rho_0\mu_0+\tilde{K}),  \quad
c_3=\frac{1}{2}\rho_0\mu_0, \quad
c_4=\frac{1}{2}\tilde{K}, 
\end{equation} 
and the bulk modulus given by $K=\tilde{K}+\frac{2}{9}(1+2\rho_0)\mu_0$ (See also Appendix~\ref{sec:app-Transtropic-3Chain}). From the energy (\ref{eq:ExtForce-nonlinsmall-Faniso}), we obtain the stress components
\begin{subequations}\label{eq:ExtForce-nonlinsmall-Faniso-sigmaepsilon} 
\begin{align} 
\sigma_{1}&= 2\mu_0\epsilon_{1} + \tilde{K}(\epsilon_{1} + \epsilon_{2}+\epsilon_{3})\\ 
\sigma_{2,3}&= 2\rho_0\mu_0\epsilon_{2,3}+ \tilde{K}(\epsilon_{1} + \epsilon_{2}+\epsilon_{3}).
\end{align} 
\end{subequations}
\end{itemize}
Similar constitutive models in this limit have also been proposed by Rosakis {\emph{et al.}} \cite{Notbohm2015} and Ronceray {\emph{et al.}} \cite{Lenz2019}, in which the compressive-softening (over critical stress, equivalently here, over critical strain) of fibrous networks due to fiber buckling upon compression  is modeled as a loss of stiffness by introducing a small \emph{compression stiffness ratio} $\rho_0\ll 1$. 

\subsubsection{Elastic responses to external shear and tensile stresses \label{sec:ExtForce-nonlinsmall-ShearTension}} 

Given the compact analytical form of the equilibrium stress-strain relation (\ref{eq:ExtForce-nonlinsmall-sigmaepsilon}), the nonlinear elastic properties of biopolymer gels can now be derived for any applied deformation. Here we consider only simple shear\cite{GardelMacKintosh2004,Lubensky2005,Janmey2007NatMat,MacKintosh2014,PalmerBoyce2008,Xu2015PRE,Meng2017,Janmey2016,MacKintoshJanmey2016,Kim2014,Kim2016} and uniaxial extension stresses\cite{Xu2015PRE,Janmey2016,MacKintoshJanmey2016} as typical examples. However, similar calculations and comparisons to experiments can also be done for other deformations such as biaxial and triaxial deformation \cite{Janmey2019Tissue}. 
%----------------- 
\begin{figure}[htbp]
  \centering
  \includegraphics[clip=true, viewport=1 1 750 310, keepaspectratio, width=0.48\textwidth]{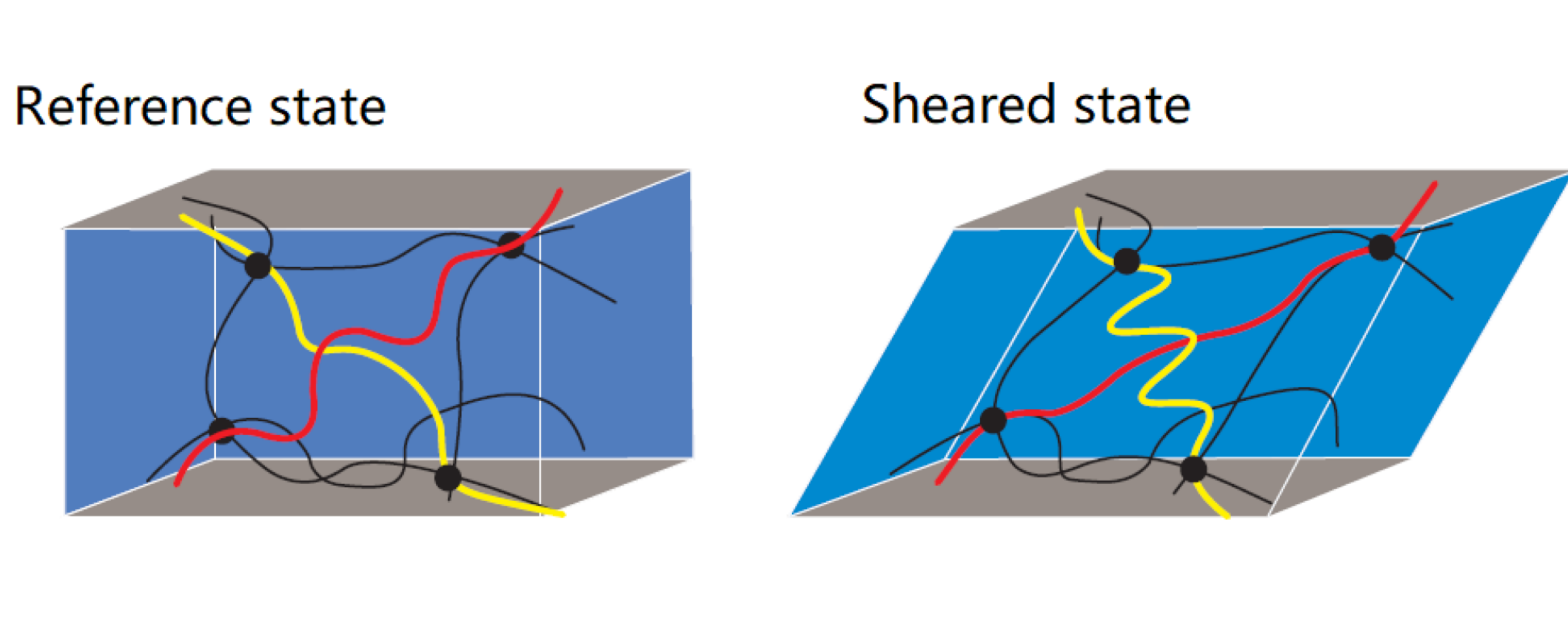}
  \caption{(Color online) Schematic diagram of the deformations of individual filaments leading to negative normal stress in an isotropic biopolymer gel under simple shear. Some filaments are elongated (red filament), whereas an equal number are compressed (yellow filament). Owing to the nonlinear asymmetric force-extension relation for semiflexible filaments, stretched filaments exert more force than the compressed ones, which leads to a negative net tension in the direction orthogonal to the shear direction. Reprinted by permission from Springer Nature: Nature Materials~\cite{Janmey2007NatMat}, COPYRIGHT(2007).}
\label{fig:ExtForce-normalstress}   
\end{figure} 

{\emph{Simple shear}} -- For biopolymer gels under simple shear stress\cite{GardelMacKintosh2004,Lubensky2005,MacKintosh2014,PalmerBoyce2008,Xu2015PRE,Meng2017,Janmey2016,MacKintoshJanmey2016,Kim2014,Kim2016} in the $x-z$ plane (as shown in Fig.~\ref{fig:ExtForce-normalstress}), we assume the gel deformation is homogeneous and the deformation tensor is given by\cite{Bower2009,MacKintosh2014}
\begin{equation}\label{eq:ExtForce-nonlinsmall-Fshear} 
{\mathbf F}=\begin{bmatrix}
1 & 0 & \gamma\\
0 & 1 & 0\\
0 & 0 & 1
\end{bmatrix}. 
\end{equation} 
In the limit of small shear strain, $\gamma \ll 1$, the three principal strain components, $\epsilon_i$, and the corresponding eigenvectors, $\hat{\mathbf{c}}^{(i)}$, are given respectively by\cite{Bower2009,MacKintosh2014}
\begin{equation}\label{eq:ExtForce-nonlinsmall-epsilon} 
\epsilon_{1}\simeq {\gamma}/{2}>0, \quad  \epsilon_{2}=0, \quad \epsilon_{3}\simeq- {\gamma}/{2}<0,
\end{equation}
\begin{equation}\label{eq:ExtForce-nonlinsmall-cvector} 
\hat{\mathbf{c}}^{(1)}\simeq\frac{1}{\sqrt{2}}(1, \, 0, \, 1), \quad
\hat{\mathbf{c}}^{(2)}\simeq (0, \, 1, \, 0), \quad
\hat{\mathbf{c}}^{(3)}\simeq\frac{1}{\sqrt{2}}(-1,\,  0, \, 1).
\end{equation} 
Here we have taken ${\gamma}>0$ for specificity without losing generality. 
The components of stress tensor are given by
\begin{equation}\label{eq:ExtForce-nonlinsmall-sigmatensor} 
    \sigma_{ij}=\sigma_1\hat{{c}}_{i}^{(1)}\hat{{c}}_{j}^{(1)}+ \sigma_2\hat{{c}}_{i}^{(2)}\hat{{c}}_{j}^{(2)}+ \sigma_3\hat{{c}}_{i}^{(3)}\hat{{c}}_{j}^{(3)},
\end{equation}
with $\sigma_i=\partial F/\partial \epsilon_i$ as given in Eq.~(\ref{eq:ExtForce-nonlinsmall-sigmaepsilon}) and particularly $\sigma_2=0$. 
% %----------------- 
\begin{figure} 
  \centering
  \includegraphics[clip=true, viewport=1 1 650 600, keepaspectratio, width=0.38\textwidth]{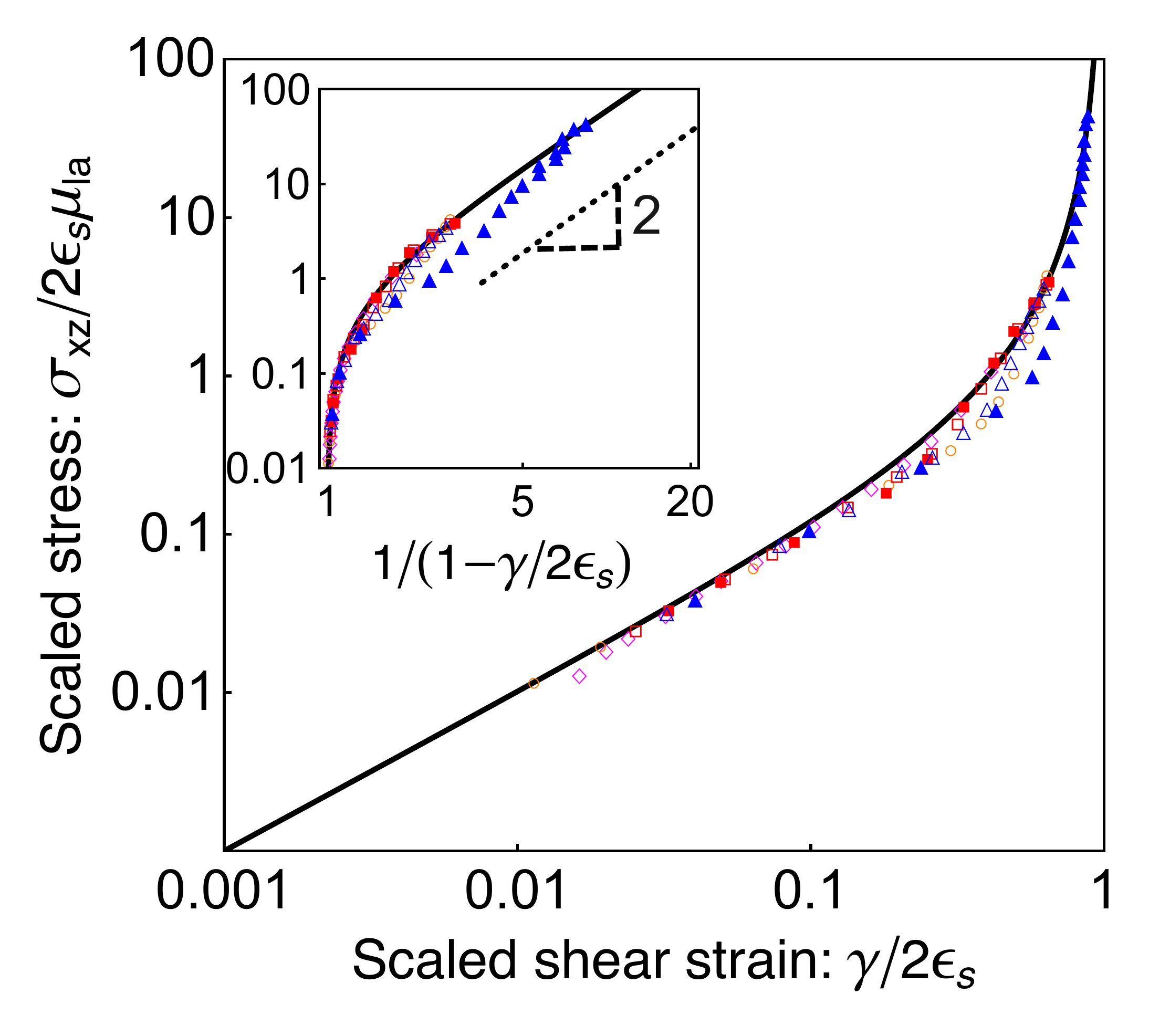} 
  \caption{(Color online) Stress-strain relation for biopolymer gels under simple shear. The discrete data points are taken from the shear experiments of different actin gels\cite{GardelMacKintosh2004,Lubensky2005,BauschHolzapfel2013,Bausch2010}. The black solid line is a theoretical curve plotted from Eq.~(\ref{eq:ExtForce-nonlinsmall-sigmaxz}). Inset: The stiffening scaling law of the normalized shear stress $\sigma_{xz}/2\epsilon_{s}\mu_{\rm la}\sim (1-\gamma/2\epsilon_{s})^{-2}$ as seen in Eq.~(\ref{eq:ExtForce-nonlinsmall-sigmaxz}) as $\gamma \to 2\epsilon_s$. Here we have taken $\epsilon_b\to 0$, $\rho_0=0$, and hence the zero-strain shear modulus is given by $\mu_{\rm la}=\mu_{0}/2$.}
\label{fig:ExtForce-stress-strain}   
\end{figure} 

\begin{figure} 
  \centering
  \includegraphics[clip=true, viewport=1 1 650 580, keepaspectratio, width=0.38\textwidth]{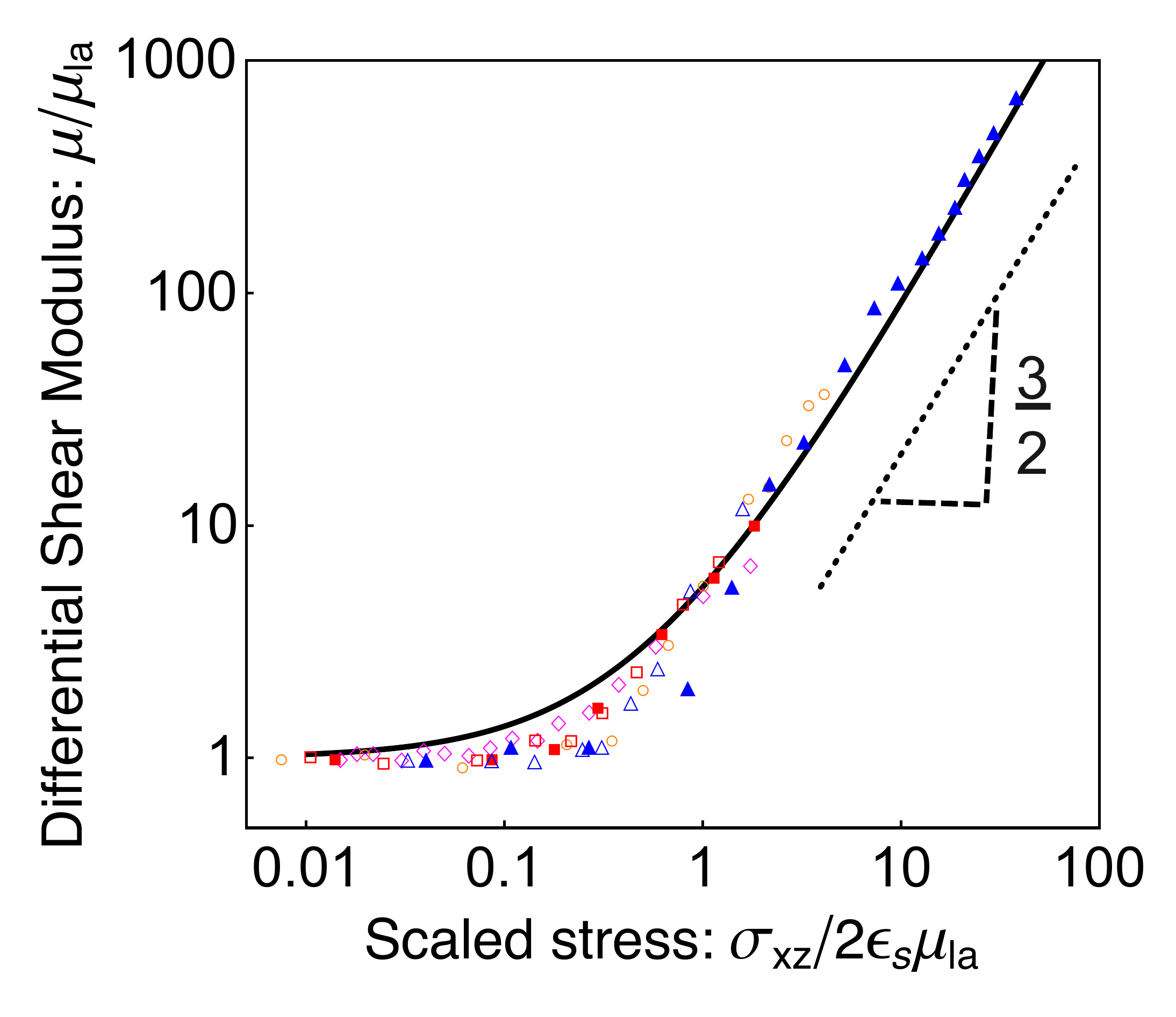}
  \caption{(Color online) The $3/2$-power stiffening law for semiflexible biopolymer gels under simple shear, \emph{i.e.}, $\mu/\mu_{\rm la}\sim (\sigma_{xz}/2\epsilon_{s}\mu_{\rm la})^{3/2}$. The discrete data points are the same as the Fig.~\ref{fig:ExtForce-stress-strain} taken from the shear experiments of different actin gels\cite{GardelMacKintosh2004,Lubensky2005,BauschHolzapfel2013,Bausch2010}. The black solid line is a theoretical curve plotted from Eq.~(\ref{eq:ExtForce-nonlinsmall-mu}). Here we have taken $\epsilon_b\to 0$, $\rho_0=0$, and hence the zero-strain shear modulus is given by $\mu_{\rm la}=\mu_{0}/2$. } 
\label{fig:ExtForce-differentialModulus-stress}   
\end{figure}

Then the shear stress is given by 
\begin{equation}\label{eq:ExtForce-nonlinsmall-sigmaxz} 
\sigma_{xz}/\mu_0=\frac{1}{2\mu_0}(\sigma_1-\sigma_3)=\frac{3}{5}\gamma\left[\frac{1-\gamma/4\epsilon_{s}}{(1-\gamma/2\epsilon_{s})^2}- \frac{1}{6}\right]+\frac{\rho}{2}\gamma+\epsilon_b(1-\rho),
\end{equation}
with $\rho(\gamma)=\rho_0+(1-\rho_0)\Theta(-\gamma/2 +\epsilon_b)$ and $\Theta$ being the Heaviside step function, from which we calculate 
% the nominal shear modulus, $G \equiv \sigma_{xz}/\gamma$, and 
the differential shear modulus 
\begin{equation}\label{eq:ExtForce-nonlinsmall-mu} 
    \mu/\mu_0\equiv \frac{\partial(\sigma_{xz}/\mu_0)}{\partial\gamma}=
    \frac{1}{2}(C_{\rm sh}+\rho), 
\end{equation}
with $C_{\rm sh}\equiv\frac{6}{5}\left[(1-\gamma/2\epsilon_s)^{-3}-1/6\right]$.
Note that near the stiffening strain $\epsilon_{\rm s}$, the shear stress $\sigma_{xz}$ 
% and the nominal shear modulus $G$ both 
diverges as $(1-\gamma/2\epsilon_s)^{-2}$, as confirmed in experiments and shown in Fig.~\ref{fig:ExtForce-stress-strain}.
% and \ref{fig:ExtForce-nominalmodulus-strain}. 
The differential shear modulus $\mu$ then diverges as $\mu \sim (1-\gamma/2\epsilon_s)^{-3}$, \emph{i.e.}, $\mu \sim \sigma_{xz}^{3/2}$, which is the well-known universal $3/2$-power strain stiffening as observed in various biopolymer gels (see Fig.~\ref{fig:introduction-GelNonlinearity}) that can be attributed to the inextensibility of individual stiff semiflexible biopolymers \cite{MacKintosh2014,GardelMacKintosh2004,Lubensky2005}. The nice fit of the above theoretical results with experimental data as shown in Figs.~\ref{fig:ExtForce-stress-strain} and \ref{fig:ExtForce-differentialModulus-stress}, provides a justification of our three-chain model using simple interpolated force-extension relations of single biopolymers in quantitatively describing the nonlinear elasticity of biopolymer gels. 

%-----------------

Note that the stiffening strain $\epsilon_s$ is experimentally known to be related to the concentration of actin, collagen, or crosslinkers in biological gels\cite{MacKintosh2014,GardelMacKintosh2004,Lubensky2005}. Preliminary estimates based on the experimental data yield values $3\%<\epsilon_s<35\%$. Theoretically, strong nonlinearity may be introduced at extremely small strain by taking arbitrarily small values of $\epsilon_s$\cite{Xu2015PRE}. 

The normal stress is given by
\begin{equation}\label{eq:ExtForce-nonlinsmall-sigmazz} 
\sigma_{zz}/\mu_0=\frac{1}{2\mu_0}(\sigma_1+\sigma_3)=\frac{3}{5}\gamma\left[\frac{1-\gamma/4\epsilon_{s}}{(1-\gamma/2\epsilon_{s})^2}- \frac{1}{6}\right]-\frac{\rho}{2}\gamma-\epsilon_b(1-\rho).
\end{equation}
It can be expanded in small shear strain $0<\gamma\ll \epsilon_s$ to the leading order as
\begin{equation}\label{eq:ExtForce-nonlinsmall-sigmazz2} 
\sigma_{zz}/\mu_0\approx (\gamma/2-\epsilon_b) [1-\rho(\gamma)] + \mathcal{O} (\gamma^2).
\end{equation}
Note that $\sigma_{zz}$ is positive for $\rho_0<1$, that is, contracting along the normal direction of the biopolymer gel under simple shear, which is usually referred to as \emph{negative Poynting effect} or \emph{negative normal stress}\cite{Janmey2007NatMat}. Normal stress (being zero in linear elastic materials) is generally regarded as a nonlinear phenomenon, since their sign cannot depend on the direction of shear, for symmetry reasons. The unusual normal stress of biopolymer gels obviously arises from their intrinsic nonlinear (stiffening-softening) elasticity. In a typical biopolymer gel with $\epsilon_b\ll \epsilon_s$ and $\rho_0\approx 0$,
if $\gamma/2<\epsilon_b$ and no gel softening, then $\rho=1$ and the positive $\sigma_{zz}$ is $\sigma_{zz}\sim \gamma^2$. If $\gamma/2>\epsilon_b$, biopolymers buckle and the gel softens (but no stiffening yet) in the principal direction  $\hat{\mathbf{c}}^{(3)}$, then $\rho \approx 0$ and $\sigma_{zz} \sim \gamma$, being the same order of magnitude as shear stress $\sigma_{xz}$ as observed in experiments by Janmey {\emph{et al.}}\cite{Janmey2007NatMat} and studied in simulations by Conti and MacKintosh \cite{MacKintosh2006PRL}. In addition, when we change the sign of shear strain from $\gamma$ to $-\gamma$ (with $\gamma>0$), the principal strains changes from Eq.~(\ref{eq:ExtForce-nonlinsmall-epsilon}) to $\epsilon_1=-\gamma/2<0$ (having compressive softening now), $\epsilon_2=0$, and $\epsilon_3=\gamma/2>0$ (having tensile stiffening now). Then as expected, the shear stress also changes sign from $\sigma_{xz}$ given in Eq.~(\ref{eq:ExtForce-nonlinsmall-sigmaxz}) to $-\sigma_{xz}$, but the normal stress will not and stay the same as given in Eq.~(\ref{eq:ExtForce-nonlinsmall-sigmazz}). 

{\emph{Uniaxial extension}} -- For biopolymer gels under uniaxial tensile stresses\cite{Xu2015PRE,Janmey2016,MacKintoshJanmey2016}, for example, along $x$-axis, we assume the gel deformation is homogeneous again and the deformation tensor is given by
\begin{equation}\label{eq:ExtForce-nonlinsmall-Fextension} 
{\mathbf F}=\begin{bmatrix}
\lambda_1 & 0 & 0\\
0 & \lambda_2 & 0\\
0 & 0 & \lambda_2
\end{bmatrix}.
\end{equation}
The three principal directions are along the three coordinate axes and the principal strain components are given by $\epsilon_{1}=\lambda_1-1>0$ (for extension), $\epsilon_2=\epsilon_3=\lambda_2-1\leq0$ (for non-negative Poisson's ratios), respectively. The three principal stress components are then given by Eq.~(\ref{eq:ExtForce-nonlinsmall-sigmaepsilon}). 

Particularly, a homogeneous uniaxial extension of a biopolymer gel with longitudinal (or axial) strain $\epsilon_1=\epsilon_1^0>0$ and transverse strains $\epsilon_2=\epsilon_3=\epsilon_2^0$ can be generated by  a uniaxial tensile stress with $\sigma_1(\epsilon_1^0, \epsilon_2^0)=\sigma_1^0$ and $\sigma_2(\epsilon_1^0, \epsilon_2^0)=\sigma_3(\epsilon_1^0, \epsilon_2^0)=0$, from which we calculate the differential Poisson ratio and longitudinal Young's modulus, respectively, as
\begin{equation}\label{eq:ExtForce-nonlinsmall-nu12E1}  
\nu_{12}\equiv - \frac{d\epsilon_2^0}{d\epsilon_1^0}=\frac{\tilde{K}}{2(\tilde{K}+\rho\mu_0)},\quad
E_1\equiv \frac{d\sigma_1^0}{d\epsilon_1^0}=2\mu_0(C_{\rm uni}+\rho\nu_{12})
\end{equation}
with $C_{\rm uni}\equiv\frac{6}{5}\left[(1-\epsilon_1^0/\epsilon_s)^{-3}-1/6\right]$.
To calculate the elastic properties in the transverse direction of the gel under given longitudinal prestress $\sigma_1^0$, we further apply an infinitesimal transverse stress $\sigma$ (for example along the direction-$2$). In this case, the new equilibrium conditions are given by
\begin{subequations}\label{eq:ExtForce-nonlinsmall-sigmaEq} 
\begin{equation}\label{eq:ExtForce-nonlinsmall-sigmaEq1} 
\sigma_1(\epsilon_1^0+\delta_1, \epsilon_2^0+\delta_2, \epsilon_2^0+\delta_3)=\sigma_1^0,
\end{equation}
\begin{equation}\label{eq:ExtForce-nonlinsmall-sigmaEq2} 
\sigma_2(\epsilon_1^0+\delta_1, \epsilon_2^0+\delta_2, \epsilon_2^0+\delta_3)=\sigma,
\end{equation}
\begin{equation}\label{eq:ExtForce-nonlinsmall-sigmaEq3} 
\sigma_3(\epsilon_1^0+\delta_1, \epsilon_2^0+\delta_2, \epsilon_2^0+\delta_3)=0.
\end{equation}
\end{subequations}
From Eqs.~(\ref{eq:ExtForce-nonlinsmall-sigmaepsilon-1}) and (\ref{eq:ExtForce-nonlinsmall-sigmaepsilon-23}), we calculate $\delta_1=\delta_1(\delta_2; \epsilon_1^0)$ and $\delta_3=\delta_3(\delta_2; \epsilon_1^0)$, and the differential Poisson ratios are $\nu_{21}(\epsilon_1^0)\equiv -{d\delta_1}/{d\delta_2}|_{\delta_2=0}$ and
$\nu_{23}(\epsilon_1^0)\equiv -{d\delta_3}/{d\delta_2}|_{\delta_2=0}$.
Substituting $\delta_1(\delta_2; \epsilon_1^0)$ and $\delta_3(\delta_2; \epsilon_1^0)$ into Eq.~(\ref{eq:ExtForce-nonlinsmall-sigmaEq2}) we obtain $\sigma=\sigma(\delta_2; \epsilon_1^0)$, from which we can calculate the transverse Young's modulus $E_2(\epsilon_1^0)\equiv {d\sigma}/{d\delta_2}|_{\delta_2=0}$:
\begin{equation}\label{eq:ExtForce-nonlinsmall-E2nuij} 
E_2 =2\mu_0(C_{\rm uni}\nu_{21}+\rho), \quad 
\nu_{21} = \frac{\rho}{C_{\rm uni}}\nu_{23}, \quad \nu_{23} = \frac{\tilde{K}}{(1+\frac{\rho}{C_{\rm uni}})\tilde{K}+2\rho\mu_0}.
\end{equation}  

To better understand the unique elastic properties of a biopolymer gel under uniaxial stretch, we consider some limiting cases as follows. 

(i) In the linear limit of $\epsilon_1^0/\epsilon_s\ll 1$ and $|\epsilon_{2,3}^0|/\epsilon_b>1$, the biogel behaves like a linear isotropic material where $C_{\rm uni}=1$, $\rho=1$, and hence the linear Young's modulus $E_1=E_2=E_0$ and the linear Poisson ratio $\nu_{12}=\nu_{21}=\nu_{23}=\nu_0$ are given by Eq.~(\ref{eq:ExtForce-liniso-E0nu0}). 
% Then, the limit of zero Poisson ratio $\nu_0=0$ corresponds to zero osmotic modulus, \emph{i.e.}, $\tilde{K}=0$, and the limit of weak compressibility with $\nu_0\approx 1/2$ corresponds to large osmotic modulus, \emph{i.e.}, $\tilde{K}\gg \mu_0$.

(ii) In the linear anisotropic limit of $\epsilon_1^0/\epsilon_s\ll 1$ and $|\epsilon_{2,3}^0|/\epsilon_b>1$, we have $C_{\rm uni}=1$, $\rho=\rho_0 \ll 1$, and hence $E_1\approx 2\mu_0$, $E_2\approx 4\rho_0\mu_0 \ll E_1$, $\nu_{12} \approx 1/2$, $\nu_{21} \approx \rho_0$, and $\nu_{23}\approx 1$. 
% the stress components are given in Eq.~(XXX) and hence
% \begin{subequations}\label{eq:fs-trans-b0-sigma}
% \begin{equation}\label{eq:fs-trans-b0-sigma1}
% 2\mu_0(\epsilon_1^0+\delta_1) + \tilde{K}(\epsilon_1^0+2\epsilon_2^0+\delta_1+\delta_2+\delta_3)=\sigma_1^0,
% \end{equation}
% \begin{equation}\label{eq:fs-trans-b0-sigma2}
% 2\rho\mu_0(\epsilon_2^0+\delta_2) + \tilde{K}(\epsilon_1^0+2\epsilon_2^0+\delta_1+\delta_2 +\delta_3)=\sigma,
% \end{equation}
% \begin{equation}\label{eq:fs-trans-b0-sigma3}
% 2\rho\mu_0(\epsilon_2^0+\delta_3)+ \tilde{K}(\epsilon_1^0+2\epsilon_2^0+\delta_1+\delta_2+\delta_3)=0,
% \end{equation}
% \end{subequations}

% $\nu_{23} =\frac{\nu_{12}}{1-(1-\rho_0)\nu_{12}}$

(iii) In the limit of strong nonlinear stiffening with $\epsilon_1^0/\epsilon_s\to 1$, the uniaxial stress $\sigma_1$ diverges as $\sigma_1\sim (1-\epsilon_1^0/\epsilon_s)^{-2}$ and $C_{\rm uni}$ diverges as $C_{\rm uni} \sim (1-\epsilon_1^0/\epsilon_s)^{-3}$, the longitudinal Young's modulus (\ref{eq:ExtForce-nonlinsmall-nu12E1}) stiffens as $E_1 \sim (1-\epsilon_1^0/\epsilon_s)^{-3}$, \emph{i.e.}, $E_1 \sim \sigma_1^{3/2}$, but the transverse Young's modulus doesn't diverge, scaling as $E_2 \sim \rho \mu_0$, and the Poisson's ratios are $\nu_{21} \to 0$ and $\nu_{12} \sim \nu_{23} \sim \nu_0$.

% Note that generally in biopolymer gels, $\nu_{21}$ is not equal to $\nu_{12}=\nu_0$ in the entire range of pre-stress $\sigma_1$ (for $\rho <1$), which is different from transverse isotropic (or hexagonal) materials although they both have hexagonal symmetry. That is, the elastic response of fibrous biopolymer gels to uniaxial stresses are characterized by \textit{three} Poisson's ratios: $\nu_{12}$, $\nu_{21}$ and $\nu_{23}$ (with $\nu_{23}>\nu_{21}$); \textit{two} Young's moduli: $E_1$ (longitudinal) and $E_2$ (transverse). 

% \begin{figure}[htbp]
%   \centering
%   \includegraphics[clip=true, viewport=1 1 600 400, keepaspectratio, width=0.4\textwidth]{SuppPoisson.eps}
%   \caption {Stretch-induced anisotropy in Poisson ratios of semiflexible polymer gels under uniaxial tensile stress $\sigma_1$ in the limit of weak compressibility. Here $\rho_0=0.1$, $\epsilon_s=0.1$ and $\nu_0=0.49$. Differential Poisson ratio $\nu_{ij} \equiv -{\partial \epsilon_j}/{\partial \epsilon_i}$ and $\tilde{\nu}_2\equiv(\nu_{21}+\nu_{23})/2$ vs. longitudinal strain $\epsilon_1$.
%   }
%   \label{Fig:SuppPoisson}
% \end{figure}

%============================%
\subsection{Continuum models of nonlinear biopolymer gels at large affine deformation \label{sec:ExtForce-nonlinlarge}}
%============================%
Up to now, we focus only on nonlinear models of biopolymer gels at small deformation. However, there are a number of entirely new effects, which are peculiar to large elastic deformations and are not to be anticipated for small-strain. In this subsection, we briefly review continuum models that are recently developed for the nonlinear elasticity at finite (large) deformations. Some of these models have also been employed to study the  transmission of internal cellular forces in fibrous biopolymer gels.
% for example, the non-zero normal stresses applied to produce a large shear strain 

\subsubsection{Storm {\emph{et al.}} full-network model}\label{sec:ExtForce-nonlinlarge-StormModel}   

In 2005, Storm {\emph{et al.}} \cite{Lubensky2005} extended the small-strain approach of MacKintosh, K\"as, and Janmey \cite{Mackintosh1995} and Morse \cite{Morse1998} to large strains, and proposed a theoretical model of equilibrium semiflexible network, where the constituting biopolymers are randomly oriented and are assumed to be deformed affinely. 

Consider the affine deformation of a crosslinked fibrous network that is described by the uniform deformation gradient tensor, $\mathbf{F} = {\partial \mathbf{x}}/{\partial \mathbf{X}}$ with $\mathbf{X}$ the reference position, and $\mathbf{x}$ the deformed position. The network density can be measured by the total filament length per unit volume, $\rho_f$, which is not conserved under deformation. After deformation, the length density of filaments per unit volume crossing a plane perpendicular to $j$ axis transforms to $(\rho_f/J)F_{jk}\hat{l}_{k}$, where $\mathbf{\hat{l}}$ is the orientation of a filament segment between two neighboring crosslinks in the undeformed network, and $J=\rm{det}(\mathbf{F})$ measures the relative volume change of the deformed network. 

The tension in a filament segment connecting two neighboring (affinely deforming) crosslinks can be denoted by $f(\epsilon_f=|\mathbf{F}\cdot\mathbf{\hat{l}}|-1)$, where $\epsilon_f$ (or $\lambda_f=|\mathbf{F}\cdot\mathbf{\hat{l}}|$) represents for the axial strain (or extension ratio) of the filament segment, and the specific force-extension relation are discussed in Sec.~\ref{sec:Single}. The $i$-th component of the tension can be calculated as $f(\epsilon_f)F_{il}\hat{l}_{l}/|\mathbf{F}\cdot\mathbf{\hat{l}}|$, and the $ij$ component of the elastic stress tensor can be obtained by following the Doi-Edwards construction of stress through adding all contributions of the filaments weighed by the amount of filament length crossing the $j$ plane\cite{Doi1988,MacKintosh2014,Lubensky2005}:
\begin{equation}\label{eq:ExtForce-nonlinlarge-Storm-sigmaij} 
\sigma_{ij}=\frac{\rho_f}{J}
\langle f(\epsilon_f)\frac{F_{il}\hat{l}_{l}F_{jk}\hat{l}_{k}}{|\mathbf{F}\cdot\mathbf{\hat{l}}|}\rangle.
\end{equation}
Particularly for simple shear with small strain $\gamma\ll 1$, deformation tensor $\mathbf{F}$ is given in Eq.~(\ref{eq:ExtForce-nonlinsmall-Fshear}), and the volume is conserved with $J=1$. In this case, the stress will be simplified as \cite{MacKintosh2014,Lubensky2005}:
\begin{equation}\label{eq:ExtForce-nonlinlarge-Storm-sigmaij2} 
\sigma_{ij}\approx \left\langle\rho_ff\left(\gamma \hat{l}_{x} \hat{l}_{z}\right) \hat{l}_{i} \hat{l}_{j}\right\rangle,
\end{equation} 
with $\gamma \hat{l}_{x} \hat{l}_{z}$ being the axial strain of a polymer along $\mathbf{\hat{l}}$ and the general nonlinear force-extension relation of individual filaments given in Eq.~(\ref{eq:Single-Stretch-xphi}). We now consider two limiting cases as follows. 

(i) In the linear limit of $\gamma\ll 2\epsilon_s$ with stiffening strain $\epsilon_s \approx 1/6c$ (or $\sigma \ll \sigma_{s}=\rho_f k_{B}T\ell_{p}/\ell_c^2$) as discussed in Sec.~\ref{sec:ExtForce-nonlinsmall}, the force-strain relation is given in Eq.~(\ref{eq:ExtForce-nonlinsmall-sigmaepsilon}) and the shear stress becomes $\sigma_{xz}=15 \mu_0 \gamma \langle \hat{l}_{x}\hat{l}_{z}\hat{l}_{x}\hat{l}_{z}\rangle$ with $\mu_0={6\rho_f k_{B}T\ell_p^2}/{\ell_c^3}$. If the filament orientation follows isotropic distribution, then $\sigma_{xz}= \mu_0 \gamma$ and hence $\mu_0$ is the linear shear modulus, which is in the same order of magnitude as that obtained from 3-chain model in Sec.~\ref{sec:ExtForce-nonlinsmall-XuSafran3chain}.

(ii) In the nonlinear stiffening limit of $\gamma\to 2\epsilon_s$ (or $\sigma \gg \sigma_{s}$), this model successfully explained the universal $3/2$-power stiffening law as observed in various biopolymer gels, \emph{i.e.}, the differential shear modulus, $\mu$, scales as $\mu \sim \sigma_{xz}^{3/2}$. 

However, using the full nonlinear force-extension relation of individual biopolymers as reviewed in Sec.~\ref{sec:Single}, it is not possible in general to arrive at a compact analytical form of the stress-strain relation of biopolymer gels even for simplest isotropic filament-orientation distribution. Moreover, the original form of the model has not discussed the elastic responses of biopolymer gels to compressive stresses at all. Furthermore, the proper modeling of the compressibility of biopolymer gels also needs further efforts.

\subsubsection{Shokef-Safran hyperelastic model}\label{sec:ExtForce-NonlinLarge-ShokefModel}  

In 2012, Shokef and Safran \cite{Sam2012a,Sam2012b} extended Knowles's hyperelastic model for incompressible elastic solids \cite{Knowles1977} to propose a phenomenological model of weakly compressible biopolymer gels where the elastic energy density functional is given by
\begin{equation}\label{eq:ExtForce-nonlinlarge-Shokef-F} 
F=\frac{\mu_0}{2b}{\left[1+\frac{b}{n}(\tilde{I}_{1}-3)\right]^n-1} + \frac{K}{2}(J-1)^2,
\end{equation}
where $\tilde{I}_{1}\equiv I_{1}/J^{2/3}$, $I_{1}=\rm{tr}(\mathbf{B})$ are the strain invariants, and $\mathbf{B}=\mathbf{F}\cdot \mathbf{F}^T$ is the left Cauchy-Green strain tensor \cite{Bower2009}. 
The dimensionless parameter $b$ and $n$ characterize the nonlinearity and the stiffening strain $\epsilon_s$ is given by $\epsilon_s=\sqrt{-n/4b}$. $\mu_0$ and $K$ are the linear shear modulus and bulk modulus, respectively. 

In the linear limit with vanishingly small $b$ or infinitely large $\epsilon_s$, the elastic energy density (\ref{eq:ExtForce-nonlinlarge-Shokef-F}) reduces to the compressible neo-Hookean form 
\begin{equation}\label{eq:ExtForce-nonlinlarge-Shokef-FNeoHookean}
F_{\rm{NH}}=\frac{\mu_0}{2}(\tilde{I}_{1}-3)+ \frac{K}{2}(J-1)^2,
\end{equation}
and the Cauchy stress tensor $\sigma_{ij}={J}^{-1}F_{ik}\frac{\partial F}{\partial F_{kj}}$ is obtained as $\bm{\sigma}_{\rm{NH}}= {\mu_0}{J^{-5/3}}(\mathbf{B}-\frac{1}{3}I_{1}\mathbf{I}) +K(J-1)\mathbf{I}$, with $\mathbf{I}$ being the unity tensor. Furthermore, for small deformations, $B_{ij}\approx \delta_{ij} +2\epsilon_{ij}$ with $\epsilon_{ij}$ being the linear strain tensor. In this case, $I_{1}\approx 3+2\rm{tr}(\bm{\epsilon})$ and $J\approx 1+\rm{tr}(\bm{\epsilon})$ and the constitutive relations reduce to Hooke's law, $\bm{\sigma}= 2\mu_0\bm{\epsilon} +(K-\frac{2}{3}\mu_0)\rm{tr}(\bm{\epsilon})\mathbf{I}$, identical to Eq.~(\ref{eq:ExtForce-liniso-sigmaepsilon}) for linear isotropic materials.

Particularly for simple shear in $xz$ plane, the shear stress is given by 
$\sigma_{xz}=\mu_0\gamma(1-\gamma^2/4\epsilon_s^2)^{n-1}$. 
In the limit of small shear with $\gamma\ll 2\epsilon_s$, the differential shear modulus $\mu\equiv {\partial \sigma_{xz}}/{\partial\gamma}=\mu_0$ is constant. For $n<0$ and $b>0$, $\sigma_{xz}$ diverges as $\sigma_{xz}\propto (1-\gamma/2\epsilon_s)^{n-1}$ when the shear strain approaches stiffening strain $\epsilon_s$, while the shear modulus stiffens as $\mu\propto\sigma^{\beta}$ with $\beta =\frac{n-2}{n-1}$. That is, $n$ determines the exponent $\beta$ that quantifies the strain-stiffening behavior of $\mu$ vs $\sigma$. The nonlinearity of semiflexible chains implies that $\beta=\frac{3}{2}$ as discussed in Sec.~\ref{sec:Single-stretch}, which is obtained by taking $n=-1$.

Shokef and Safran \cite{Sam2012a,Sam2012b} used this model to show that nonlinear strain stiffening elasticity facilitates the transmission of internal forces, which are caused either by material defects and inhomogeneities or by active forces that molecular motors generate in living cells. Exponential scaling laws relate the far-field renormalized strain to the near-field strain applied by the inclusion or active force. However, this hyperelastic model shows strain stiffening for both extension and compression from the symmetry in the energy functional (\ref{eq:ExtForce-nonlinlarge-Shokef-F}), which, therefore, can't represent the softening behaviors of biopolymer gels upon compression. Furthermore, strain-induced fiber alignment and elastic anisotropy have not been considered either. A more realistic model that takes into account of both stretch-stiffening and compressive-softening is necessitated\cite{PalmerBoyce2008,Xu2015PRE,Meng2017} as reviewed in the previous section. 

\subsubsection{Wang {\emph{et al.}} fiber-reinforced material model} \label{sec:ExtForce-NonlinLarge-WangModel} 

In 2014, Wang {\emph{et al.}} \cite{Shenoy2014b} developed a constitutive model of fibrous matrices. 
They assume that when a fibrous matrix undergoes stretch, there are two families of fibers: the set of fibers that align with the direction of the maximum principal stretch as the material is loaded, and the set of fibers that do not align with the applied load and thus display an isotropic mechanical response.

To capture the presence of these two distinct families of aligned and isotropic fibers, Wang {\emph{et al.}} assume that the overall energy density, $F$, of the collagen network consists of two contributions \cite{Shenoy2014b}:
\begin{equation}\label{eq:ExtForce-nonlinlarge-Wang-F} 
F= F_{\rm{NH}}(\tilde{I}_{1},J)+\sum_{i=1}^3f(\lambda_i),
\end{equation}
with $\lambda_{1,2,3}$ being the principal extensions. 
Here the first term $F_{\rm{NH}}$ captures the isotropic response, which takes the compressible neo-Hookean form as given in Eq.~(\ref{eq:ExtForce-nonlinlarge-Shokef-FNeoHookean}), and the second term is the contribution from the aligned fibers. The stress tensor is then given by
\begin{equation}\label{eq:ExtForce-nonlinlarge-Wang-dfdlambdai}
\begin{aligned}
 \bm{\sigma}= 
    \bm{\sigma}_{\rm{NH}}+ 
    \frac{1}{J}\sum_{i=1}^{3}\frac{\partial f(\lambda_{i})}{\partial\lambda_{i}}\lambda_{i}(\mathbf{\hat{c}}^{(i)}\otimes\mathbf{\hat{c}}^{(i)})
\end{aligned}
\end{equation}
where $\bm{\sigma}_{\rm{NH}}$ is the isotropic neo-Hookean stress tensor as given near Eq.~(\ref{eq:ExtForce-nonlinlarge-Shokef-FNeoHookean}), $\mathbf{\hat{c}}^{(i)}$ are the unit vector along the principal strain directions, and the energy $f(\lambda_{i})$ takes the following piecewise form
\begin{equation}
\frac{\partial f(\lambda_{i})}{\partial\lambda_{i}}=
\begin{cases}
     0   &  \lambda_{i}<\lambda_L, \\
    \frac{E_f}{n+1}\left(\frac{\lambda_{i}-\lambda_{L}}{\lambda_R-\lambda_L}\right)^n(\lambda_{i}-\lambda_L),  &  \lambda_L\leq\lambda_{i}<\lambda_R,\\
    E_{f}\left[\frac{\lambda_R-\lambda_L}{n+1}+\frac{(1+\lambda_{i}-\lambda_R)^{m+1}-1}{m+1}\right],  & \lambda_{i}\geq \lambda_R. 
\end{cases}
\end{equation}
with $\lambda_L=\lambda_c-\lambda_t/2$ and $\lambda_R=\lambda_c+\lambda_t/2$. Note that the principal filamentous stress in the second term of Eq.~(\ref{eq:ExtForce-nonlinlarge-Wang-dfdlambdai}) vanishes below the critical (tensile) principal stretch, $\lambda_c$, and stiffens above $\lambda_c$ in the direction of tensile principal stretches, as observed in experiments and discrete fiber simulations\cite{Shenoy2014a,Shenoy2014b}. The stiffening behavior is assumed  phenomenologically and is characterized by the modulus $E_f$ and a stiffening exponent, $m>0$. A smooth interpolation function is introduced between transition region
$(\lambda_L,\,\lambda_R)$ to ensure that the stress continuity near the transition point around $\lambda_c$ with transition width $\lambda_{\rm{t}}$ and transition exponent $n>0$. 

Wang {\emph{et al.}} \cite{Shenoy2014b} used this model with finite element simulations to investigate systematically the impact of cells and their contractility on their matrices. They showed that tension-driven collagen-fiber alignment plays a crucial role in force transmission. Small critical stretch for fiber alignment, large fiber stiffness and fiber strain-stiffening behavior enable long-range interaction. Furthermore, the range of collagen-fiber alignment for elliptical cells with polarized contraction is much larger than that for spherical cells with diagonal contraction.  In addition, recently, they further extended this model \cite{Shenoy2019} by including the coupling of the multiaxial deformations and the Poisson effect. They showed that the fibrous nature of the extracellular matrix leads to strong coupling between these modes due to bending, buckling, and stretching of the fibers. However, Wang {\emph{et al.}} \cite{Shenoy2014b} has not quantified the various scaling regimes for the transmission of \textcolor{black}{internal active cellular forces} in networks composed of different types of fibers. Particularly, this model can't explain the extremely slow decay of displacements, $u$, or forces as experimentally observed \cite{NotbohmLesman2015,Mark2020} in the vicinity of contracting cells that is embedded in 3D fibrous matrices with $u\propto 1/r^n$ with $n<0.5$ as shown in Fig.~\ref{fig:IntForce-CellMatrixScaling}.

\subsubsection{Meng-Terentjev 3-chain model} \label{sec:ExtForce-NonlinLarge-MengModel} 

In 2016, Meng and Terentjev \cite{Meng2016} proposed a general 3-chain model for biopolymer gels at large affine deformation. The free energy density is given by $F=\frac{1}{3}n_f\sum_{i=1}^3 w_{\rm{chain}}(\lambda_i)$, in which the free energy of an individual biopolymer can be calculated by using $w_{\rm{chain}} (\lambda)= \ell_c\int_{x_r}^{\lambda x_r} f(x') dx'$ from the force-extension relations in Eq.~(\ref{eq:Single-Stretch-Terentjev}). The free energy density $F$ can be rewritten in terms of strain invariants as 
\begin{equation}\label{eq:ExtForce-nonlinlarge-Meng-F}
F=\frac{n_fk_B T}{3}\left[\frac{\pi^2c}{2}(3-x_0^2I_1)+ \frac{2(3-2I_1x_0^2+I_2x_0^4)}{\pi c(1-I_{1}x_0^2+I_{2}x_0^4-I_{3}x_0^6)}\right],
\end{equation}
where $c = l_{p}/\ell_c$ and the pre-tension $x_r= \xi \ell_c$ with $\xi$ being the mesh size at reference (undeformed) state, $I_{1,2,3}$ are the three strain invariants: 
\begin{equation}\label{eq:ExtForce-nonlinlarge-Meng-I123}
I_{1}=\lambda_{1}^{2}+\lambda_{2}^{2}+\lambda_{3}^{2}, \quad I_{2}=\lambda_{1}^{2} \lambda_{2}^{2}+\lambda_{1}^{2} \lambda_{3}^{2}+\lambda_{2}^{2} \lambda_{3}^{2}, \quad I_{3}=\lambda_{1}^{2} \lambda_{2}^{2} \lambda_{3}^{2}.
\end{equation}
This model is applicable in a large range of polymer stiffnesses $c$. It is similar but more complicated and less tractable (for example in studying the decay of internal cellular forces) than Xu-Safran 3-chain model for stiff semiflexible biopolymers with $c\gsim 1$ at small affine deformation.

Given the compact analytical form of the elastic energy $F$, the stress-strain relations and some of the nonlinear elastic properties of fibrous network have been calculated and justified, \emph{e.g.}, the $3/2-$power law stiffening (with differential modulus $\mu \sim \sigma^{3/2}$) and the negative normal stress in biopolymer gels under simple shear. The stability condition has also been obtained from the non-negativeness of linear shear modulus:
\begin{equation}\label{eq:ExtForce-nonlinlarge-Meng-stability}
c\le \frac{2}{\pi^{3/2}} \frac{\sqrt{1+x_r^2}}{(1-x_r^2)^{3/2}}. 
\end{equation}
In addition, Meng and Terentjev \cite{Meng2016} have also shown that both 3-chain and 8-chain models fit the shear-experiment data equally well. However, more stringent tests are needed to distinguish which of these two models is working better in fibrous networks. Furthermore, we notice that this general model has not been used to study the transmission of internal cellular forces in fibrous networks. In this case, a weak compressibility has to be considered explicitly, and a full analytical analysis using this general model will pose huge challenges in comparison to the simpler Xu-Safran 3-chain model. The extension of this model to compressible biopolymer gels still needs further exploration. 

Finally, we would like to point out that the list of continuum models for biopolymer gels may not be complete and for other models, we suggest the nice reviews on semiflexible polymers networks by Broedersz and MacKintosh \cite{MacKintosh2014} and by Meng and Terentjev \cite{Meng2017}. 

%==========================================================%
% \section{Decay of displacements induced by a spherically contracting cell in biopolymer gels}
\section{Continuum models for the transmission of internal cellular forces \label{sec:IntForce} }
%==========================================================%
In this section, we use the continuum models of linear materials and biopolymer gels as reviewed previously to study the decay of displacements induced by a contractile cell that is well adhered to elastic materials or biopolymer gels. Particularly, we show how the nonlinear elasticity of fibrous biopolymer gels \textcolor{black}{impacts} the transmission of internal cellular forces or displacements.

\begin{figure}[h]
\centering
\includegraphics[height=0.8\linewidth]{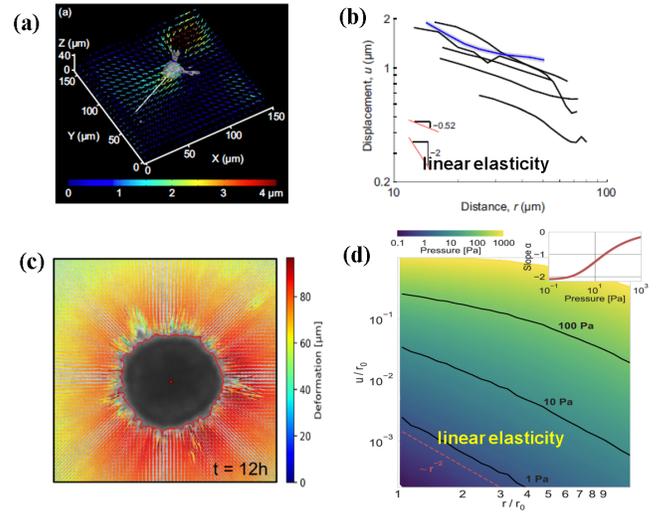}
  \caption{(Color online) Experimentally measured slow decay of displacement induced by contracting cells embedded in three-dimensional biopolymer gels. (a) The displacement vector fields induced by an isolated fibroblast cell in a fibrin gel. Paths (white) are chosen proceeding outward from the cell body. (b) The decay of displacement magnitudes along the paths are averaged for multiple time points and plotted. Each curve is for a different cell. An effective near-field power-law exponent $n\approx 0.5$ is obtained. Reproduced from Notbohm \emph{et al.} \cite{NotbohmLesman2015} with permission from Royal Society. (c) The displacement magnitude fields induced by a spheroid containing 7,500 (U87-MG) glioblastoma cells embedded in a collagen gel (12 h after the gel has polymerized). (d) Power-law scaling decay of near-field displacements as a function of the distance $r$ away from the cell (normalized by the cell radius $r_0$), obtained from simulations with spherical inclusions of different contractility (or inbound pressure). Experimental measurements are consistent with these simulation results. The inset shows the near-field ($r/r_0<2$) power-law exponent $-n$ of the deformation field as a function of the cell contractility, in which $n$ can be as small as $0.2$. Reproduced from Mark \emph{et al.} \cite{Mark2020} with permission from eLife Sciences Publications Ltd. } 
\label{fig:IntForce-CellMatrixScaling}
\end{figure}
% due to a stiffening of the collagen fibers. This is in line with reported theoretical models (Xu and Safran, 2015; Grimmer and Notbohm, 2018 and experimental findings (Burkel and Notbohm, 2017; Han \emph{et al}., 2018). Note that a much smaller power-law exponent $n$ in biopolymer gels than $2$ for linear elasticity indicates the long-range force transmission. 
%============================%
\subsection{Boundary value problem for the cell contraction in an infinite elastic material}\label{sec:IntForce-BVP} 
%============================%
An adherent cell can apply active contractions to their surrounding matrix, which can be modeled as a contractile force dipole \cite{Sam2013a}.
In the simplest case, we here review the scaling laws for the decay of displacements that are induced by a spherically contracting cell in a three-dimensional infinite extracellular matrix as done in experiments shown in Fig.~\ref{fig:IntForce-CellMatrixScaling} and schematically shown in Figs.~\ref{fig:IntForce-Soften}-\ref{fig:IntForce-StiffenSoften}. In this case, the active cell contraction can be characterized by a boundary condition of fixed radial displacement $-u_c$ at the cell boundary $r=R_c$, \emph{i.e.},
\begin{equation}\label{eq:IntForce-BVP-uc}
u(r=R_c)=-u_c
\end{equation}
with $u_c>0$ for cell contraction\cite{Sam2012a,Sam2012b,Sam2015,Xu2015PRE}. 

A fibrous extracellular matrix with strong crosslinkers and of pore sizes much smaller than cell dimensions can be modeled as a continuum elastic material as described by the energetic models presented in Sec.~\ref{sec:ExtForce}. Then in any matrix element of size much larger than the average pore size of the matrix gel but much smaller than cell dimensions, the matrix deformation can be assumed to be homogeneous and affine. Therefore, the cell-matrix mechanical interaction yields an elastic boundary value problem with inhomogeneous deformation in the extracellular gel that depends on the radial coordinate, $r$, in spherical geometry. 
Note that in this case the principal directions of strain and stress tensor of each gel element are along the radial and the two perpendicular angular directions. This spherical symmetry significantly simplifies the calculations\cite{Sam2012a,Sam2012b,Sam2015,Xu2015PRE,Shenoy2014b,Lenz2019,Notbohm2015}. 

The equation of mechanical equilibrium $\nabla \cdot \bm{\sigma}=0$ in spherical coordinates simplifies to \cite{Sam2012a,Sam2012b,Sam2015,Xu2015PRE,Shenoy2014b,Lenz2019,Notbohm2015}
\begin{equation}\label{eq:IntForce-BVP-equilibrium}
\frac{d \sigma_1}{d r}+\frac{2}{r}\left(\sigma_1-\sigma_2\right)=0,
\end{equation}
with $\sigma_1=\sigma_{rr}$ and $\sigma_2=\sigma_{\theta \theta}$ being the two principal components of stress tensor along the radial and angular directions, respectively. The principal strains are given by
\begin{equation}\label{eq:IntForce-BVP-epsilon123}
\epsilon_1=u^{\prime} \equiv \frac{d u}{d r}>0, \quad \epsilon_2=\epsilon_3=\frac{u}{r}<0. 
\end{equation}
with $\epsilon_1=\epsilon_{rr}$ and $\epsilon_2=\epsilon_{\theta \theta}$, and $\epsilon_3=\epsilon_{\varphi \varphi}$ being the three principal strain components along the radial and two angular directions, respectively.

Note that in a two-dimensional matrix with a circularly contracting cell, the strain is planar with cylindrical symmetry and all fibers remain in the two-dimensional plane\textcolor{black}{~\cite{Xu2020BJ,Shenoy2014a}}. In this case, the equilibrium equation takes a similar form as Eq.~(\ref{eq:IntForce-BVP-equilibrium}) but the coefficient $2/r$ becomes $1/r$.

We would like to point out that the continuum elastic models reviewed here on the transmission of internal cellular forces in biopolymer gels are most relevant for environments sparse in cells, such as connective tissues and engineered tissue scaffolds. In particular, they would apply to cells in scaffolds in the first few hours following seeding, when cells normally assume a spherical shape, and matrix degradation and synthesis are absent\cite{Koenderink2013,Mao2017}. The predictions provide  new strategies in controlling force transmission in biomaterials by altering the material structure, particularly elastic anisotropy. 

% \begin{figure}[h]
% \centering
% \includegraphics[height=3cm]{contractionCell}
% \caption{An illustration of a contractile cell embedded in an linear homogeneous
% fibrous network.}
% \label{fig:contractionCell}
% \end{figure}

%============================%
\subsection{Linear isotropic materials}\label{sec:IntForce-liniso} 
%============================%
Linear isotropic materials are characterized by the constitutive relation (\ref{eq:ExtForce-liniso-sigmaepsilon}). Substituting Eqs.~(\ref{eq:ExtForce-liniso-sigmaepsilon}) and (\ref{eq:IntForce-BVP-epsilon123}) into the equilibrium equation Eq.~(\ref{eq:IntForce-BVP-equilibrium}), we obtain the equilibrium equation for displacement field $u(r)$ as 
\begin{equation}\label{eq:IntForce-liniso-ueqn}
    \frac{d^2u}{dr^2} +\frac{2}{r}\frac{du}{dr} -\frac{2u}{r^2} =0,
\end{equation}
whose general solution is $u(r)=C_1 r + C_2 r^{-2}$. For a cell contracting in an infinite elastic medium with boundary condition (\ref{eq:IntForce-BVP-uc}) and the natural boundary condition $u(r\to +\infty)=0$, we obtain
\begin{equation}\label{eq:IntForce-liniso-urscaling}
\tilde{u}(\tilde{r})=\tilde{r}^{-2},
\end{equation}
in which we have introduced the normalized displacement $\tilde{u}=-u/u_c$, and the normalized radius $\tilde{r}=r/R_c$.  

%============================%
\subsection{Linear anisotropic materials}\label{sec:IntForce-linaniso} 
%============================%
Linear anisotropic materials are characterized by the constitutive relation (\ref{eq:ExtForce-linaniso-sigmaepsilon}). Substituting Eqs.~(\ref{eq:ExtForce-linaniso-sigmaepsilon}) and (\ref{eq:IntForce-BVP-epsilon123}) into the equilibrium equation~(\ref{eq:IntForce-BVP-equilibrium}), we obtain 
\begin{equation}\label{eq:IntForce-linaniso-ueqn}
    \frac{d^2u}{dr^2} +\frac{2}{r}\frac{du}{dr} -g\frac{2u}{r^2} =0
\end{equation}
where we have used Eq.~(\ref{eq:ExtForce-linaniso-cEnu}); the dimensionless variable $g\equiv {E_{2}(1-\nu_{12})}/{E_{1}(1-\nu_{23})}$, $\nu_{12}$ and $\nu_{23}$ are the Poisson ratios in the radial and transverse planes, respectively. The general solution of Eq.~(\ref{eq:IntForce-linaniso-ueqn}) is $u(r) = C_1 r^{n-1} + C_2 r^{-n}$ with $n$ given by
\begin{equation}\label{eq:IntForce-linaniso-n}
n=\frac{1}{2}\left(1+\sqrt{1+8g}\right).
\end{equation}
For a cell contracting in an infinite elastic medium with boundary condition (\ref{eq:IntForce-BVP-uc}) and the natural boundary condition $u(r\to\infty)=0$, we obtain
\begin{equation}\label{eq:IntForce-linaniso-urscaling}
\tilde{u}(\tilde{r}) =\tilde{r}^{-n}.
\end{equation}
In comparison, in a two-dimensional matrix\textcolor{black}{~\cite{Xu2020BJ,Shenoy2014a}}, $n_{\rm{2D}}=\sqrt{E_{2}/E_{1}}$.
Interestingly, note that in the isotropic limit of $E_1=E_2=E_0$ and $\nu_{12}=\nu_{23}=\nu_0$, we have $g=1$, the solution (\ref{eq:IntForce-linaniso-n}) reduces to its usual isotropic form with $n=2$ in Eq.~(\ref{eq:IntForce-liniso-urscaling}) (or $n=1$ in two dimension) \cite{Shenoy2014a, Sam2012a}.
In the strong anisotropic limit, firstly if $E_1\gg E_2$, we have $g\to 0$ with $n \to 1$ (or $n\to 0$ in two dimension) \cite{Shenoy2014a, Lekhnitskii1981, Lesman2014}, which indicates that displacement decays much more slowly as $r$ increases compared with the isotropic case. This leads to a longer range of force transmission \cite{Sam2012a, Sander2011, Sander2013, Qi2013, Hart2013, Shenoy2014a, Shenoy2014b}. On the other hand, if $E_1\ll E_2$, then $g\gg 1$ and hence $n\gg 2$ (or $n\gg 1$ in two dimension), indicating fast decay of cell-induced displacements and smaller-range of force transmission. 

In addition, as mentioned in Sec.~\ref{sec:ExtForce-linaniso}, Goren \emph{et al.} \cite{Xu2020BJ} has constructed, in their two-dimensional finite element simulations, an intrinsically anisotropic network of linear fibers with predefined tunable elastic anisotropy as discussed in Sec.~\ref{sec:ExtForce-linaniso} and shown in Fig.~\ref{fig:ExtForce-LinearAnisotropy}(a). They studied the decay of displacements induced by a contractile cell in such intrinsically anisotropic networks. Even for very small cell contractions, they found that the decay of cell-induced displacements follows a power-law with exponent shown a very good linear proportionality to the elastic anisotropy, $n\sim\sqrt{E_{2}/E_{1}}$ for all values of $n$ as predicted by the above simple linear anisotropic elasticity (see Fig.~\ref{fig:ExtForce-LinearAnisotropy}(b)). When the network is stiffer along the radial direction ($E_{1}>E_{2}$), it means $n<1$, \emph{i.e.}, the displacements decay slowly, and the range of cell-cell communications mediated by the matrix is considered enhanced. In contrast, when the network is stiffer along the angular direction ($E_{2}>E_{1}$), it means $n>1$, \emph{i.e.}, the displacements decay faster than in linear isotropic elastic medium and the range of cell-cell communication is restricted. This indicates that the transmission of cellular forces in fibrous networks and hence the efficiency of matrix-mediated cell-cell communications can be reprogrammed by modifying the network anisotropic elastic properties.

\begin{figure}[h]
\centering
\includegraphics[height=9cm]{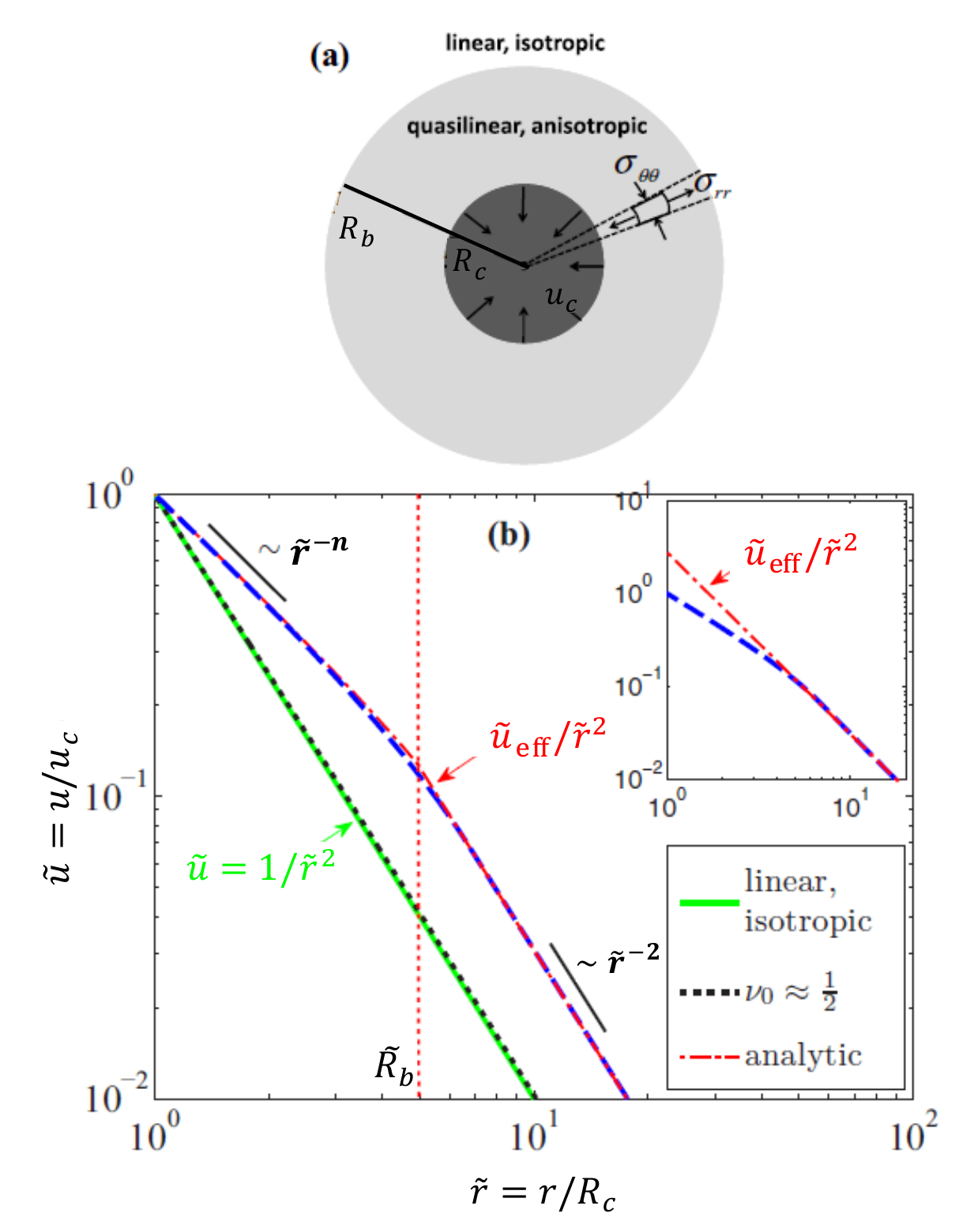}
\caption{(Color online) (a) Schematic illustration of a spherical cell contracting in a biopolymer gel with nonlinear compressive-softening elasticity. Two regimes separated at  $\tilde{r}=\tilde{R}_b$ are identified.
(b) The decay of the displacement induced by the contractile cell for $\rho_{0}=0.1$. Case 1: $v_{0}=0$ (blue dashed line). In the near-field ($1<\tilde{r} \ll \tilde{R}_b$), $\tilde{u}\sim \tilde{r}^{-1.2}$ as in Eq.~(\ref{eq:IntForce-linaniso-urscaling}) for linear anisotropic materials, and in the far-field ($\tilde{r} \gg \tilde{R}_b$), $\tilde{u}\sim \tilde{r}^{-2}$ as in Eq.~(\ref{eq:IntForce-liniso-urscaling}) for the linear isotropic materials. Case 2: $v_{0}=0.49$ (black dotted line), similar to that of linear isotropic materials. Inset: Normalized effective far-field displacement $\tilde{u}_{\mathrm{eff}}$ as explained in Eq.~(\ref{eq:IntForce-Soften-ufar}). Adapted with permission from Xu and Safran \cite{Xu2015PRE}. Copyright(2015) by the American Physical Society.}
\label{fig:IntForce-Soften}
\end{figure}

\subsection{Linear stretch and nonlinear compressive-softening materials \label{sec:IntForce-Soften}}

As mentioned in Sec.~\ref{sec:Single-Compress}, stiff filaments such as athermal rods and semiflexible biopolymers buckle and lose stiffness if the compression force exceeds a critical value. Consequently, crosslinked biopolymer gels or networks comprised of such stiff filaments softens (due to filament buckling) upon increasing compression \cite{Janmey2016,MacKintoshJanmey2016,Kim2014,Kim2016,Xu2017PRE}. Such compressive-softening nonlinear elasticity can be modeled by a piecewise quadratic elastic energy as proposed by Rosakis {\emph{et al.}} \cite{Notbohm2015} and more generally by Xu and Safran in their 3-chain model \cite{Xu2015PRE} (see Eq.~(\ref{eq:ExtForce-nonlinsmall-F2})). 

In this case, the elastic energy can be regarded as a limit of the general energy in Eq.~(\ref{eq:ExtForce-nonlinsmall-F2}) with $\epsilon_s\to \infty$ (\emph{i.e.}, no stiffening behaviors), and then the stress-strain relations can be obtained from Eq.~(\ref{eq:ExtForce-nonlinsmall-sigmaepsilon}). Substituting them into the equilibrium equation (\ref{eq:IntForce-BVP-equilibrium}), we obtain 
\begin{equation}\label{eq:IntForce-Soften-ueqn}
\frac{d^2u}{dr^2}+\frac{2}{r}\frac{du}{dr}-g\frac{2u}{r^2}+(1-g)\frac{2\epsilon_b}{r}=0
\end{equation}
where $g(\rho, \nu_0)=\rho + \nu_0(1-\rho)/(1-\nu_0)$, $\rho(u(r))=\rho_0+(1-\rho_0)\Theta(u/r +\epsilon_b)$, and $0 \le g\le 1$ depending on the Poisson ratio $0\le \nu_0\le 1/2$ and the softening parameter $0\leq \rho_0 \leq 1$. Note that in the limit of $\epsilon_b \to 0$, Eq.~(\ref{eq:IntForce-Soften-ueqn}) reduces to Eq.~(\ref{eq:IntForce-linaniso-ueqn}) for linear anisotropic materials. In addition, $g=1$ at any positions $r$ for incompressible materials with $\nu_0=1/2$ and at positions with small angular compressive strain, $-u/r<\epsilon_b$, for $0\le \nu_0 < 1/2$. In this case, the equilibrium equation (\ref{eq:IntForce-Soften-ueqn}) reduces to Eq.~(\ref{eq:IntForce-liniso-ueqn}) for linear isotropic materials. Furthermore, $g=0$ is taken only when $\rho_0=\nu_0=0$, that is, when the softened gel loses stiffness almost completely and the material is highly compressible where bulk modulus $K$ is comparable to shear modulus $\mu_0$ (or the modified bulk modulus $\tilde{K}$ tends to zero). 

\begin{figure}[h]
\centering
\includegraphics[width=0.95\linewidth]{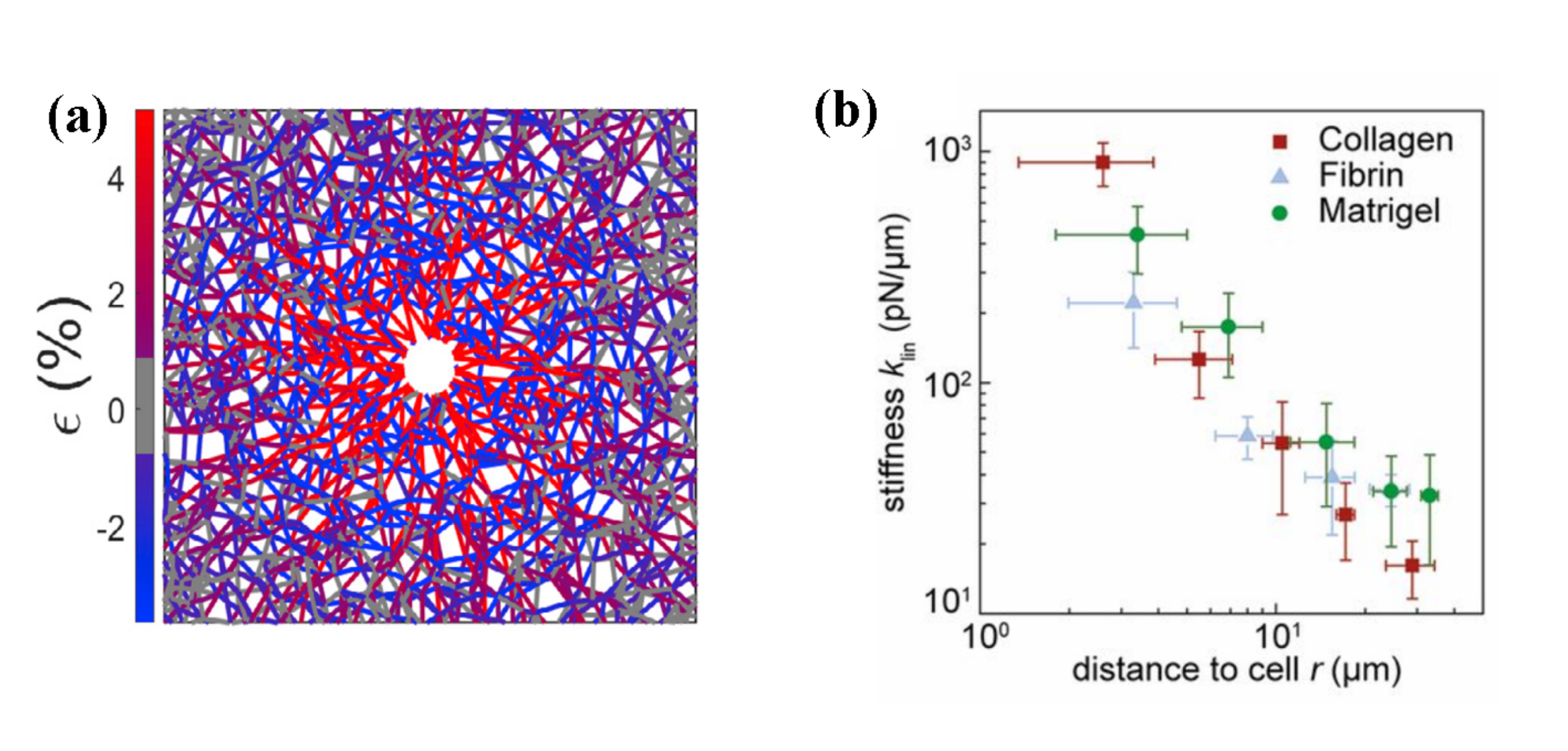}
\caption{(Color online) Simulation and experimental evidence for cell-induced filament buckling and stiffening in fibrous biopolymer gels. (a) Finite element simulations for a circular contracting inclusion in a two-dimensional fibrous network of nonlinear filaments. The negative strain (blue lines) along the angular direction indicates compression and filament buckling, resulting in angular softening. The positive (red) strain in the radial directions indicates stretch and stiffening. Reproduced from Goren \emph{et al.} \cite{Xu2020BJ} with permission from Elsevier. (b) Stiffening of various biopolymer gel matrices by contracting (MDA-MB-231) breast cancer cells. Local linear stiffness $k_{\rm lin}$ is plotted against the distance to the cell $r$ along its principal contraction direction in collagen (red squares), fibrin (blue triangle), and Matrigel (green circle). All three different ECM model gels exhibit a strong cell-induced stiffening gradient.
Reproduced from Han \emph{et al.} \cite{Lenz2018} with permission from National Academy of Sciences.}
\label{fig:IntForce-SoftenStiffenEXP}
\end{figure}

The solution of Eq.~(\ref{eq:IntForce-Soften-ueqn}) depends on the strength of (normalized) cell contraction $\mathcal{A}_b\equiv u_c/R_c\epsilon_{\rm{b}}$, which is a dimensionless parameter that measures the nonlinearity of the cell-contracted network. Here $\epsilon_b$ is the critical strain over which the network softens either due to microbuckling of the constituent filaments \cite{MacKintosh2014,NotbohmLesman2015,Notbohm2015,Lenz2015,Xu2015PRE,Xu2017PRE} or reorientation of filaments away from the compressed directions \cite{Xu2017PRE,Janmey2016}. If $\mathcal{A}_b<1$, the cell-contracted network behaves like a linear isotropic material, the equilibrium Eq.~(\ref{eq:IntForce-Soften-ueqn}) reduces to Eq.~(\ref{eq:IntForce-liniso-ueqn}), and the decay of cell-induced displacement follows Eq.~(\ref{eq:IntForce-liniso-urscaling}) as $\tilde{u}(\tilde{r})=\tilde{r}^{-2}$. However, if $\mathcal{A}_b>1$, the nonlinearity of the network becomes significant and two power-law regimes for the decay of cell-induced displacement can be identified according to the magnitude of $-u(r)/r\epsilon_{\rm{b}}=\mathcal{A}_b\tilde{u}/\tilde{r}$ as follows. The two regimes are separated by a length scale $R_b$ that characterizes the nonlinear compressive-softening elasticity of the material. 
\begin{itemize}
\item {\emph {Far-field regime}} -- Far away from the contracting cell with $\tilde{r} > \tilde{R}_b$ with $\tilde{R}_b\equiv R_b/R_c$ such that $\mathcal{A}_b\tilde{u}/\tilde{r}<1$, we have $\rho=1$ and hence $g=1$. Therefore, the far-field network behaves like a linear isotropic material and the cell-induced displacement decays as
\begin{equation}\label{eq:IntForce-Soften-ufar}
\tilde{u}_{\rm far} = {\tilde{u}_{\rm eff}}/{\tilde{r}^2},
\end{equation}
which fulfils the natural boundary condition $\tilde{u}(\tilde{r}\to \infty)=0$. Note that $\tilde{u}_{\rm eff}$ is the normalized effective far-field displacement and it measures the effects of near-field nonlinearity on the far-field displacements. In the linear isotropic case, $\tilde{u}_{\rm eff}=1$ (or ${u}_{\rm eff}=-u_c$). However, in nonlinear cases, $\tilde{u}_{\rm eff}=1$ can be much larger than $1$, which means that in the far-field regime, although the decay of displacements (respectively, the strains) still follows the power-law $\sim 1/{\tilde{r}^2}$ (respectively, $\sim 1/{\tilde{r}^3}$) as in linear isotropic materials, the amplitudes of far-field displacements or strains are much larger than expected in linear isotropic materials.

\item {\emph {Near-field regime}} -- Close to the cell with $\tilde{r} < \tilde{R}_b$ and $\mathcal{A}_b\tilde{u}/\tilde{r}>1$, there are significant filament buckling as shown in Fig.~\ref{fig:IntForce-SoftenStiffenEXP}(a) and hence we have $\rho=\rho_0$ and hence $g=\rho_0 + \nu_0(1-\rho_0)/(1-\nu_0)$. In this case, the solution of Eq.~(\ref{eq:IntForce-Soften-ueqn}) is 
\begin{equation}\label{eq:IntForce-Soften-unear}
\tilde{u}_{\rm near} = C_1 \tilde{r}^{n-1} + C_2 \tilde{r}^{-n} +\mathcal{A}_b^{-1} \tilde{r},
\end{equation}
in which $n=\frac{1}{2}\left(1+\sqrt{1+8g}\right)$ and hence $1\leq n\leq 2$ with $n=2$ for $\rho_0=1$ or $\nu_0=1/2$, and $n=1$ for $\rho_0=\nu_0=0$. 
\end{itemize} 
The above far-field and the near-field solutions match at $\tilde{r}=\tilde{R}_{b}$ where the angular compressive strain equals to the critical strain $\epsilon_b$, \emph{i.e.}, $\mathcal{A}_b\tilde{u}(\tilde{R}_{b})/\tilde{R}_{b}=1$, or $\tilde{u}_{\rm{eff}} / \tilde{R}_{b}^{3}=\mathcal{A}_b^{-1}$. From the matching continuity conditions for displacements and stresses (or strains) and the condition in Eq.~(\ref{eq:IntForce-BVP-uc}) at the cell boundary (\emph{i.e.}, $\tilde{u}(\tilde{r}=1)=1$), we obtain $\tilde{u}_{\rm{eff}}=\mathcal{A}_b^{-1} \tilde{R}_{b}^{3}$ and the matching radius satisfies
$\tilde{R}_{b}^{n+1}- \tilde{R}_{b}^{2-n}-\frac{2n-1}{3}(\mathcal{A}_b-1)=0$. 
The normalized effective far-field displacement $\tilde{u}_{\rm{eff}}$ can be much larger than that of linear isotropic materials for which $\tilde{u}_{\rm{eff}}=1$ (with $n=2$). Particularly, in the limit of strong nonlinearity with $\mathcal{A}_b \gg 1$ and $\tilde{R}_{b} \gg 1$, we obtain 
\begin{equation}\label{eq:IntForce-Soften-Rbueff}
\tilde{R}_{b} \sim \mathcal{A}_b^{{1}/({n+1})}, \quad \tilde{u}_{\rm{eff}} \sim \mathcal{A}_b^{{(2-n)}/{(n+1)}}.
\end{equation}
Similarly in a two-dimensional matrix, we obtain $\tilde{u}_{\rm{eff}}\sim \mathcal{A}_b^{{(1-n)}/{(n+1)}}$, which has been shown, in the limit of $\mathcal{A}_b \gg 1$ where $n\to 1$, to agree with recent finite element simulations for crosslinked networks of athermal rod-like filaments \cite{Xu2020BJ}. 

Note that if further $\mathcal{A}_b\gg 1$ or $\epsilon_b \to 0$, the equilibrium equation (\ref{eq:IntForce-Soften-ueqn}) reduces to the form of Eq.~(\ref{eq:IntForce-linaniso-ueqn}) for linear anisotropic materials. The far-field linear isotropic regime becomes irrelevant and there is only one power-law regime as shown in Eq.~(\ref{eq:IntForce-linaniso-urscaling}), $\tilde{u} =\tilde{r}^{-n}$. 
In this limit, Rosakis {\emph{et al.}} \cite{Notbohm2015} have also showed from their piecewise quadratic continuum model that a contracting cell would induce an elastic anisotropy with angular modulus getting smaller than radial modulus due to filament microbuckling \cite{NotbohmLesman2015,Notbohm2015,Lenz2015}. In this case, as shown and discussed above, the cell-induced displacement would decay slower than in linear isotropic matrices, which has been related to the experimental measurement of the slow decay of displacements in fibrin networks by a contracting fibroblast cell \cite{NotbohmLesman2015,Notbohm2015}. In contrast, an expanding cell would compress the network in the radial direction and softens while the angular direction is stretched, which induce an inverse elastic anisotropy with radial modulus smaller than angular modulus. In this case, the cell-induced displacement would decay faster than in linear elastic matrices. In addition, Janmey {\emph{et al.}} \cite{Janmey2016Tissue,Janmey2019Tissue} have found that some living tissues show compressive stiffening elasticity, in which case an embedded contracting cell should induce a faster decaying displacement fields.

Before ending this subsection, we would like to point out that in this subsection we have assumed that the dependence of material elastic modulus follows a simple one-step function form, that is, the modulus decreases from the linear constant $E_0$ to a smaller constant $\rho_0E_0$ when the compression is over the critical strain $\epsilon_b$. However, in most cell-contracting-matrix experiments and cell-contracting-network simulations, the elastic anisotropy induced by cells is more complicated, usually increasing with increased cell contraction. In this case, since the power-law-decaying exponent $n$ is inversely proportional to the elastic anisotropy, $n$ will decrease with increasing cell contraction; or the decay of displacement may follows different law according to the specific function form of the softening nonlinearity.  

\begin{figure}[h]
\centering
\includegraphics[height=9cm]{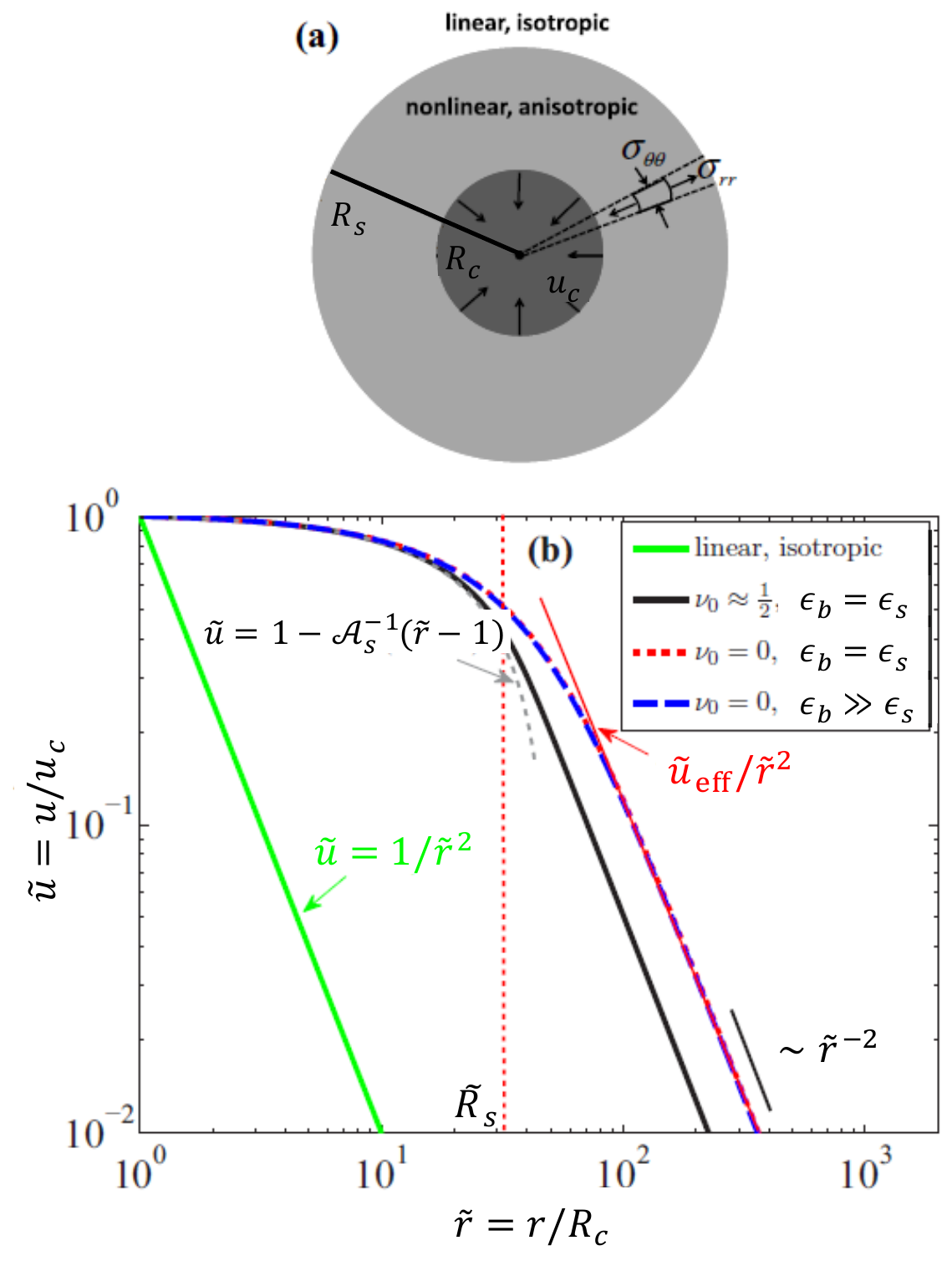}
\caption{(Color online)
(a) Schematic illustration of a spherical cell contracting in a biopolymer gel with nonlinear stiffening-softening elasticity in the limit of $\epsilon_b \sim \epsilon_s \ll 1$ or $\epsilon_b \gg \epsilon_s$. Two regimes separated by $\tilde{r}=\tilde{R}_s$ are identified. 
(b) The decay of the displacement induced by the contractile cell in the highly nonlinear limit of ${\cal A}_s=u_c/R_c\epsilon_s = 50\gg 1$ and $\rho_{0}=0.1$. Case 1: $v_{0}=0$ (red dotted line and blue dashed line). In the near-field ($1<\tilde{r} \ll \tilde{R}_s$), $\tilde{u}\sim 1-\mathcal{A}_s^{-1}(\tilde{r}-1)$ in Eq.~(\ref{eq:IntForce-StiffenSoften-unear}), and in the far-field ($\tilde{r} \gg \tilde{R}_s$), $\tilde{u}\sim \tilde{r}^{-2}$ as in Eq.~(\ref{eq:IntForce-liniso-urscaling}) for linear isotropic materials. Case 2: $v_{0}=0.49$ (black dotted line). Note that the two cases of $\epsilon_b=\epsilon_s$ and $\epsilon_b \gg \epsilon_s$ lie on the same line. Case 2: $v_{0}=0.49$ (black solid line), qualitatively similar to Case 1. Adapted with permission from Xu and Safran \cite{Xu2015PRE}. Copyright(2015) by the American Physical Society. }
\label{fig:IntForce-Stiffen}
\end{figure} 

\subsection{Nonlinear stretch-stiffening and nonlinear compressive-softening biopolymer gels \label{sec:IntForce-StiffenSoften}}

We now turn to the more general case: a spherical contractile cell embedded in a biopolymer gel with both nonlinear strain-stiffening and compressive-softening elasticity as modeled by the nonlinear elastic energy in Eq.~(\ref{eq:ExtForce-nonlinsmall-F2}) \cite{Xu2015PRE}. Substituting the Eqs.~(\ref{eq:ExtForce-nonlinsmall-sigmaepsilon}) and (\ref{eq:IntForce-BVP-epsilon123}) into the equilibrium equation Eq.~(\ref{eq:IntForce-BVP-equilibrium}), we obtain  
\begin{align}\label{eq:IntForce-StiffenSoften-ueqn}
&C_0\left[\tilde{U}-g\frac{2\tilde{u}}{\tilde{r}^2}+(1-g)\frac{2\epsilon_b}{\tilde{r}}\right] \\ \nonumber 
&-\tilde{U}+(1+\mathcal{A}_s\tilde{u}')^{-3}\left(\tilde{U}+ \frac{3\mathcal{A}_s\tilde{u}'^2}{\tilde{r}}+\frac{\mathcal{A}_s^2\tilde{u}'^3}{\tilde{r}} \right)=0,
\end{align}
with the boundary conditions at cell boundary $\tilde{u}(1)=1$ and natural boundary condition at infinity $\tilde{u}(\tilde{r}\to \infty)=0$. Here $\tilde{U}\equiv \tilde{u}''+{2\tilde{u}'}/\tilde{r}$, $\tilde{u}' \equiv {d\tilde{u}}/{d\tilde{r}}$, $C_0={5(1-\nu_0)}/{6(1-2\nu_0)}$, $g$ is defined as in Sec.~\ref{sec:IntForce-linaniso}, and ${\rho}= \rho_0 +(1-\rho_0)\Theta(1-\mathcal{A}_b\tilde{u}/\tilde{r})$.
The dimensionless parameters $\mathcal{A}_s\equiv u_c/R_c\epsilon_s$ and $\mathcal{A}_b\equiv u_c/R_c\epsilon_b$ (both positive for contractile cells with $u_c<0$) characterize the strengths of the nonlinearity of radial stiffening and angular softening, respectively. We now discuss the implications of Eq.~(\ref{eq:IntForce-StiffenSoften-ueqn}) in several limiting cases.

Firstly, in the limit of $\mathcal{A}_s, \, \mathcal{A}_b  \to 0$ (\emph{i.e.}, $\epsilon_s,\,\epsilon_b \to \infty$), the elastic nonlinearities of stretch-stiffening and compressive-softening do not manifest explicitly and the cell-contracted material behaves like a linear isotropic medium, the equilibrium Eq.~(\ref{eq:IntForce-StiffenSoften-ueqn}) reduces to Eq.~(\ref{eq:IntForce-liniso-ueqn}), and the decay of cell-induced displacement follows Eq.~(\ref{eq:IntForce-liniso-urscaling}) as $\tilde{u}(\tilde{r})=\tilde{r}^{-2}$. 

Secondly, in the limit of $\mathcal{A}_s \to 0$ and $\mathcal{A}_b > 1$ (\emph{i.e.}, $\epsilon_s \to \infty$ and $\epsilon_b \ll 1 $), the stretch-stiffening nonlinearity does not manifest, but compressive-softening becomes important and the cell-contracted material behaves like a linear anisotropic medium, the equilibrium Eq.~(\ref{eq:IntForce-StiffenSoften-ueqn}) reduces to Eq.~(\ref{eq:IntForce-linaniso-ueqn}), and two power-law regimes for the decay of cell-induced displacement, as shown in Fig.~\ref{fig:IntForce-Soften}, are identified where $\tilde{u}(\tilde{r})=\tilde{r}^{-2}$ in the far field and $\tilde{u}(\tilde{r})=\tilde{r}^{-n}$ with $1 \le n\le 2$ in the near field close to the contracting cell.  

Thirdly, in the nonlinear limit of $\mathcal{A}_s \gg 1$ (\emph{i.e.}, $\epsilon_s \ll 1$), the effects of strain-stiffening becomes critical, in which case the following two different cases can be found depending on the relative magnitudes of $\mathcal{A}_s$ and $\mathcal{A}_b$. 

(i) In the first case with either $\mathcal{A}_s \gg \mathcal{A}_b\gg 1$ (\emph{i.e.}, $\epsilon_s \ll \epsilon_b \ll 1$) or $\mathcal{A}_b\sim \mathcal{A}_s\gg 1$ (\emph{i.e.}, $\epsilon_b \sim \epsilon_s \ll 1$), the effects of strain-stiffening dominate over compressive-softening. We can identify two scaling regimes as shown in Fig.~\ref{fig:IntForce-Stiffen}, which are separated by a new length scale $R_{s}$ that characterizes the the nonlinear strain-stiffening elasticity of biopolymer gels. 
\begin{itemize}
    \item {\emph {Far-field regime}} -- In the far field, $\tilde{r} \gg \tilde{R}_s$ with $\tilde{R}_s=R_s/R_c$, the stresses are small such that neither strain-stiffening nor compressive softening are significant. The biopolymer gel, therefore, behaves like a linear isotropic material and the decay of cell-induced displacement follows the power-law, $\tilde{u}_{\rm far}=\tilde{u}_{\rm {eff }} / \tilde{r}^{2}$, as shown in Eq.~(\ref{eq:IntForce-liniso-urscaling}).
    \item {\emph {Near-field regime}} -- In the near field, $1<\tilde{r} \ll \tilde{R}_s$, the stresses are large and both strain-stiffening and compressive softening are important as shown in Fig.~\ref{fig:IntForce-SoftenStiffenEXP}. The near-field displacement is simply expanded as 
    \begin{equation}\label{eq:IntForce-StiffenSoften-unear}
    \tilde{u}_{\rm near}=1-\mathcal{A}_s^{-1}(\tilde{r}-1),   
    \end{equation}
    to leading order in $1 / \mathcal{A}_s$, \emph{i.e.}, the displacement decays almost linearly, much more slowly than that in either linear isotropic or linear anisotropic media as discussed above. 
\end{itemize}
The above far-field and the near-field solutions match at $\tilde{r}=\tilde{R}_s$. From the matching continuity conditions for displacements and stresses (or strains), we obtain 
\begin{equation}\label{eq:IntForce-StiffenSoften-Rsueff}
\tilde{R}_s \sim \mathcal{A}_s, \quad \tilde{u}_{\mathrm{eff}} \sim \mathcal{A}_s^{2}.
\end{equation}
This means that the nonlinear strain stiffening (quantified by $\mathcal{A}_s$) of the system can significantly amplify the magnitude of the strain at long distances. Note that the scaling laws found here are different from $\tilde{R}_s \sim e^{2\mathcal{A}_s/ \sqrt{3}}$ and $\tilde{u}_{\mathrm{eff}} \sim \mathcal{A}_s^{-1} e^{2 \sqrt{3} \mathcal{A}_s}$ obtained by using Shokef-Safran hyperelastic model \cite{Sam2012a,Sam2012b} where an isotropically stiffened gel model is employed with compression stiffening as opposed to softening.

\begin{figure}[h]
\centering
\includegraphics[height=9cm]{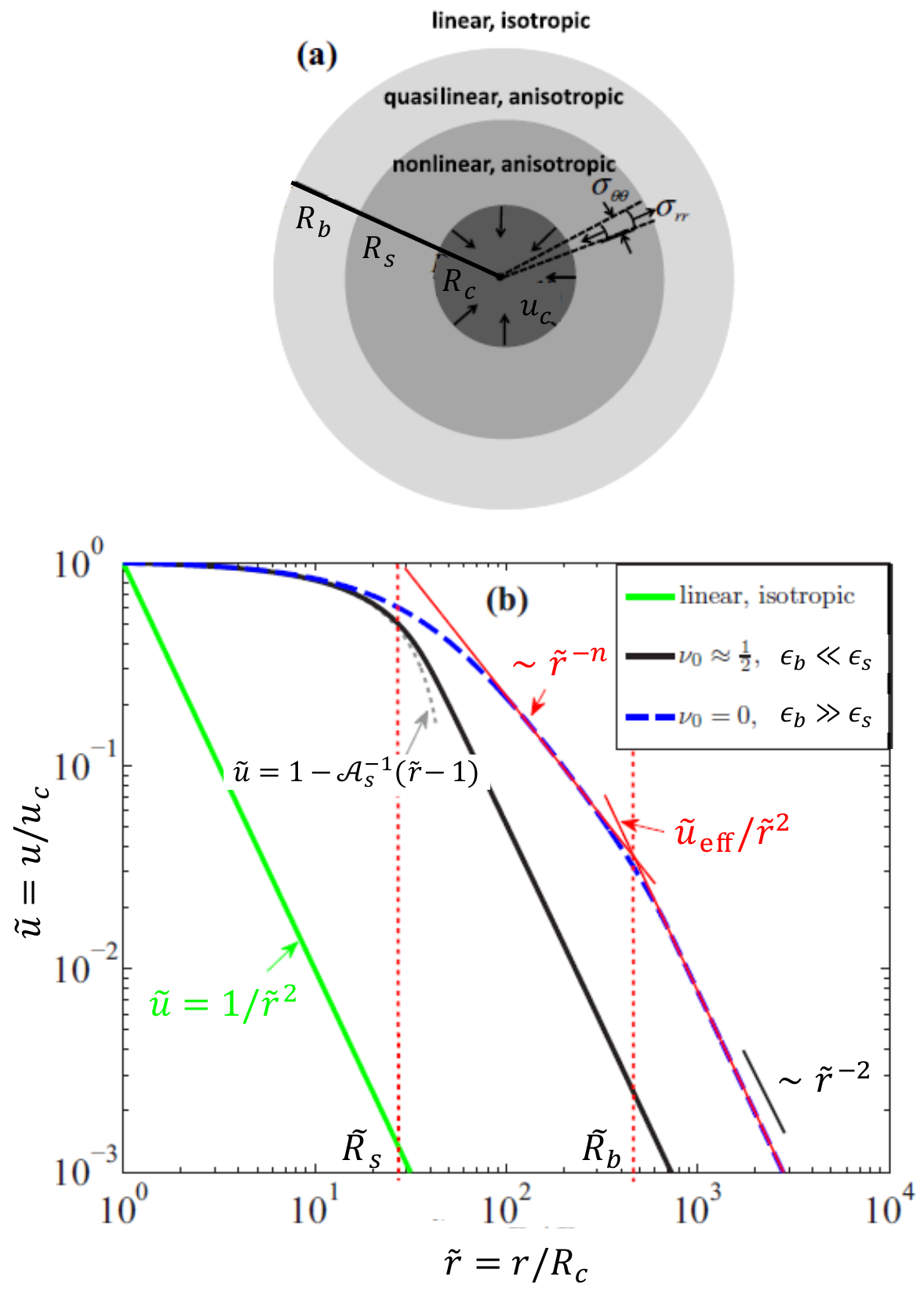}
\caption{(Color online)
(a) Schematic illustration of a spherical cell contracting in a biopolymer gel with nonlinear stiffening-softening elasticity in the limit of $\epsilon_b \ll \epsilon_s \ll 1$. Three regimes separated by $\tilde{r}=\tilde{R}_s$ and $\tilde{r}=\tilde{R}_b$ are identified. 
(b) The decay of the displacement induced by the contractile cell in the highly nonlinear limit of ${\cal A}_s=u_c/R_c\epsilon_s = 50\gg 1$ and $\rho_{0}=0.1$. Case $1: v_{0}=0$ (blue dashed line). In the near-field ($1<\tilde{r} \ll \tilde{R}_s$), $\tilde{u}\sim 1-\mathcal{A}_s^{-1}(\tilde{r}-1)$ in Eq.~(\ref{eq:IntForce-StiffenSoften-unear}). In the intermediate regime ($\tilde{R}_s\ll \tilde{r} \ll \tilde{R}_b$), $\tilde{u}\sim \tilde{r}^{-1.2}$ as in Eq.~(\ref{eq:IntForce-linaniso-urscaling}) for linear anisotropic materials. In the far-field ($\tilde{r} \gg \tilde{R}_b$), $\tilde{u}\sim \tilde{r}^{-2}$ as in Eq.~(\ref{eq:IntForce-liniso-urscaling}) for linear isotropic materials. Case 2: $v_{0}=0.49$ (black solid line). There are only two regimes, as shown in Fig.~\ref{fig:IntForce-Stiffen}. That is, the effects of compressive-softening are important only for compressible gels. Adapted with permission from Xu and Safran \cite{Xu2015PRE}. Copyright(2015) by the American Physical Society. }
\label{fig:IntForce-StiffenSoften}
\end{figure}  

(ii) In the second cases with $\mathcal{A}_b\gg\mathcal{A}_s \gg 1$ (\emph{i.e.}, $\epsilon_b \ll \epsilon_s \ll 1$), the effects of strain-stiffening and compressive-softening are both important and one dominates the other in different distances (regions) away from the cell boundary. Three distinct scaling regimes are identified as shown in Fig.~\ref{fig:IntForce-StiffenSoften}, which are separated by two length scale $R_s$ and $R_b$ that characterizes the nonlinear strain-stiffening and compressive softening elasticity of biopolymer gels, respectively.

\begin{itemize}
    \item {\emph {Far-field power-law regimes}} -- In the far field region with $\tilde{r} \gg$ $\tilde{R}_b$, the stiffening and softening nonlinearity becomes unimportant. The biopolymer gel behaves like a linear isotropic material and the decay of the cell-induced displacement follows $\tilde{u}_{\rm{far}}=\tilde{u}_{\mathrm{eff}} / \tilde{r}^{2}$ as shown in Eq.~(\ref{eq:IntForce-liniso-urscaling}). 
    
    \item {\emph {Intermediate power-law regime}} -- In the intermediate region with $\tilde{R}_s \ll \tilde{r} \ll \tilde{R}_b$, the nonlinear strain-stiffening is negligible, however, compressive-softening in the angular direction is still significant such that the gel responds like a linear anisotropic material as discussed in previous subsection. In this region, the decay of the cell-induced displacement follows $\tilde{u}_{\rm int} = C_1 \tilde{r}^{n-1} + C_2 \tilde{r}^{-n} +\mathcal{A}_b^{-1} \tilde{r}$, as shown in Eq.~(\ref{eq:IntForce-Soften-unear}) with  $n=\frac{1}{2}\left(1+\sqrt{1+8g}\right)$ and hence $1\leq n\leq 2$. 
    
    \item {\emph {Near-field linearly-decaying regime}} -- In the near field with $1<\tilde{r} \ll \tilde{R}_s$, the gel stiffens strongly (\emph{i.e.}, ${d\tilde{u}}/{d\tilde{r}} \to \epsilon_s$) in the radial direction such that the displacement decays almost linearly and is approximated by Eq.~(\ref{eq:IntForce-StiffenSoften-unear}), $\tilde{u}_{\rm near}=1-\mathcal{A}_s^{-1}(\tilde{r}-1)$, as in the previous first case.
\end{itemize}
The above far-field and the near-field solutions match at $\tilde{r}=\tilde{R}_s$ and $\tilde{r}=\tilde{R}_b$, respectively. From the matching continuity conditions for displacements and stresses (or strains) and the condition in Eq.~(\ref{eq:IntForce-BVP-uc}) at the cell boundary (\emph{i.e.}, $\tilde{u}(\tilde{r}=1)=1$), we obtain  
\begin{equation} \label{eq:IntForce-StiffenSoften-RsRbueff}
\tilde{R}_s\sim \mathcal{A}_s, \quad \tilde{R}_b/\tilde{R}_s \sim(\mathcal{A}_b/\mathcal{A}_s)^{1/(n+1)}, \quad
\tilde{u}_{\mathrm{eff}} \sim \mathcal{A}_s^{3n/(n+1)}\mathcal{A}_b^{(2-n)/(n+1)}.
\end{equation}
In particular, for the case of $\nu_0,\,\rho_0\to 0$, we have $n=1$ and hence $\tilde{R}_s\sim \mathcal{A}_s$, $\tilde{R}_b/\tilde{R}_s \sim(\mathcal{A}_b/\mathcal{A}_s)^{1/2} \gg 1$, $\tilde{u}_{\mathrm{eff}} \sim \mathcal{A}_s^{3/2} \mathcal{A}_b^{1/2} \gg 1$. This means that the nonlinear strain stiffening (quantified by $\mathcal{A}_s=u_c / R_c \epsilon_{s}$) and the nonlinear compressive softening (quantified by $\mathcal{A}_b= u_c / R_c \epsilon_{b}$) of the biopolymer gel can both significantly amplify the magnitude of the displacements, strains, or stresses at long distances away from the contracting cell. 

In summary, it has been predicted from the above affine-deformation theory that biopolymer gels stiffen with larger elastic modulus and becomes elastically anisotropic (due to filament buckling and stiffening) in the vicinity of strongly contracting cells. Such stiffening and anisotropic behaviors have already been observed and measured (i) in finite element simulations for crosslinked networks composed of athermal stiff fibers \cite{Xu2020BJ,Lenz2016} as shown in Fig.~\ref{fig:IntForce-SoftenStiffenEXP}(a); (ii) in single-cell experiments for fibroblast cells in fibrin gels as shown in Figs.~\ref{fig:IntForce-CellMatrixScaling}(a)--~\ref{fig:IntForce-CellMatrixScaling}(b)  and for human breast (epithelial) cancer cells in 3D ECM model systems like collagen, fibrin, and Matrigel \cite{Lenz2018} as shown in Fig.~\ref{fig:IntForce-SoftenStiffenEXP}(b); and (iii) recently in multicellular experiments for a spherically contracting spheroid containing thousands of glioblastoma cells in collagen gels as shown in Figs.~\ref{fig:IntForce-CellMatrixScaling}(c)--~\ref{fig:IntForce-CellMatrixScaling}(d). In addition, as commented in the previous subsection, an expanding cell would induce stiffening in the angular direction and an inverse elastic anisotropy with radial modulus smaller than angular modulus. In this case, the cell-induced displacement would decay faster than in linear elastic matrices. Moreover, if the contracting cell is embedded in a compressive-stiffening gel \cite{Janmey2016Tissue,Janmey2019Tissue} would also induce a faster decaying displacement fields \cite{Xu2020BJ,Xu2015PRE}.

\begin{figure}[htbp]
  \centering
  \includegraphics[width=0.85\linewidth]{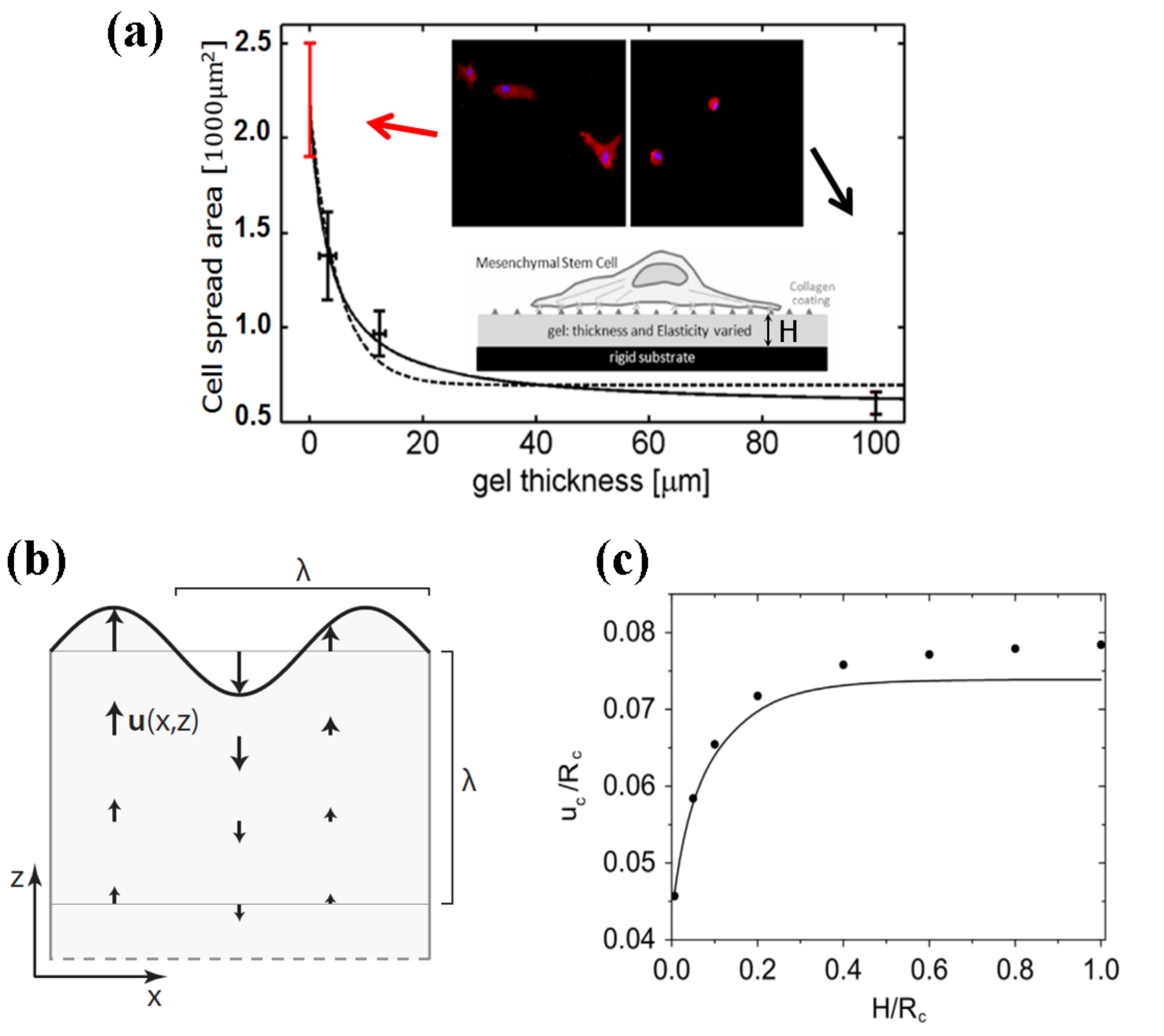}
  \caption{How deeply do cells feel? (a) Mesenchymal stem cells were cultured on soft brain-like PA gels of varying thickness. As the gel thickness $H$ increases, the cell contraction increases (larger $u_c$) and cell spreading is suppressed. Reproduced from Buxboim \emph{et al.} \cite{Discher2010} with permission from IOP Publishing. All rights reserved (b) Penetration depth of surface deformation into the linear elastic substrate is the same order of magnitude as the characteristic wavelength $\lambda$ of the surface deformation. Copyright\textsuperscript{\textcopyright}2017 by Siber~\emph{et al.}~\cite{Siber2017}.Reproduced by permission of Taylor and Francis Group, LLC. (c) Theoretical calculations for a cell disk contracting on an linear elastic substrate of thickness $H$ \cite{Gao2014}. As $H$ increases, the displacement at the cell periphery increases and reaches a plateau when $H/R_c>1$. This is consistent with the experiments in (a). Reproduced from He \emph{et al.} \cite{Gao2014} with permission from Elsevier.} 
\label{fig:IntForce-Substrate}   
\end{figure} 

Finally, we would like to mention the transmission of cellular forces in a different geometry where cells are adhered to a thin film of gels as shown in Fig.~\ref{fig:IntForce-Substrate}(a) as in many \emph{in vitro} cell experiments \cite{Mohammadi2014,Sam2013a,Discher2010}. Firstly, consider a linear isotropic elastic half-space with an imposed deformation at the top surface described by $\mathbf{u}_0(x)=u_0(x)\hat{\mathbf{e}}_z$ as shown in Fig.~\ref{fig:IntForce-Substrate}(b) \cite{Siber2017}. The displacement field $\mathbf{u}(x,z)$ follows the equilibrium equation $\nabla^2\mathbf{u}+(1-2\nu_0)^{-1}\nabla \nabla\cdot \mathbf{u}=0$ in the absence of body force \cite{Landau1986}, in which $\mathbf{u}(x,z)$ can be decomposed as $\mathbf{u}=\mathbf{u}_t+\mathbf{u}_l$ with transverse part $\mathbf{u}_t$ satisfying $\nabla \cdot \mathbf{u}_t=0$ and longitudinal part $\mathbf{u}_l$ satisfying $\nabla \times \mathbf{u}_l=0$. In this case, we have $\frac{\partial^2\mathbf{u}_t}{\partial x^2}+\frac{\partial^2\mathbf{u}_t}{\partial z^2}=0$. If the surface deformation takes the form of $\exp(iqx)$ of characteristic length $\lambda=2\pi/q$, then $\mathbf{u}_t \sim \exp(qz)\exp(iqx)$ decays exponentially with distance $z$ away from the surface and the penetration depth takes the order of magnitude of $\lambda$ as shown in Fig.~\ref{fig:IntForce-Substrate}(b). The same behavior also holds for the longitudinal mode. Detailed calculations for an adhesion of size $R_a$ that exerts a force on the surface of a semi-infinite substrate has been done by \textcolor{black}{Nicolas \emph{et al.}~\cite{Sam2004}} and they find that the decay length indeed scales with $R_a$. Calculations and simulations for a cell disk contracting on an linear elastic substrate by Wang \emph{et al.}~\cite{Shenoy2014b} and by He \emph{et al.}~\cite{Gao2014} also give similar conclusion that the penetration depth of cell-induced displacements is around cell radius as shown in Fig.~\ref{fig:IntForce-Substrate}(c). This result is consistent with experimental observations as shown in Fig.~\ref{fig:IntForce-Substrate}(a) and suggests that a substrate thicker than the cell radius can be approximated as a semi-infinite substrate \cite{Gao2014}. 

%============================%%============================%
\section{Matrix-mediated long-range cell-cell interactions}\label{sec:CellCellInt}
%============================%%============================%

\subsection{A generic theory for matrix-mediated cell-cell interactions}\label{sec:CellCellInt-GenericTheory}

\begin{figure}
    \centering
    \includegraphics[width=0.8\linewidth]{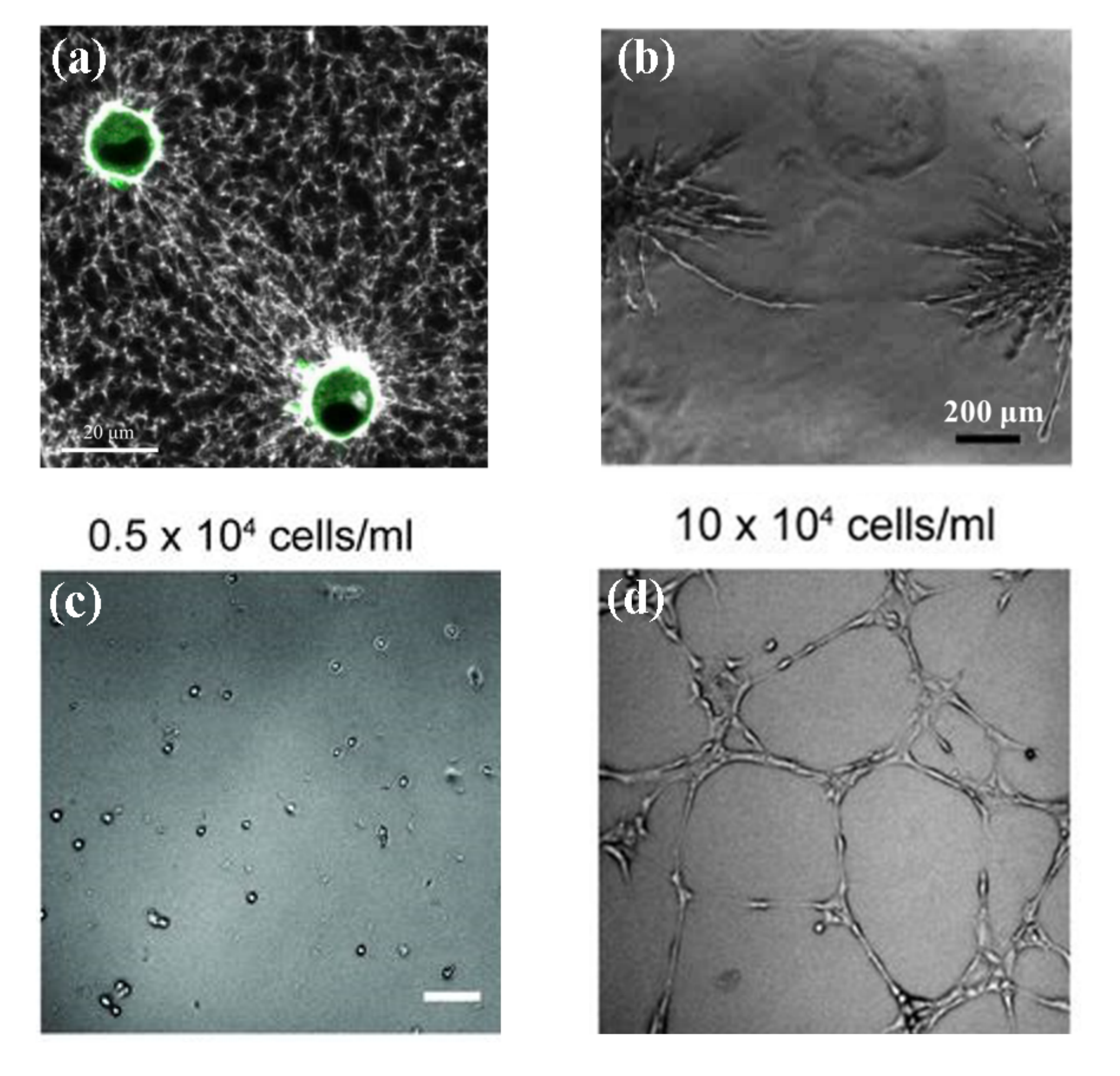}
    \caption{Experiments of matrix-mediated cell-cell interactions. (a) Two fibroblast (GFP-actin) cells (gray) embedded in fluorescently labeled fibrin gel (white) are shown \cite{Lesman2018}. Cells deform the fibrous matrix in a highly directional manner toward neighbouring cells, creating highly remodeled matrix ``bands" between pairs of neighbouring cells. Reproduced from Sopher \emph{et al.}\cite{Lesman2018} with permission from Elsevier. (b) Directional sprouting of capillary-like structures towards each other, originating from two endothelial cell spheroids embedded in a collagen gel with a distance over tens times of cell radius. Reproduced from Korff \emph{et al.}\cite{Korff1999} with permission from Journal of cell science. (c) and (d): Strong dependence of cellular self-organization on cell density (or cell-cell distance) in arrays of Human endothelial (HUVECs) cells seeded on Matrigel. At low cell density (c), HUVECs stayed separate or formed small group of cells. At higher cell density (d), HUVECs formed tubes. Scale bar: $100\, \mu m$. Reproduced from R\"{u}diger~\emph{et al}~\cite{Stefan2019} with permission from Elsevier. }
    \label{fig:CellCellInt-CellCellEXP}
\end{figure}

\begin{figure}[htbp]
  \centering
  \includegraphics[width=0.85\linewidth]{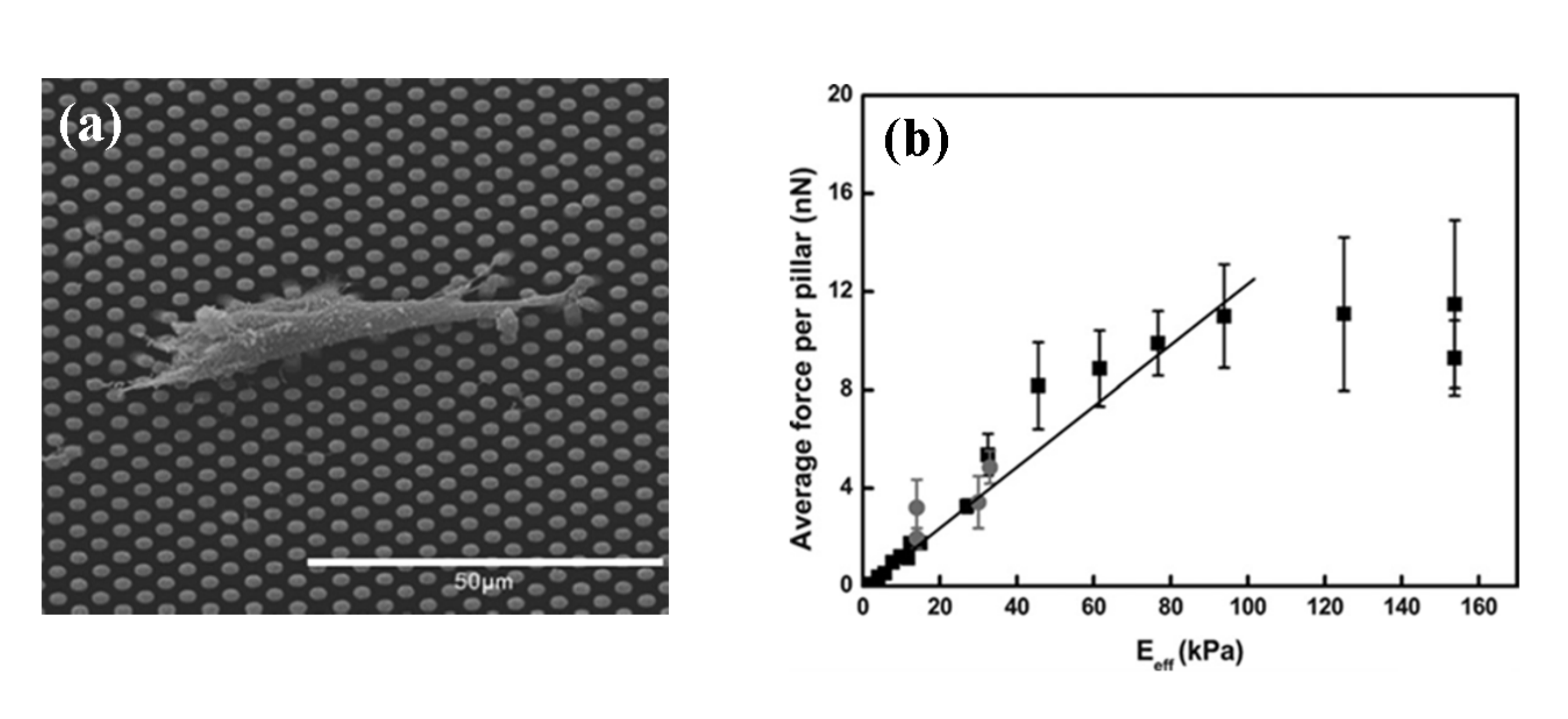}
  \caption{Experimental evidence for mechanical homeostasis of cells on soft substrates.  (a) Scanning electron micrograph of a 3T3 fibroblast on an array of oval pillars. Scale bar: $50 \, \mathrm{\mu m}$. (b) Plot of the mean force exerted by fibroblasts as a function of the equivalent Young’s modulus, $E_{\rm eff}$. For small $E_{\rm eff}$, cell force increase linearly, indicating a constant deformation (or displacement) at the cell periphery. For $E_{\rm eff}$ larger than some critical value $E^* \sim 90 \, \mathrm{kPa}$, cell force saturates, indicating a constant stress at the cell periphery.  Reproduced from Ghibaudo \emph{et al.}~\cite{Ladoux2008} with permission from the Royal Society of Chemistry.} 
\label{fig:ExtForce-Homeostasis}   
\end{figure}

We now consider how the transmission of forces in extracellular matrix as reviewed in the previous section can mediate the mechanical interactions between adherent cells in a distance as shown in Fig.~\ref{fig:CellCellInt-CellCellEXP}. Many types of animal cells have been found to actively adhere to their surrounding matrix by some discrete focal adhesions. These adherent cells contract by intracellular actomyosin stress fibers and transmit forces to the matrix by adhesions. In this way, cells probe, sense and respond actively to the mechanical or geometrical signals of their surrounding matrix, for example cells actively adjust their contractility by remodeling their stress fibers and focal adhesion \cite{Sam2013a} as shown in Fig.~\ref{fig:introduction-cellmatrixcell}. Such remodeling in subcellular scales usually shows up at cellular scales as changes in shape and orientation, polarization and directional migration \cite{Sam2013a,Mohammadi2014}. Furthermore, many experiments show that active cell responses tend to maintain their local homeostasis (see Fig.~\ref{fig:ExtForce-Homeostasis}), in which the contractile apparatus of the cell is biologically programmed to preserve certain mechanical properties of the cell, such as stresses, strains or displacements at the cell boundary, even in the presence of perturbations in their mechanical environment \cite{Ladoux2012}. The mechanical homeostasis of adherent cells has been observed in experiments for cells on soft pillar substrates~\cite{Ladoux2008} and for cells under quasi-static stretch~\cite{Eastwood1998}, which has been further used to explain cell orientation in response to cyclic stresses~\cite{Sam2007}. Based on this coarse-grained concept of cell homeostatic feedback effects, Ben-Yaakov \emph{et al.} \cite{Sam2015} proposed a generic and unified theory to explain how adherent cells respond to mechanical perturbations of their surrounding matrix such as the presence of neighbouring cells, slowly applied stretch, or gradients of matrix rigidity. Here we briefly review this theory and the most important results particularly for matrix-mediated cell-cell interactions. 

The local dynamics of adhesions of an adherent cell on elastic substrates is proposed to follow a phenomenological Langevin equation:
\begin{equation}\label{eq:CellCellInt-GenericTheory-uceqn}
\zeta \frac{d {\mathbf u}_c}{dt}=\mathbf{f}_c+\mathbf{f}_e+\mathbf{f}_p.
\end{equation}
Here $\zeta$ is proportional to the dynamic friction between the adhesion and substrate. $\mathbf{f}_c$ is the contractile cellular force and $\mathbf{f}_p$ is the stochastic protrusive force, which in general can both be a function of the adhesion position at the cell boundary. $\mathbf{f}_e$ is the elastic matrix force that balances both $\mathbf{f}_c$, $\mathbf{f}_p$, and the frictional force applied on the adhesion. The protrusion forces $\mathbf{f}_p$ function as a type of noise that allows the cell to explore its surroundings so that the friction can be overcome to allow motion of the adhesions. In their work, Ben-Yaakov \emph{et al.} \cite{Sam2015}, however, didn't consider the full adhesion dynamics but only focused on the predictions of the direction and magnitude of the displacement of an adhesion in a non-motile, well-adhered cell in the presence of mechanical perturbations of its elastic environment. They treat the biological activity of the cell via a homeostatic, mechanical boundary condition at the cell-matrix interface; this presents a well defined boundary value problem of elasticity. 
Note that in this model the focal adhesions are coarse-grained as material points and are displaced from its reference position in the absence of mechanical perturbations of the matrix. It cannot resolve the spacing between the adhesions and the membrane. Furthermore, while the details of the above forces depend on the interactions of many molecular components involved, the protrusive forces $\mathbf{f}_p$ are denoted as the net forces that tend to move an adhesion in the locally forward direction (away from the cell nucleus) and the contractile forces $\mathbf{f}_c$ are denoted as the net forces that tend to move the adhesion in the opposite manner (backward to cell nucleus). However, the advantage of this treatment is its independence of the details of how the cytoskeletal structure, myosin activity and adhesion size and density are regulated by the cell in achieving the local deformation or stress dictated by the cell's genetic program.

\subsection{Mechanical interactions in one-dimensional cell-matrix system}\label{sec:CellCellInt-1DCellMatrix}

\begin{figure}
    \centering
    \includegraphics[width=0.9\linewidth]{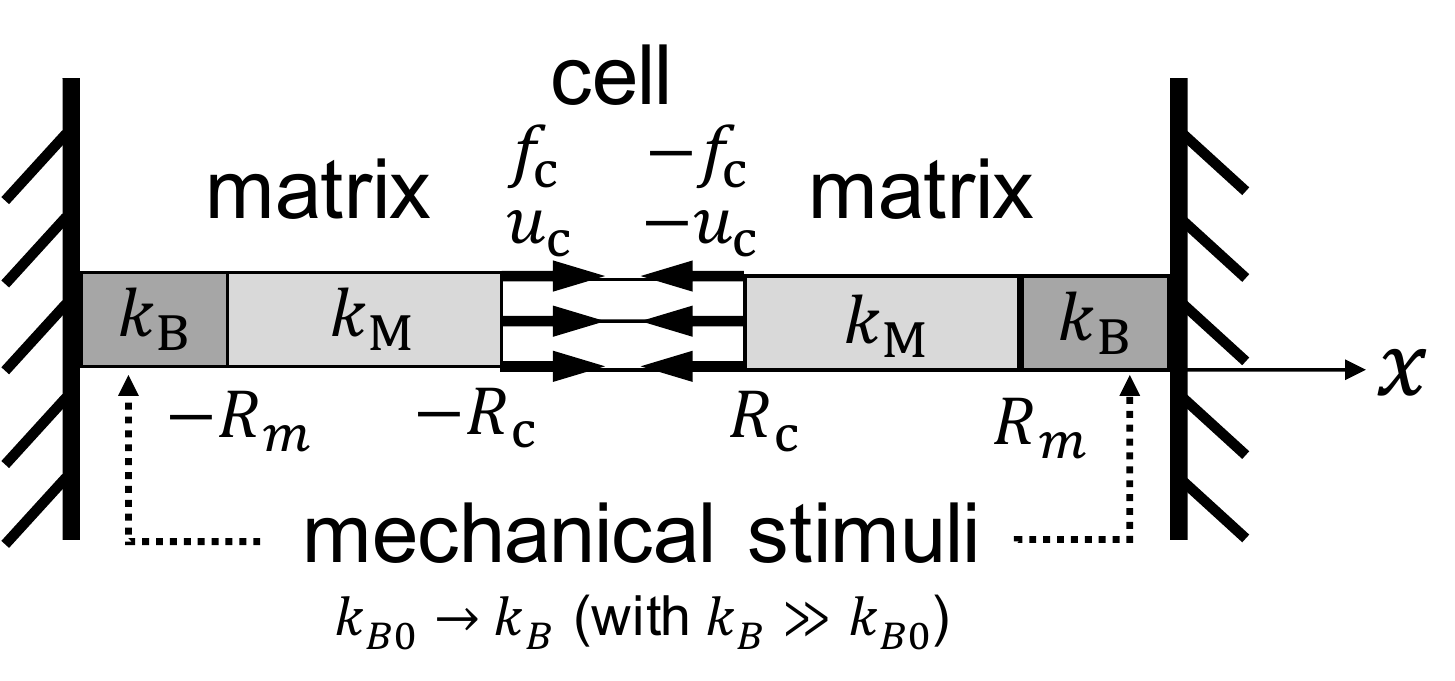}
    \caption{A one-dimensional toy system where a cell is actively adhered to an elastic matrix. The matrix is composite and comprises two parts: a soft matrix part with spring constant $k_M$ and a boundary matrix part with spring constant $k_B$. The boundary matrix part is stiffened by some external stimuli and the spring constant is increased from $k_{B0}$ to $k_B$ with $k_{B}\gg k_{B0}$.}
    \label{fig:CellCellInt-1DCellMatrix}
\end{figure}

We first consider the active responses of adherent cells to mechanical perturbations in their surrounding matrix in a one-dimensional toy cell-matrix system as shown in Fig.~\ref{fig:CellCellInt-1DCellMatrix}. The composite matrix are divided into two parts: a soft matrix part with spring constant $k_M$ and a boundary matrix part with spring constant $k_B$. At the unperturbed state, the spring constant of the boundary matrix is small given by $k_B=k_{B0}\ll k_M$. The cell applies a contractile force, $f_c=f_{c0}>0$, to the matrix at its boundary $x=\pm R_c$. In this case, the total energy is given by
\begin{equation}\label{eq:CellCellInt-1DCellMatrix-F0}
{\cal F}_{0}=k_{\rm eff0}u_{c}^{2}-2 f_{c0} u_{c},    
\end{equation} 
in which the effective spring constant of the composite matrix $k_{\rm eff0}=k_Mk_{B0}/(k_M+k_{B0})$ scales as $k_{B0}$ for $k_M\gg k_{B0}$. Minimization of ${\cal F}_{0}$ with respect to $u_c$ gives the force balance equation, $f_e-f_{c0}=\frac{1}{2}\frac{d{\cal F}_{0}}{du_c}=0$, from which we obtain the equilibrium displacement, $u_{c0}=f_{c0}/k_{\rm eff0}$. Here $f_e=k_{\rm eff0}u_{c}$ is the elastic restoring force of the matrix and the factor ${1}/{2}$ comes from cell contraction on its two sides in 1-dimension. Cellular protrusions induce deviation of its boundary displacement $u_c$ from $u_c^e$, the total restoring force that opposes the protrusions is 
$\delta f=k_{\rm eff0}(u_c-u_{c0})$.

Suppose that the boundary matrix part is stiffened by some stimuli as shown in Fig.~\ref{fig:CellCellInt-1DCellMatrix}, in which the spring constant of the boundary matrix is increased from $k_{B0}$ to $k_B$ with $k_{B}\gg k_{B0}$, and hence $k_{\rm eff0}$ is increased to $k_{\rm eff}=k_Mk_{B}/(k_M+k_{B})\gg k_{\rm eff0}$ scaling as $k_M$ for $k_{B} \gg k_M \gg k_{B0}$. This mechanical perturbation of boundary stiffening and the resulted small displacements in the boundary matrix also mimics the situation when two cells are contracting the 1D matrix identically on its two sides, in which case the displacements near the mid-plane are close to zero by symmetry. To see how cell contractility and homeostasis result in active cell responses to this particular mechanical perturbation and matrix-mediated cell-cell interactions, we consider the following two homeostatic conditions of cells upon mechanical stimuli. 
\begin{itemize}
    \item \emph{Homeostatic displacement with $u(R_c)=-u_c\to -u_{c0}<0$ and $u(-R_c)=u_c\to u_{c0}>0$ being fixed under the perturbation of boundary stiffening.} -- One can introduce a phenomenological ``potential" energy to characterize cell homeostasis that keeps its boundary displacement $u_{c0}$ to the set-point magnitude $u_{c0}$ at unperturbed states. The change of total energy at the perturbed state is then given by
    \begin{equation}\label{eq:CellCellInt-1DCellMatrix-Fu}
    \Delta {\cal F}_u =\left(k_{\rm eff}u_{c}^{2}-k_{\rm eff0}u_{c0}^{2}\right)-2 f_{c 0}\left(u_{c}-u_{c0}\right)+\gamma_u\left(u_{c}-u_{c0}\right)^{2}, 
    \end{equation} 
    where the constant $\gamma_u$ characterizes the tendency of cells to attain the unperturbed set-point value of their boundary displacement. The equilibrium adhesion displacement, $u_{c}^e$, can be obtained by minimizing $\Delta {\cal F}_u$ with respect to $u_c$ (\emph{i.e.}, $\frac{\partial \Delta {\cal F}_u}{\partial u_c}|_{u_c=u_c^e}=0$), which is equivalent to force balance equations. For cells with high tendency of homeostatic displacements (\emph{i.e.}, $\gamma_u\gg k_{\rm eff}$), $u_{c}^e$ is very close to $u_{c0}$ even in the presence of mechanical perturbations. The stiffness of the total energy near equilibrium displacement $u_{c}^e$, defined by $\frac{\partial^2 \Delta {\cal F}_u}{\partial u_c^2}|_{u_c=u_c^e}$, increases from $k_{\rm eff0}$ to $k_{\rm eff}+\gamma_u \gg k_{\rm eff0}$ 
    \begin{equation}\label{eq:CellCellInt-1DCellMatrix-ku}
    k_u=\left.\frac{d^2\Delta {\cal F}_f}{du_c^2}\right|_{u_c=u_c^e}=2k_{\rm eff}(1+\gamma_u/k_{\rm eff}) \gg k_{\rm eff0},
    \end{equation}        
    after the boundary matrix is stiffened as shown in Fig.~\ref{fig:CellCellInt-1DCellMatrix}. Physically, in the symmetric scenario considered here, the interactions (repulsion or attraction) between cells are hinted at the change in the energy stiffnesses, $k = {d^2\Delta {\cal F}_f}/{du_c^2}$. A larger energy stiffness at the perturbed state means that for a given deviation of cell boundary displacement from $u_{c}^e$ (induced by stochastic lamellipodia protrusion), the restoring force that opposes the protrusion force is larger under the mechanical perturbation of a stiffened boundary. This indicates that the cell has a tendency to stay away from the stiffened boundary, or there exists an effective \emph{repulsion} of the cell from the stiffened boundary. 
    
    \item \emph{Homeostatic force with $f(R_c)=-f_c\to -f_{c0}<0$ and $f(-R_c)=f_c\to f_{c0}>0$ being fixed under the perturbation of boundary stiffening.} -- One can also introduce a phenomenological energy that characterize cell homeostasis that keeps the contraction force $f_c$ to its set-point magnitude, $f_{c0}$, at unperturbed states. Then the change of total energy at the perturbed state is given by
    \begin{equation}\label{eq:CellCellInt-1DCellMatrix-Ff}
    \Delta \mathcal{F}_f=\left(k_{\rm eff} u_{c}^{2}-k_{\rm eff0}  u_{c0}^{2}\right)-2 f_{c}\left(u_{c}-u_{c0}\right)+\gamma_f\left(f_{c}-f_{c0}\right)^{2}, 
    \end{equation}
    where the constant $\gamma_f$ characterizes the tendency of cells to attain the unperturbed set-point value of its contraction force. Minimization of $\Delta {\cal F}_f$ with respect to $u_c$ and $f_c$ gives the equilibrium displacement $u_{c}^e$ and force $f_c^e$, respectively. For cells with high tendency of homeostatic force (\emph{i.e.}, $\gamma_f\gg 1/k_{\rm eff}$), $f_c^e$ is very close to $f_{c0}$ even in the presence of mechanical perturbations and $u_{c}^{e} \sim u_{c0} k_{\rm eff0}/k_{\rm eff}\ll u_{c0}$. This means that cell contracts less after the boundary matrix is stiffened as shown in Fig.~\ref{fig:CellCellInt-1DCellMatrix}, which indicates a forward adhesion movement and predicts an effective \emph{attraction} of the cell to the stiffened boundary. Note that if a homeostatic force can be achieved during the protrusion, then the cellular force relaxes quickly to $f_c^e (u_c)$, which is obtained from $\partial \Delta {\cal F}_f/\partial f_c=0$. Substituting it into Eq.~(\ref{eq:CellCellInt-1DCellMatrix-Ff}) and obtain $\Delta {\cal F}_f$ as a function of $u_c$ only. The stiffness of the energy near $u_c=u_c^e$ is then given by
    \begin{equation}\label{eq:CellCellInt-1DCellMatrix-kf}
    k_f=\left.\frac{d^2\Delta {\cal F}_f}{du_c^2}\right|_{u_c=u_c^e}=2k_{\rm eff}(1-1/\gamma_f k_{\rm eff}),
    \end{equation}    
    which scales as $k_f\sim 2k_{\rm eff}$ for large $\gamma_f \gg 1/k_{\rm eff}$, also greater than the stiffness $2k_{\rm{eff0}}$ given near Eq.~(\ref{eq:CellCellInt-1DCellMatrix-F0}) at the unperturbed state.
\end{itemize}

\subsection{Mechanical interactions in an array of contractile cells embedded in three-dimensional matrix}\label{sec:CellCellInt-3DCellMatrix}

Now we consider the matrix-mediated interactions between contractile cells that are embedded in a three-dimensional matrix. A direct calculation of the interaction between two such cells including homeostatic conditions of fixed displacement or stress at the boundary of each cell in the presence of the other, is complex and requires one to consider an infinite series of ``induced force dipoles". Golkov et.al \cite{Shokef2017,Shokef2019} modelled cells as spherical active force dipoles surrounded by an infinite elastic matrix, and analytically evaluated the interaction energy for different homeostatic behaviors. They start from the mechanical equilibrium of displacements around each force dipole generated by spherical force dipoles. This case is analogous to a force dipole moment on one cell that is induced by the stress field due to its neighbor, which yields the elastic analogy of the van der Waals interaction. They found that the interactions between two such cells, each of which maintains a fixed displacement at its boundary even in the presence of the other, is \emph{repulsive}. In contrast, the interaction between cells that maintain a fixed stress at their boundary is \emph{attractive}. The interaction energy decreases with cell separation $d$, as $1/d^6$, which is reminiscent of the van der Waals interaction; both effects are due to induced polarization of the dipoles in the two bodies. Furthermore, for a three-dimensional array of cells with typical distance $d$ between neighbors, one can integrate this interaction energy over all the interacting pairs in the system and find that the incremental force scales as $1/d^3$. 

\begin{figure}
    \centering
    \includegraphics[width=0.9\linewidth]{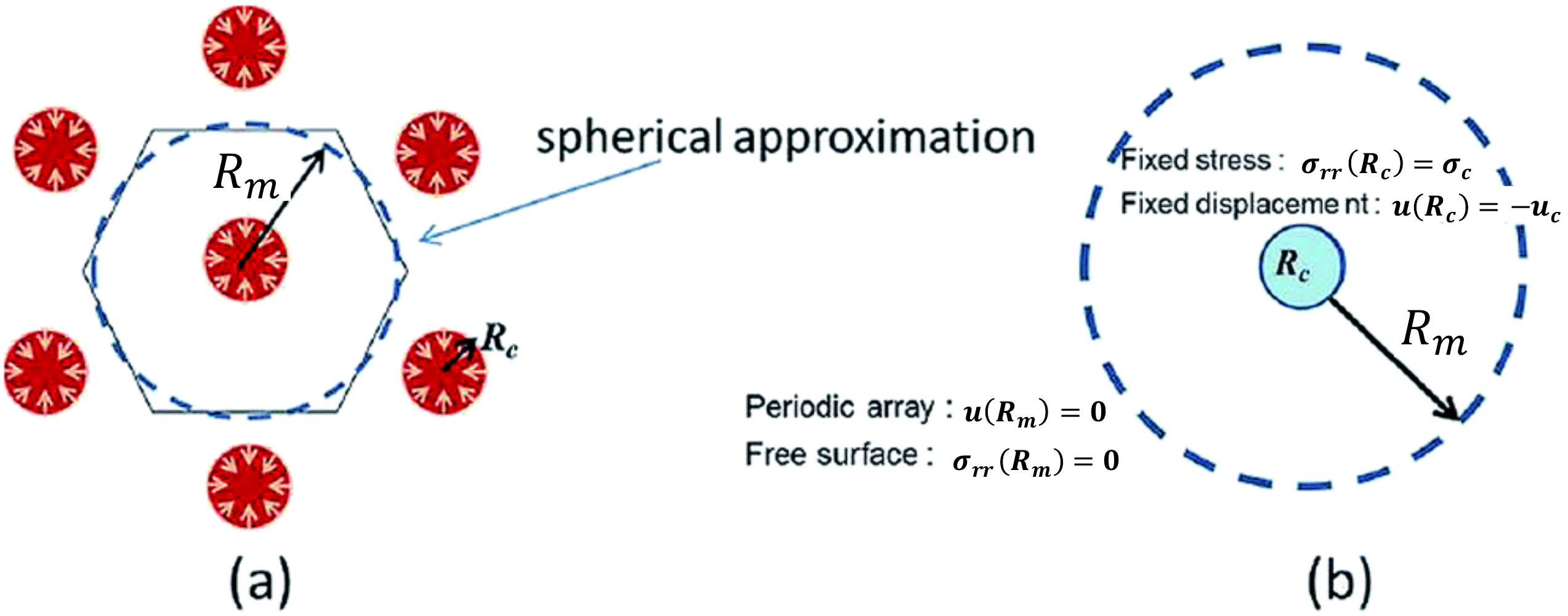}
    \caption{(a) A periodic array of spherical contractile cells embedded in a three-dimensional biopolymer gel. The periodic Wigner-Seitz unit volume is approximated by a sphere of radius $R_{m}$.
    (b) A spherical contractile cell of radius $R_{c}$ embedded in a matrix sphere of radius $R_{m}$. Homeostasis implies that the displacement, or strain, or stress that the cell applies at its boundary $r=R_{c}$ is fixed. The boundary conditions at $r=R_m$ for free surface (unperturbed state) and periodic array (perturbed state) are represented by $\sigma_{rr}(R_{m})=0$ and $u(R_{m})=0$, respectively. Reproduced from Ben-Yaakov \emph{et al.} \cite{Sam2015} with permission from the Royal Society of Chemistry.}
    \label{fig:CellCellInt-WScell}
\end{figure}

Here we focus on the simpler and more tractable geometry as studied by Ben-Yaakov \emph{et al.} \cite{Sam2015} and shown in Fig.~\ref{fig:CellCellInt-WScell}. Consider the interactions of a periodic array of spherically contracting cells in an elastic matrix. At the mid plane between cells – each of which pulls in opposite directions – the matrix displacement in the direction perpendicular to the mid plane is zero. The smallest volume contained by the intersection of all the mid plane boundaries (termed in condensed-matter physics, the Wigner-Seitz unit cell or bounding volume) is the region in which the elastic problem must be solved. For simplicity, they further replace this Wigner-Seitz unit volume by a spherical matrix of diameter $d=2R_m$ (see Fig. ~\ref{fig:CellCellInt-WScell}), which is proportional to the center-to-center distance between nearest-neighbor spherically contracting cells. 

In this cell-matrix composite system with spherical symmetry, the total elastic energy stored in the matrix due to cell contractility is given by ${\cal F}_e= \int_{\Omega} d^3{\mathbf{r}} F(\epsilon_{ik})$ with integration over the whole matrix space, denoted by ${\Omega}$, and $F(\epsilon_{ik})$ being the strain energy density. Particularly in linear isotropic materials with $F(\epsilon_{ik})=\frac{1}{2}\sigma_{ik}\epsilon_{ik}$, we have
\begin{equation}\label{eq:CellCellInt-3DCellMatrix-Fe}
{\cal F}_e = \frac{1}{2}\int_{\Omega}d^3{\mathbf{r}} \sigma_{ik}\nabla_{i}u_{k}
= \frac{1}{2}\oint_{\partial \Omega}dA(\hat{n}_i\sigma_{ik}u_{k})=2\pi R_c^2 u_c\sigma_c, 
\end{equation}
in which for contractile cells $u(R_c)=-u_c<0$ and $\sigma_{rr}(R_c)=\sigma_c>0$, $\hat{n}_i$ is the outward unit normal vector of matrix surfaces, and ${\partial \Omega}$ denotes the boundary surfaces of the matrix domain. 
Here we have used the symmetry of the stress tensor, integration by parts, and the equilibrium condition $\nabla \cdot \bm{\sigma}=0$. Furthermore, two particular boundary conditions are considered at the outer boundary of the matrix, $r=R_m$: (i) unperturbed state with $\sigma_{rr}(r=R_m)=0$, representing a free boundary or an elastically incoherent boundary, and (ii) perturbed state with $u(r=R_m)=0$, representing a clamped boundary or the mid plane of periodic array of cells. In both cases, the surface integral in Eq.~(\ref{eq:CellCellInt-3DCellMatrix-Fe}) vanishes at the outer matrix boundary and we therefore obtain the last equality in Eq.~(\ref{eq:CellCellInt-3DCellMatrix-Fe}).

At the unperturbed state with $\sigma_{rr}(r=R_m)=0$, the cell applies a homogeneous contractile stress, $\sigma_{c0}>0$, at its spherical boundary $r=R_c$ to the matrix. In this case, the elastic energy stored in the matrix is written in terms of the cell-boundary displacement $u_c$ as
\begin{equation}\label{eq:CellCellInt-3DCellMatrix-Fe0}
{\cal F}_{e0}=8\pi\mu_{\rm eff0}R_cu_{c}^{2},
\end{equation} 
with the effective modulus approximated by $\mu_{\rm eff0}\approx \mu [1-\frac{3K+4\mu}{3K}(R_c/R_m)^3]$ for $R_m\gg R_c$, and the total energy is given by
\begin{equation}\label{eq:CellCellInt-3DCellMatrix-F0}
{\cal F}_{0}= {\cal F}_{e0}-f_{c0}u_{c},
\end{equation} 
in which $f_{c0}=\sigma_{c0}4\pi R_c^2>0$ is the magnitude of the active force applied by contracting cell on the matrix. Minimization of ${\cal F}_{0}$ with respect to $u_c$ gives the force balance condition
\begin{equation}\label{eq:CellCellInt-3DCellMatrix-fefc0}
f_e-f_{c0}=\frac{d{\cal F}_{0}}{d u_c}=16\pi\mu_{\rm eff0}R_cu_{c}-\sigma_{c0}4\pi R_c^2=0,
\end{equation} 
from which we obtain the equilibrium displacement $u_{c0}=\sigma_{c0}R_c/4\mu_{\rm eff0}\approx (\sigma_{c0}R_c/4\mu) [1+\frac{3K+4\mu}{3K}(R_c/R_m)^3]$ for $R_m\gg R_c$. Here $f_e=-\frac{\partial {\cal F}_{e0}}{\partial u(R_c)}=\frac{d{\cal F}_{e0}}{du_c}=16\pi\mu_{\rm eff0}R_cu_{c}$ is the elastic restoring force of the matrix. 
The total restoring force that opposes the (forward) protrusions is $\delta (f_e-f_{c0})=k_{\rm eff0} \left(u_c-u_{c0}\right)$, in which the stiffness of the total energy around $u_{c0}$ is given by
\begin{equation}\label{eq:CellCellInt-3DCellMatrix-k0}
k_0=\left.\frac{d^2{\cal F}_{0}}{d u_c^2}\right|_{u_c=u_{c0}}=\left.\frac{d(f_e-f_{c0})}{du_c}\right|_{u_c=u_{c0}}=16\pi\mu_{\rm eff0}R_c.
\end{equation}  
Assuming that the protrusion forces are ineffective (due to friction and contractility) in causing cell motility at unperturbed states in the absence of the mechanical perturbations.

% the matrix elastic force $f_e$ and the cellular contractile force $f_c$ can quickly reach mechanical equilibrium and adapt displacement to equilibrium $u_{c0}$ with different specific mechanical perturbations. The stochastic protrusions introduce a deviation of cell boundary displacement $u_c$ from $u_c^{\rm{e}}$. 
% The time scale of protrusive forces ($\sim 1 {\rm s}$) is longer than the equilibration times ($\sim 1 {\rm \mu s}$) of contractile forces and mechanical perturbations. 

When some mechanical perturbation such as the presence of other cells, rigidity gradients or external stretch is applied, the cellular contractile force $f_c$ regulates and the matrix elastic force $f_e$ adapts quickly to reach new mechanical equilibrium. The protrusion forces $f_p$ introduce additional deviation of cell boundary displacement from equilibrium. Here we consider the perturbation by the presence of other cells, and in the simplified geometry with spherical symmetry the displacement at the outer boundary of the matrix is changed to be zero, \emph{i.e.}, $u(r=R_m)=0$, as shown in Fig.~\ref{fig:CellCellInt-WScell}.  In this case, the elastic energy stored in the matrix takes the same form as Eq.~(\ref{eq:CellCellInt-3DCellMatrix-Fe0}) for unperturbed states: 
\begin{equation}\label{eq:CellCellInt-3DCellMatrix-DFe}
{\cal F}_e=8\pi\mu_{\rm eff}R_cu_{c}^{2},
\end{equation} 
but the effective modulus $\mu_{\rm eff}$ is here approximated by $\mu_{\rm eff}\approx \mu [1+\frac{3K+4\mu}{4\mu}(R_c/R_m)^3]>\mu_{\rm eff0}$ for $R_m\gg R_c$. Note that the elastic energy of the spherical-unit-volume approximation of the many-cell, periodic system scales with $1/R_m^3 \sim 1/d^3$ which is consistent with the elastic energy calculated explicitly for two neighbouring cells by Golkov et.al \cite{Shokef2017,Shokef2019}. We now analyze the matrix mediated cell-cell interactions for three different homeostasis mechanisms as follows. 
\begin{itemize}
    \item \emph{Homeostatic displacement with $u(R_c)=-u_c\to-u_{c0}<0$ being fixed in the presence of other cells.} -- As in the previous 1D model, such a cell homeostatic condition can be characterized by a phenomenological ``potential" energy and the change of total energy is given by
    \begin{equation}\label{eq:CellCellInt-3DCellMatrix-Fu}
    \Delta {\cal F}_u ={\cal F}_{e}(u_c)-{\cal F}_{e0}(u_{c0})-\sigma_{c0}4\pi R_c^{2}(u_{c}-u_{c0})+\frac{1}{2}\gamma_u(u_{c}-u_{c0})^{2}, 
    \end{equation}  
    in comparison to Eq.~(\ref{eq:CellCellInt-1DCellMatrix-Fu}) for the 1D model. Minimization of $\Delta {\cal F}_u$ with respect to $u_c$ gives the force balance equation $f_e-f_c=0$ with $f_e=16\pi\mu_{\rm eff}R_cu_{c}$ and $f_c=\sigma_{c0}4\pi R_c^2-\gamma_u(u_{c}-u_{c0})$. From this equation, one obtains the equilibrium $u_{c}^e$, which can be very close to $u_{c0}$ for large $\gamma_u\gg R_c\mu_{\rm eff}$. That is, the adhesion displacement is fixed at a constant value independent of the boundary conditions far from the cell, including the presence of other cells in distance. In this case, the cell contraction force is regulated to balance the elastic force. The stiffness of the energy near $u_{c}^e$ is then given by
    \begin{equation}\label{eq:CellCellInt-3DCellMatrix-ku}
    k_u=\left.\frac{d^2\Delta {\cal F}_u}{du_c^2}\right|_{u_c=u_c^e}=16\pi\mu_{\rm eff}R_c+\gamma_u,
    \end{equation}
    which is greater than the stiffness $k_0$ given in Eq.~(\ref{eq:CellCellInt-3DCellMatrix-k0}) at the unperturbed state. This means that for a given deviation of cell boundary displacement from $u_{c}^e$ (induced by stochastic protrusions), the restoring force $\delta (f_e-f_c)$ that opposes the protrusion force is larger in the presence of other cells. This indicates that the cell has a tendency to stay away from other cells and there exists an effective \emph{repulsion} of the cell from others.
    % , which is consistent with the total elastic energy of this system which increases as the cells are brought closer together.

    \item \emph{Homeostatic volumetric strain with $\epsilon_c= \left(\frac{du}{dr}+\frac{2u}{r}\right)_{r=R_c} \to\epsilon_{c0}$ being fixed in the presence of other cells.} -- One can also introduce a phenomenological energy to characterize this strain homeostasis and the change of total energy is given by
    \begin{equation}\label{eq:CellCellInt-3DCellMatrix-Fepsilon}
    \Delta {\cal F}_{\epsilon}={\cal F}_{e}(u_c)-{\cal F}_{e0}(u_{c0})-\sigma_{c0}4\pi R_c^{2}(u_{c}-u_{c0})+\frac{1}{2}\gamma_{\epsilon} (\epsilon_c-\epsilon_{c0})^{2}. 
    \end{equation}
    At unperturbed and perturbed states, the volumetric strains are given by $\epsilon_{c0}=-\frac{4\mu}{3K}\frac{u_{c0}R_c^2}{R_m^3}$ and $\epsilon_c=\frac{u_cR_c^2}{R_m^3}$, respectively. The last term in the energy (\ref{eq:CellCellInt-3DCellMatrix-Fepsilon}) can then be replaced by $\frac{1}{2}\gamma_{\epsilon}\frac{R_c^4}{R_m^6} (u_c+\frac{4\mu}{3K}u_{c0})^{2}$, which takes similar form as Eq.~(\ref{eq:CellCellInt-3DCellMatrix-Fu}) for displacement homeostasis. As before, minimization of $\Delta {\cal F}_{\epsilon}$ with respect to $u_c$ gives the force balance equation $f_e-f_c=0$ with $f_e=16\pi\mu_{\rm eff}R_cu_{c}$ and $f_c=\sigma_{c0}4\pi  R_c^2-\gamma_{\epsilon}\frac{R_c^4}{R_m^6}(u_{c}+\frac{4\mu}{3K}u_{c0})$, from which one obtains the equilibrium $u_{c}^e$. Note that for large $\gamma_{\epsilon}\gg \mu_{\rm eff}R_m^6/R_c^3$, the equilibrium displacement $u(R_c)=-u_{c}^e$ at cell boundary is very close to $\frac{4\mu}{3K}u_{c0}>0$. This positive (\emph{i.e.}, along the radial direction) cell displacement indicates a forward adhesion movement  and an effective \emph{attraction} of cells to each other. In addition, in this case the stiffness of the energy near $u_{c}^e$ is given by
    \begin{equation}\label{eq:CellCellInt-3DCellMatrix-kepsilon}
    k_{\epsilon}=\left.\frac{d^2\Delta {\cal F}_{\epsilon}}{du_c^2}\right|_{u_c=u_c^e}=16\pi\mu_{\rm eff}R_c+\gamma_{\epsilon}\frac{R_c^4}{R_m^6},
    \end{equation}
    which is greater than the stiffness $k_0$ given in Eq.~(\ref{eq:CellCellInt-3DCellMatrix-k0}) at the unperturbed state. 
    
    \item \emph{Homeostatic stress with $\sigma_{rr}(R_c)=\sigma_c\to\sigma_{c0}>0$ being fixed in the presence of other cells.} -- A phenomenological energy can be introduced to characterize cell homeostasis \cite{Sam2007,Sam2009} that keeps the contraction stress $\sigma_c$ to its set-point magnitude, $\sigma_{c0}$, at unperturbed states and the change of total energy is given by
    \begin{equation}\label{eq:CellCellInt-3DCellMatrix-Fsigma}
    \Delta {\cal F}_{\sigma}={\cal F}_{e}(u_c)-{\cal F}_{e0}(u_{c0})-\sigma_{c}4\pi R_c^{2}(u_{c}-u_{c0})+\frac{1}{2}\gamma_{\sigma}(\sigma_{c}-\sigma_{c0})^{2}.
    \end{equation}   
    Minimization of $\Delta {\cal F}_{\sigma}$ with respect to $u_c$ and $\sigma_c$ gives the equilibrium displacement $u_{c}^e$ and stress $\sigma_c^e$. For large $\gamma_{\sigma} \gg \pi R_c^3/\mu_{\rm eff}$, $\sigma^e$ is very close to the cellular stress $\sigma_{c0}$ at unperturbed states, as indicated by the stress homeostatic condition. Moreover, in this limit, $u_{c}^e \sim \sigma_{c0}R_c/4\mu_{\rm eff}$ which is smaller than the cell boundary displacement $u_{c0}=\sigma_{c0}R_c/4\mu_{\rm eff0}$ with $\mu_{\rm eff}>\mu_{\rm eff0}$. That is, cell contracts less in the presence of other cells, which indicates a forward adhesion movement and an effective \emph{attraction} of cells to each other. 
% is consistent with the total elastic energy which decreases as the cells are brought closer together.     
    Note that if homeostatic stress can be achieved during the protrusion, that is, stress relaxes quickly to $\sigma_c^e (u_c)$ from $\partial \Delta {\cal F}_{\sigma}/\partial \sigma_c=0$, one substitutes it into Eq.~(\ref{eq:CellCellInt-3DCellMatrix-Fsigma}) and obtain $\Delta {\cal F}_{\sigma}$ as a function of $u_c$ only. The stiffness of the energy near $u_c=u_c^e$ is then given by
    \begin{equation}\label{eq:CellCellInt-3DCellMatrix-ksigma}
    k_{\sigma}=\left.\frac{d^2\Delta {\cal F}_{\sigma}}{du_c^2}\right|_{u_c=u_c^e}=16\pi\mu_{\rm eff}R_c\left(1-\frac{\pi R_c^3}{\gamma_{\sigma}\mu_{\rm eff}}\right),
    \end{equation}    
    which scales as $k_{\sigma}\sim 16\pi\mu_{\rm eff}R_c$ for large $\gamma_{\sigma} \gg \pi R_c^3/\mu_{\rm eff}$, also greater than the stiffness $k_0$ given in Eq.~(\ref{eq:CellCellInt-3DCellMatrix-k0}) at the unperturbed state.
\end{itemize}
Here we have summarized some of the results predicted from the generic theory proposed by Ben-Yaakov \emph{et al.} \cite{Sam2015} for the active responses of cells to mechanical perturbations of the surrounding matrix. We only focus on how cell contractility and homeostatic feedback effects together can explain the matrix-mediated cell-cell interactions in both one-dimensional toy model system and in a three-dimensional system with spherical geometry. Depending on the mechanical properties of the matrix and cell types, cells show different tendency of mechanical homeostatic behaviors, \emph{e.g.}, displacement homeostasis and stress homeostasis. In both one- and three-dimensional systems, the same conclusions about attractive or repulsive cell-cell interactions have been drawn for each specific homeostatic state. In the presence of other cells, for homeostatic displacement, cells have a tendency to stay away from other cells, that is, matrix mediates an effective repulsive cell-cell interactions. However, for homeostatic stress, cell adhesions are displaced towards neighboring cells and cells are attracted to nearby cells. In the theoretical work by Ben-Yaakov \emph{et al.} \cite{Sam2015} and later works by Golkov et.al \cite{Shokef2017,Shokef2019}, the authors have also discussed cell responses to other mechanical perturbations such as rigidity gradients or externally applied stretch. Finally, note that in reality, systems are typically not spherically symmetric so that the change in the force balance induced by the presence of other cells, or rigidity gradients, external stress only occur in specific directions. In that case, the protrusion forces will tend to move or reorient the cell in those directions where the local contractility forces have been sufficiently decreased by the mechanical perturbations of the matrix. The theoretical predictions of \textcolor{black}{cell-cell} attractions for either homeostatic stress or strain can be compatible with experimental observations~\cite{Reinhart2008} which showed attractions on soft substrates but short-ranged repulsion upon contact on rigid substrates. The predictions for homeostatic stress or strain (but not displacement) are also consistent with typical experiments on durotaxis~\cite{Lo2000} indicating that cells are attracted to the more rigid regions of their surroundings (corresponding to the outer fixed boundary with zero displacement in the spherical geometry shown in Fig.~\ref{fig:CellCellInt-WScell}(b)). 

\section{Concluding Remarks} \label{sec:Conclusion}
% The conclusions section should come in this section at the end of the article, before the Conflicts of interest statement.

In summary, we have reviewed some recent continuum elastic models for the transmission of forces in biopolymer gels with focus placed on models that are based on the assumption of small and affine deformations. Biopolymer gels are composed of crosslinked stiff semiflexible biopolymers with stiffness parameter $c=\ell_p/\ell_c>1$, that is, the persistence length $\ell_p$ is larger than the polymer contour length $\ell_c$. We start from a very brief review of the nonlinear elasticity of individual biopolymers that has been understood quite well and reviewed elsewhere \cite{MacKintosh2014,Meng2017}. We summarize the force-strain relations of stiff biopolymers upon both tension and compression. We compared several useful interpolations of the exact force-strain relations and concluded with a piecewise interpolation that is more tractable to the study of force transmissions in biopolymer gels. 

We next reviewed some popular constitutive models for the elastic responses of biopolymer gels to externally applied forces after a short review of the theory of linear isotropic and anisotropic elasticity. We focused on the 3-chain model of nonlinear stiffening-and-softening biopolymer gels that is proposed by Xu and Safran \cite{Xu2015PRE} based on the small-affine-deformation assumption. We showed that this simple model can well fit the experimental data of simple shear on some typical biopolymer gels. Furthermore, this model gives some predictions on the normal stress in simple shear of biopolymer gels that are consistent with previous experiments and simulations. This indicates that the phenomenon of negative Poynting effects or negative normal stress does not require large or non-affine deformations and is resulted from the asymmetric nonlinear elasticity of biopolymer gels to extension (stiffening) and compression (softening). We have also briefly reviewed some continuum models of biopolymer gels at large affine deformation. 

We then reviewed continuum models for the transmission of \textcolor{black}{internal active forces} induced by a spherically contracting cell that is embedded in three-dimensional biopolymer gels. The various scaling regimes for the decay of cell-induced displacements are reviewed for linear isotropic and anisotropic materials, for nonlinear compressive-softening and stiffening-and-softening biopolymer gels, respectively. We showed that the normalized cell contraction ${\cal A}_{b,\, s}= u_c/R_c\epsilon_{b,\, s}$ is an essential ``emergent" dimensionless parameter involved in the long-range transmission of cellular forces; it measures the degree of nonlinearity in the deformed gel. Once the cell-induced elastic anisotropy and nonlinearity is set by ${\cal A}_{b,\, s}= u_c/R_c\epsilon_{b,\, s}$ the decay of the displacements will be dictated accordingly, independent of the specific mechanical properties of individual fibers. 

After that, we considered how the transmission of forces in nonlinear biogel matrix can mediate the long-range mechanical interactions between adherent cells in a distance. We reviewed the generic and unified theory proposed by Ben-Yaakov \emph{et al.} \cite{Sam2015} to explain how adherent cells respond to mechanical perturbations of their surrounding matrix such as the presence of neighbouring cells, slowly applied stretch, or gradients of matrix rigidity. We considered matrix-mediated cell-cell interactions in two different systems: one-dimensional cell-matrix system that is simplified to be a toy cell-spring system, and three-dimensional array of cells that is approximated by a spherical unit of cell-matrix volume. Calculations using an energy approach instead of the original force approach in Ben-Yaakov \emph{et al.} \cite{Sam2015} showed that the cell-cell interactions depends sensitively on the specific homeostatic behaviors of cells that are determined collectively by complex signaling pathways through groups of proteins under various matrix conditions. For cells with homeostatic stresses or (volumetric) strains, cells tend to attract to each other, while for cells with homeostatic displacements, an effective repulsion is found between cells.

Below we make a few general remarks and outlook.

(i) \emph{Failure of the assumption of continuum and affine deformations.} As in other elastic continuum theories, each material element should be macroscopically small such that material properties and deformation are homogeneous, but microscopically large in comparison to the characteristic dimensions of materials' microstructure and constituent elements such that statistical averaging is meaningful and fluctuations are negligible. Therefore, when the characteristic structural length $\xi$ such as pore sizes, fiber-segment lengths, or lengths characterizing the heterogeneity of biopolymer gels, are large and comparable to cell sizes $R_c$ or cell-to-cell distances $d$, \emph{i.e.}, when $\xi\sim R_c, \, d$, continuum concepts will not be applicable, discrete fibrous nature of the gels becomes essential, and fluctuations are significant in determining the behaviors of force transmissions and self-organization of the cell aggregates. Furthermore, in this case, the displacement induced even by a contracting ``perfect" spheroid are found in experiments to be highly heterogeneous and nonaffine~\cite{Notbohm2017,Notbohm2018}. Therefore, it is necessary to combine multiple approaches (except for continuum theories) such as discrete fiber network simulation approaches~\cite{Lesman2018}, vertex-model simulations~\cite{Alt2017} and other coarse-grained multicellular simulations methods~\cite{Fujimori2019} to study the nonlinear mechanics of biopolymer gels, their interactions with cells and the self-organization of cell aggregates or tissues.

(ii) \emph{Constitutive models of biopolymer gels applicable to external multiaxial loads.} We have only reviewed the successes of the listed constitutive models of biopolymer gels to uniaxial extension or simple shear. Recent experiments \cite{Shenoy2019} showed that the transmission of external forces is anomalous for biaxial and triaxial loads because of strong coupling between various deformation modes due to bending, buckling, and stretching of the fibers. A phenomenological model has been proposed to explain these anomalies by Ban \emph{et al}. \cite{Shenoy2019} by generalizing Wang \emph{et al}. fiber-reinforced material model \cite{Shenoy2014b}. However, the applicability of other constitutive models has not been explored at all. 
 
(iii) \emph{Effects of elastic nonlinearity of biopolymer gels on cell-cell interactions.} In Sec.~\ref{sec:CellCellInt}, we have only considered how cell contractility and homeostasis can determine the interactions between cells that are embedded in linear isotropic materials. However, cells \emph{in vivo} are usually adhered to nonlinear extracellular matrix. It is worth exploring systematically how nonlinear elasticity can impact cell-cell interactions either for cells embedded in three-dimensional matrix \cite{NotbohmLesman2015}, or for cells adhered to the surfaces of nonlinear matrix \cite{Gao2014}. In these cases, competitions of multiple length scales are involved, \emph{e.g.}, cell radii $R_c$, cell-to-cell distances $d=2R_m$, and some length scales of the nonlinear matrix such as $R_b$ and $R_s$ due to buckling-induced softening and inextensibility-associated stiffening, respectively. 

(iv) \emph{Dynamics of cellular force transmission.} We have only considered the continuum elastic models for force transmission, which is relevant to short-time scale behaviors, for example, during the short period after seeding the cell in the biopolymer gel. The dynamic transmission of forces involves more physics, including the presence of various dissipation mechanisms \cite{Sam2013b} and transient soft crosslinkers \cite{MacKintosh2014,Meng2017}, etc.  

(v) \emph{Constitutive models for living tissues as active cell-matrix composite gels.} 
The theory of ``active gel" for cell-matrix composite systems has been found to be very successful in describing the structure and dynamics of living animal tissues during wound healing, tissue morphogenesis and embryo development \cite{Prost2015}. However, such theories are phenomenological and usually adopt a purely macroscopic point of view \cite{Lenz2019}, in which one of the major players, active stress, is usually assumed to be constant and cellular homeostatic feedback effects are often neglected completely. Furthermore, recent experiments have showed that although the semiflexible polymer networks that comprise the extracellular matrix soften under compression and stiffen under tension or shear, yet intact living tissues respond to external forces in a completely different manner: stiffen in compression but not in shear or extension \cite{Janmey2016Tissue,Janmey2019Tissue}. Therefore, a systematic study in both theories and experiments on the nonlinear constitutive relations characterizing the active cell-matrix composites (\emph{i.e.}, their mechanical responses to externally applied forces) will enable further applications of the active gel framework into nonlinear regimes that are inaccessible in their current form, and deepen our understanding of tissue rheology that emerges from an interplay between strain-stiffening polymer networks and volume-conserving cells within them \cite{Janmey2016Tissue,Janmey2019Tissue}.  

\appendix
\section*{Appendices} 
\addcontentsline{toc}{section}{Appendices}
\renewcommand{\thesubsection}{\Alph{subsection}} 
\renewcommand{\theequation}{\Alph{subsection}\arabic{equation}}
  % redefine the command that creates the equation no.
\setcounter{equation}{0}  % reset counter 
  
\subsection{Transtropic fiber networks with orientation-dependent fiber stiffness}\label{sec:app-Transtropic}

\subsubsection{Relations between sets of independent elastic coefficients}\label{sec:app-Transtropic-ElasticCoeff} 

For transtropic materials, the energy density function is given in terms of principal strains by Eq.~(\ref{eq:ExtForce-linaniso-Fc2}) as
\begin{align*} 
F= 
\frac{1}{2}c_1\epsilon_{1}^2
+ 2c_2(\epsilon_{2}+\epsilon_{3})^2 +c_3(\epsilon_{2}-\epsilon_{3})^2+2c_4\epsilon_{1}(\epsilon_{2}+\epsilon_{3}).
\end{align*}
The Young's moduli and Poisson's ratios of the transtropic fiber network can also be calculated from Eqs.~(\ref{eq:ExtForce-linaniso-sigmaepsilon}) and (\ref{eq:ExtForce-linaniso-epsilonsigma}) or directly from Eq.~(\ref{eq:ExtForce-linaniso-cEnu}) as 
\begin{subequations}\label{eq:app-Transtropic-ElasticCoeff-cE}
\begin{align}
   E_1&= c_1-c_4^2/c_2, \, 
   &E_2= \frac{16c_2c_3(c_1-c_4^2/c_2)}{2c_2(c_1-c_4^2/c_2)+c_1c_3}, \\
  & \nu_{12}=\frac{c_4}{4c_2}, \, 
   &\nu_{23}= \frac{2c_2(c_1-c_4^2/c_2)-c_1c_3}{2c_2(c_1-c_4^2/c_2)+c_1c_3},
\end{align} 
\end{subequations}
and $\nu_{21}=\nu_{12} E_2/E_1$, from which we obtain the parameter $g$ defined near Eq.~(\ref{eq:IntForce-linaniso-ueqn})
\begin{equation}\label{eq:app-Transtropic-ElasticCoeff-g}
g\equiv\frac{E_2(1-\nu_{12})}{E_1(1-\nu_{23})}=\frac{8c_2-2c_4}{c_1}.
\end{equation}

\subsubsection{Network model with uniform fiber orientations} \label{sec:app-Transtropic-UniFiberOrient}
It is interesting to note that a fiber network with (frozen) anisotropic elastic properties of transtropic materials can be constructed at least in finite element simulations. Goren \emph{et al.} \cite{Xu2020BJ} have constructed a two-dimensional transtropic fibrous network composed of linear fibers that are uniformly distributed in orientation and have orientation-dependent stiffness, $k_f$:   
\begin{equation*}
k_f=k_1\cos^2\theta +k_2\sin^2\theta
\end{equation*}
where $\theta\in[0,\pi)$ is the angle of fiber with respect to the axis of symmetry (say, $\hat{\mathbf{x}}_1$-axis). $k_1$ and $k_2$ are the two extrema of $k_f$ along longitudinal directions (\emph{i.e.}, along the symmetry axis with $\theta=0$) and transverse directions (with $\theta=\pi/2$), respectively. Such a network is anisotropic in elasticity but not in geometry (without collective fiber alignment).

If the fibers in the network are deformed affinely, the strain of a fiber with orientation $\theta$ and $\varphi$ is
\begin{equation}\label{eq:app-Transtropic-UniFiberOrient-epsilonf} 
\epsilon_f= 
\sqrt{\lambda_1^2\cos^2\theta + \lambda_2^2\sin^2\theta\cos^2\varphi+\lambda_3^2\sin^2\theta\sin^2\varphi}-1,
\end{equation}
where $\lambda_i\equiv 1+\epsilon_i$ ($i=1,2,3$) are the three principal extensions along the symmetry axis $\hat{\mathbf{x}}_1$, $\hat{\mathbf{x}}_2$, and $\hat{\mathbf{x}}_3$ in transverse isotropic plane, respectively. Then the deformation energy of the fiber is ${\cal F}_f(\theta,\tau,\epsilon_i)
=\frac{1}{2}k_f l_0^2\epsilon_f^2$ , with $i=1,2,3$,  $l_0$ being the contour length of fibers at undeformed networks. For small strains, fiber angle distribution stays to be more or less uniform. We then expand ${\cal F}_f(\theta,\varphi,\epsilon_i)$ in $\epsilon_i$ to quadratic order, integrate it for all fibers over fiber angles $\theta$  and $\varphi$ , and obtain the deformation energy density as a function of $\epsilon_i$: 
\begin{align}\label{eq:app-Transtropic-UniFiberOrient-F}
&F 
=\int_{0}^{2\pi}\int_{0}^{\pi} {\cal F}_f(\theta,\varphi,\epsilon_i)\frac{n_f}{4\pi}\sin\theta d\theta d\varphi
\approx \frac{n_fl_0^2}{35}\left[(\frac{5}{2}k_1+k_2)\epsilon_{1}^2\right. \nonumber \\
&\left.+(\frac{1}{2}k_1+\frac{2}{3}k_2) (3\epsilon_{2}^2+2\epsilon_{2}\epsilon_{3}+3\epsilon_{3}^2)+(\frac{1}{3}k_1+2k_2)\epsilon_{1}(\epsilon_{2}+\epsilon_{3})\right],
\end{align} 
with $n_f=N_f/V_0$ and $N_f$ being the volume density and total number of fibers. Note that the deformation energy density (\ref{eq:app-Transtropic-UniFiberOrient-F}) takes the form of Eq.~(\ref{eq:ExtForce-linaniso-Fc2}) and comparisons give 
\begin{equation}\label{eq:app-Transtropic-cmu} 
c_1=5\tilde{\mu}_1+2\tilde{\mu}_2, \,
c_2=c_3=\frac{1}{2}\tilde{\mu}_1+\frac{2}{3}\tilde{\mu}_2, \,
c_4=\frac{1}{6}\tilde{\mu}_1+\tilde{\mu}_2,  
\end{equation}
where we defined $\tilde{\mu}_i\equiv n_f k_il_0^2/35$, $i=1,2$.
From $F\left(\epsilon_{1}, \epsilon_{2}\right)$, we calculate the two principal (linear) elastic stiffnesses:
\begin{equation}\label{eq:app-Transtropic-UniFiberOrient-E1E2} 
E_{1}=\frac{\partial^{2} F}{\partial \epsilon_{1}^{2}}=\frac{n_{f} l_0^2}{16}\left(5 k_{1}+k_{2}\right), \quad
E_{2}=\frac{\partial^{2} F}{\partial \epsilon_{2}^{2}}=\frac{n_{f} l_0^2}{16}\left(k_{1}+5 k_{2}\right) 
\end{equation} 
from which we obtain
\begin{equation}\label{eq:app-Transtropic-UniFiberOrient-E2E1ratio}
\frac{E_{2}}{E_{1}}=\frac{\left(c_E k_{1}+k_{2}\right)}{\left(k_{1}+c_Ek_{2}\right)}
\end{equation} 
with $c_E=5.0$. Direct numerical simulations gives $c_E\approx 4.1$ by least-square fitting as shown in Fig.~\ref{fig:ExtForce-LinearAnisotropy}(a). 

Note that for fibers behaving as thin elastic rods of cross-section radius $a$, $k\sim E_fa^2/l_0$ and hence $\tilde{\mu}_i \sim E_f\phi_f$ with $E_f$ and $\phi_f\equiv n_fa^2l_0$ being the Young‘s modulus and volume fraction of fibers\cite{Xu2017PRE}, respectively. For fibers behaving as semiflexible polymers\cite{MacKintosh2014} of persistence length $l_p$, $k\sim T l_p^2/l_0^4$ and hence $\tilde{\mu}_i \sim n_f T l_p^2/l_0^2$. 

We now consider some particular limits of the elastic parameters. 

(i) In the isotropic limit of $k_1=k_2$ and hence $\tilde{\mu}_1=\tilde{\mu}_2$, we have $c_1=7\tilde{\mu}_1$, $c_2=c_3=c_4=7\tilde{\mu}_1/6$, and hence from Eqs.~(\ref{eq:app-Transtropic-ElasticCoeff-cE}) and (\ref{eq:app-Transtropic-ElasticCoeff-g}) we obtain: $E_1=E_2=35\tilde{\mu}_1/6$, $\nu_{12}=\nu_{21}=\nu_{23}=1/4$, and $g=1$. From Eq.~(\ref{eq:IntForce-linaniso-n}), we have $n=2$. 

(ii) In the limit of $k_1\gg k_2$ and hence $\tilde{\mu}_1\gg \tilde{\mu}_2$, we have $c_1\approx 5\tilde{\mu}_1$, $c_2=c_3\approx\tilde{\mu}_1/2$, $c_4\approx \tilde{\mu}_1/6$, and hence from Eqs.~(\ref{eq:app-Transtropic-ElasticCoeff-cE}) and (\ref{eq:app-Transtropic-ElasticCoeff-g}) we obtain: $E_1\approx 5\tilde{\mu}_1$, $E_2\approx 8\tilde{\mu}_1/3$, $\nu_{12}\approx 1/12$, $\nu_{21}\approx 2/45$, $\nu_{23}\approx 1/3$, and $g\approx 11/15$. From Eq.~(\ref{eq:IntForce-linaniso-n}), we have $n\approx 1.8$, which is smaller than $n=2$ in isotropic materials, indicating slower decay of cell-induced displacement. 

(iii) In the limit of $k_1\ll k_2$ and hence $\tilde{\mu}_1\ll \tilde{\mu}_2$, we have $c_1\approx 2\tilde{\mu}_2$, $c_2=c_3\approx2\tilde{\mu}_2/3$, $c_4\approx \tilde{\mu}_2$, and hence from Eqs.~(\ref{eq:app-Transtropic-ElasticCoeff-cE}) and (\ref{eq:app-Transtropic-ElasticCoeff-g}) we obtain: $E_1\approx \tilde{\mu}_2/2$, $E_2\approx 16\tilde{\mu}_2/9$, $\nu_{12}\approx 3/8$, $\nu_{21}\approx 4/3$, $\nu_{23}\approx -3/4<0$, and $g\approx 80/63$. From Eq.~(\ref{eq:IntForce-linaniso-n}), we have $n\approx 2.2$, which is larger than $n=2$ in isotropic materials, indicating faster decay of cell-induced displacement. 

\subsubsection{Special network model: 3-chain models}\label{sec:app-Transtropic-3Chain}

The 3-chain model was firstly proposed to study incompressible and isotropic elasticity of rubber with $\epsilon_{1}+\epsilon_{2}+\epsilon_{3}=0$,
in which case the the deformation energy density is:
\begin{equation}\label{eq:app-Transtropic-3Chain-F}
F = \frac{n_fl_0^2}{6}\left[k_1 \epsilon_{1}^2 + k_2 (\epsilon_{2}^2+\epsilon_{3}^2)\right]. 
\end{equation} 
This deformation energy can be modified to study weakly compressible materials usually by the two following methods.

(i) We added an additional energy contribution from volumetric deformation to the 3-chain energy (\ref{eq:app-Transtropic-3Chain-F}) by the simplest quadratic form as
\begin{equation}\label{eq:app-Transtropic-3Chain-Fcomp}
F = \frac{n_fl_0^2}{6}\left[k_1 \epsilon_{1}^2 + k_2 (\epsilon_{2}^2+\epsilon_{3}^2)\right]+\frac{\tilde{K}}{2}\left(\epsilon_1 +\epsilon_2+\epsilon_3\right)^2, 
\end{equation}
which takes the general energy form of Eq.~(\ref{eq:ExtForce-linaniso-Fc2}) and a direct comparison gives
\begin{equation}\label{eq:app-Transtropic-3Chain-cmuK}
c_1=2\tilde{\mu}_1+\tilde{K}, \quad
c_2=\frac{1}{4}(\tilde{\mu}_2+\tilde{K}),  \quad
c_3=\frac{1}{2}\tilde{\mu}_2, \quad
c_4=\frac{1}{2}\tilde{K}, 
\end{equation} 
with $\tilde{\mu}_i\equiv n_f k_il_0^2/6$, $i=1,2$. From Eq.~(\ref{eq:app-Transtropic-3Chain-Fcomp}), we calculate the bulk modulus given by $K=\tilde{K}+\frac{2}{9}(\tilde{\mu}_1+2\tilde{\mu}_2)$ and from Eqs.~(\ref{eq:app-Transtropic-ElasticCoeff-cE}), we obtain the Young's moduli and Poisson's ratios: 
\begin{align}\label{eq:app-Transtropic-3Chain-Enu}
   E_1&=\frac{2\tilde{\mu}_1\tilde{\mu}_2+(2\tilde{\mu}_1+\tilde{\mu}_2)\tilde{K}}{\tilde{\mu}_2+\tilde{K}}, &
  & E_2= \frac{2\tilde{\mu}_1\tilde{\mu}_2+(2\tilde{\mu}_1+\tilde{\mu}_2)\tilde{K}}{\tilde{\mu}_1+\frac{1}{2}(1+\tilde{\mu}_1/\tilde{\mu}_2)\tilde{K}}, \nonumber \\  
   &\quad \nu_{12}= \frac{\tilde{K}/2}{\tilde{\mu}_2+\tilde{K}}, &
   &\nu_{23}= \frac{\tilde{K}/2}{\tilde{\mu}_2+\frac{1}{2}(1+\tilde{\mu}_2/\tilde{\mu}_1)\tilde{K}},
\end{align} 
and $\nu_{21}=\nu_{12} E_2/E_1$. 
Particularly, in the isotropic limit with $k_1=k_2$ and hence $\tilde{\mu}\equiv \tilde{\mu}_1=\tilde{\mu}_2$, we obtain the classical Young's modulus and Poisson's ratio of isotropic materials as 
\begin{equation}\label{eq:app-Transtropic-3Chain-Enuiso}
E= \frac{9K\mu}{3K+\mu}, \quad
\nu=\frac{3K-2\mu}{2(3K+\mu)},
\end{equation} 
with the shear modulus, $\mu$, and bulk modulus, $K$, given by $\mu=\tilde{\mu}$ and $K=\tilde{K}+2\tilde{\mu}/3$, respectively.
Furthermore, from Eq.~(\ref{eq:app-Transtropic-3Chain-Enuiso}) we find the parameter $g$ (defined near Eq.~(\ref{eq:IntForce-linaniso-ueqn})):
\begin{equation}\label{eq:app-Transtropic-3Chain-g}
g=\frac{\tilde{\mu}_2+\tilde{K}/2}{\tilde{\mu}_1+\tilde{K}/2},
\end{equation} 
from which we can see that 
(a) $g \to 1$ in the incompressible limit with $\tilde{K}\gg \tilde{\mu}_1, \, \tilde{\mu}_2$, and  the decay of cell-induced displacement follows the classical scaling law $u\sim r^{-2}$ as in linear isotropic materials; 
(b) $g \to \tilde{\mu}_2/\tilde{\mu}_1$ in the limit of zero Poisson ratio (\emph{i.e.}, $\tilde{K}\to 0$), the decay of cell-induced displacement follows $u\sim r^{-m}$ where $m=\frac{1}{2}(1+\sqrt{1+8g}$ with $1<m<2$ if $\tilde{\mu}_2<\tilde{\mu}_1$, indicating slower decay and longer range force transmission (w.r.t. linear isotropic materials), and $m$ can be much larger than $2$ if $\tilde{\mu}_2>\tilde{\mu}_1$, indicating faster decay and shorter range force transmission. 

(ii) Alternatively, we can decompose the strain tensor into deviatoric part and volumetric part. The energy for the former part is obtained from the 3-chain model by considering the deviatoric strain of each chain. The energy for the latter part is taken as the simplest quadratic form. In this case, the total deformation energy is then given by 
\begin{equation}\label{eq:app-Transtropic-3Chain-Fcomp2}
F = \frac{n_fl_0^2}{6}\left[k_1 \tilde{\epsilon}_{1}^2 + k_2 (\tilde{\epsilon}_{2}^2+\tilde{\epsilon}_{3}^2)\right]+\frac{K}{2}\left(\epsilon_1 +\epsilon_2+\epsilon_3\right)^2, 
\end{equation}
with $\tilde{\epsilon}_i\equiv \epsilon_i-(\epsilon_1 +\epsilon_2+\epsilon_3)/3$ being the deviatoric (principal) strains. 

Note that the energy (\ref{eq:app-Transtropic-3Chain-Fcomp2}) also take the general energy form of Eq.~(\ref{eq:ExtForce-linaniso-Fc2}) and comparisons give 
\begin{equation}\label{eq:app-Transtropic-3Chain-cmuK2}
c_1=\frac{4}{3}\mu_{a}+K, \,
c_2=\frac{1}{4}\left(\frac{1}{3}\mu_{a}+K\right), \,
c_3=\frac{1}{2}\tilde{\mu}_2, \,
c_4=\frac{1}{2}\left(-\frac{2}{3}\mu_{a}+ K\right),
\end{equation} 
with $\mu_{a}\equiv (2\tilde{\mu}_1+\tilde{\mu}_2)/3$. 
From Eqs. 
 (\ref{eq:app-Transtropic-ElasticCoeff-cE}), we obtain the Young's moduli and Poisson's ratios: 
\begin{align}\label{eq:app-Transtropic-3Chain-Enu2}
   E_1&= \frac{9\mu_a K}{\mu_a+3K}, \quad 
   E_2= \frac{9\tilde{\mu}_2 K}{\tilde{\mu}_2+3K(3/4+\tilde{\mu}_2/4\mu_a)} \\
   \nu_{12}&= \frac{-2\mu_a+3K}{2(\mu_a+3K)} , \quad  
   \nu_{23}=\frac{-2\mu_a+3K(3\tilde{\mu}_a/2\mu_2-1/2)}{2\left[{\mu}_a+3K(3\tilde{\mu}_a/4\mu_2+1/4)\right]},
\end{align} 
and $\nu_{21}=\nu_{12} E_2/E_1$. Particularly, in the isotropic limit with $k_1=k_2$, we also obtain the Young's modulus and Poisson's ratio of isotropic materials in Eq.~(\ref{eq:ExtForce-liniso-E0nu0}) with $\mu=\mu_a= \tilde{\mu}_1=\tilde{\mu}_2$ being the shear modulus, and $K$ being the bulk modulus, respectively. 
Furthermore, from Eq.~(\ref{eq:IntForce-linaniso-ueqn}) we find $g=1$, which indicates that the decay of cell-induced displacement follows the classical scaling law $u\sim r^{-2}$ as in linear isotropic materials.

\section*{Conflicts of interest}
There are no conflicts to declare. 
%In accordance with our policy on %\href{http://www.rsc.org/journals-books-databases/journal-authors-reviewe%rs/author-responsibilities/#code-of-conduct}{Conflicts of interest} %please ensure that a conflicts of interest statement is included in your %manuscript here.  Please note that this statement is required for all %submitted manuscripts. If no conflicts exist, please state that ``There %are no conflicts to declare''.

\section*{Acknowledgements}
We thank Samuel Safran from Weizmann Institute of Science, Israel for his useful comments and suggestions. This work was supported in part by Grants No.~12004082 of the National Natural Science Foundation of China (NSFC), by Guangdong Province Universities and Colleges Pearl River Scholar Funded Scheme (2019), by 2020 Li Ka Shing Foundation Cross-Disciplinary Research Grant (No.~2020LKSFG08A), and by Featured Innovative Projects (No.~2018KTSCX282) and Youth Talent Innovative Platforms (No.~2018KQNCX318) in Universities in Guangdong Province. 

%%%END OF MAIN TEXT%%%


% \balance


%%%REFERENCES%%%
\bibliography{ForceTrans} %You need to replace "rsc" on this line with the name of your .bib file
\bibliographystyle{rsc} %the RSC's .bst file

\end{document}